\documentclass[pteplogo]{ptephy_v1}
\pdfoutput=1
\usepackage{lineno}
\usepackage{xcolor}
\usepackage{verbatim}
\usepackage{hyperref}
\usepackage{graphicx}
\usepackage{caption}
\usepackage{subcaption}
\usepackage{verbatim}
\usepackage{hepunits}
\usepackage{amsmath,amssymb}
\usepackage[symbol]{footmisc}
\usepackage{ulem}

\def\nuebar{\bar\nu_e}

\def\GEANTfour{\textsc{Geant4}}

\def\GC{$\gamma$-catcher~}
\begin{document}

\title{
RENE experiment for the sterile neutrino search
using reactor neutrinos% at Yeonggwang

%RENE (Reactor Experiment for Neutrinos and Exotics) experiment for the sterile neutrino search
%using reactor neutrinos at Yeonggwang

%Reactor Experiment for Neutrinos and Exotics (RENE) Experiment\\Technical Design Report (TDR)
} 
%\author[1]{{\color{red}PLEASE ADD OR UN-COMMENT YOUR NAME IN AUTHORS.TEX}\\}

\author[1]{Byeongsu Yang}

\author[1]{Da Eun Jung}
\author[1]{Dong Ho Moon}
\author[1]{Eungyu Yun}
\author[1]{HyeonWoo Park}
\author[1]{Jae Sik Lee}
\author[1]{Jisu Park}
\author[1]{Ji Young Choi}
\author[1]{Junkyo Oh}
\author[1]{Kyung Kwang Joo}
\author[1]{Ryeong Gyoon Park}
\author[1]{Sang Yong Kim}
\author[1]{Sunkyu Lee}

\author[2]{Insung Yeo}
\author[2]{Myoung Youl Pac}

\author[3]{Jee-Seung Jang}

\author[4]{Eun-Joo Kim}

\author[5]{Hyunho Hwang}
\author[5]{Junghwan Goh}
\author[5]{Wonsang Hwang}

\author[6]{Jiwon Ryu}
\author[6]{Jungsic Park}
\author[6]{Kyu Jung Bae}
\author[6]{Mingi Choe}
\author[6]{SeoBeom Hong}

\author[7]{Jubin Park}
\author[7]{Myung-Ki Cheoun}

\author[8]{Hyunsoo Kim}

\author[9]{Han Il Jang}

\author[10]{Dojin Kim}
\author[10]{Jonghee Yoo}
\author[10]{Seok-Gyeong Yoon}
\author[10]{Wonjun Lee}

\author[11]{Intae Yu}

\affil[1]{
    Center for Precision Neutrino Research (CPNR), Department of Physics, 
    Chonnam National University, 77 Yongbong-ro, Buk-gu, Gwangju 61186, Republic of Korea
    %\email{dhmoon@chonnam.ac.kr}
}

\affil[2]{
    Department of Radiology, Dongshin University,
    67 Dongshindae-gil,
    Naju, Jeollanam-do 58245,
    Republic of Korea
}
\affil[3]{
    Department of Physics and Photon Science, Gwangju Institute of Science and Technology,
    123 Cheomdangwagi-ro, Buk-gu,
    Gwangju 61005,
    Republic of Korea,
}
\affil[4]{
    Division of Science Education, Jeonbuk National University,
    567 Baekje-daero, Deokjin-gu,
    Jeonju, Jeollabuk-do 54896,
    Republic of Korea,
}
\affil[5]{
    Department of Physics, Kyung Hee University,
    26 Kyungheedae-ro, Dongdaemun-gu,
    Seoul 02447, Republic of Korea
    %\email{jhgoh@cern.ch}
}
\affil[6]{
    Department of Physics, Kyungpook National University,
    80 Daehak-ro, Buk-gu,
    Daegu 41566, Republic of Korea
    %\email{jungsic@knu.ac.kr}
}

\affil[7]{
    Origin of Matter and Evolution of Galaxy (OMEG) Institute, Department of Physics, Soongsil University, 369 Sangdo-ro, Dongjak-gu,  Seoul, 06978, Republic of Korea
}

\affil[8]{
    Department of Physics and Astronomy, Sejong University,
    209 Neungdong-ro, Gwangjin-gu,
    Seoul 05006, Republic of Korea
}
\affil[9]{
    Department of Fire Safety, Seoyeong Univsersity,
    1 Seogang-ro, Buk-gu,
    Gwangju 61268, Republic of Korea
}
\affil[10]{
    Department of Physics and Astronomy, Seoul National University,
    1 Gwanak-ro, Gwanak-gu,
    Seoul 08826, Republic of Korea
}

\affil[11]{
    Department of Physics, Sungkyunkwan University, Seobu-ro, Jangan-gu, Suwon-si, Gyeong Gi-do, 31206, Republic of Korea
    \email{dhmoon@chonnam.ac.kr, jhgoh@khu.ac.kr, jungsicpark@knu.ac.kr, kyujung.bae@knu.ac.kr}
}

%\affil[10]{
%    Center for Precision Neutrino Research,
%    Gwangju 61186, Republic of Korea
%}

%\emailAdd{jungsicpark@knu.ac.kr}

\begin{abstract}
This paper summarizes the details of the Reactor Experiment for Neutrinos and Exotics (RENE) experiment. It covers the detector construction, Monte Carlo (MC) simulation study, and physics expectations. 
The primary goal of the RENE project is to investigate the sterile neutrino oscillation at 
    $\Delta{m}^{2}_{41}\sim 2\,{\rm{eV}^{2}}$. 
which overlap with the allowed region predicted by the Reactor 
Antineutrino Anomaly (RAA). 
On the other hand, the STEREO and PROSPECT experiments have excluded certain regions of the parameter space with 95 \% confidence level (C.L.), while the joint study conducted by RENO and NEOS suggests possible indications of sterile neutrinos at 
    $\Delta{m}^{2}_{41}\sim2.4\,{\rm{eV}^{2}}$ and 
    $\sim{1.7}{\,\rm{eV}^{2}}$ with sin$^{2}\theta_{41} < 0.01$.
Accordingly, a more meticulous investigation of these remaining regions continues to be a scientifically valuable endeavor.
This paper reports the technical details of the detector and physics objectives.
\end{abstract}

\maketitle

\newpage
%\tableofcontents

\newpage
\section{Introduction}

\subsection{Historical overview for sterile neutrino search}
%Neutrino physics provides new insights into the fundamental nature of elementary particles based on the standard model (SM). 
Neutrinos are fundamental particles within the Standard Model (SM) that play a key role in advancing our understanding of the universe and its underlying principles. 
Research on neutrinos has steadily progressed, contributing to new developments in particle physics. 
Advancements in neutrino detection technology have revealed that the three neutrino flavors ($\nu_e, \nu_\mu, \nu_\tau$) oscillate as they propagate through space. This discovery suggests that neutrinos possess masses and 
provides strong evidence for the existence of physics beyond the SM. 

%%% edited by KJB
%{\color{red} 
Within the SM framework, the neutrino masses can be only introduced through the dimension-5 Weinberg operator since the weak charges of neutrinos prohibit simple mass terms at the renormalizable level~\cite{Intro1}. It is thus robust to consider a new physics origin of the neutrino masses. Furthermore, the neutrino mass eigenstates do not necessarily align with the weak interaction eigenstates, 
%and this misalignment cuases neutrino flavor oscillations when neutrinos are produced from the nuclear interactions. 
and consequently neutrinos oscillate into other flavors during their propagation. 
Once the neutrinos are detected at a distant place from the production point, 
it is possible to observe the nonzero probability of the appearance or disappearance of certain neutrino flavors.
Therefore, precision measurements of neutrino oscillations are essential for elucidating the neutrino masses and thoroughly investigating the physics beyond the SM.

The probability of the neutrino flavor change is given by
\begin{eqnarray}
P_{\nu_{\alpha}\to\nu_{\beta}}
%&=&\sum_{k,j}^n U_{\alpha k}U^*_{\beta k}U_{\alpha j}U^*_{\beta j}
%\exp\left(-i\frac{\Delta m_{kj}^2L}{2E}\right)\nonumber\\
&=&\delta_{\alpha\beta}-4\sum_{k<j}^n{\rm Re}[U_{\alpha k}U^*_{\beta k}U_{\alpha j}U^*_{\beta j}]\sin^2\left(\frac{\Delta m_{kj}^2L}{4E}\right)\nonumber\\
&&+2\sum_{k<j}^n{\rm Im}[U_{\alpha k}U^*_{\beta k}U_{\alpha j}U^*_{\beta j}]\sin\left(\frac{\Delta m_{kj}^2L}{2E}\right)\,,
\end{eqnarray}
where $U$ is the Pontecorvo–Maki–Nakagawa–Sakita (PMNS) mixing matrix,
$\Delta m^2_{kj}=m^2_k-m^2_j$ is the difference of mass square,
$L$ is the baseline distance, and $E$ is the neutrino energy.
The sum runs over all neutrino states, so $n$ is 3 in the standard 3-$\nu$ paradigm.
For given $L/E$, the measurements of $P_{\nu_{\alpha}\to\nu_{\beta}}$ lead to the determination of the differences of mass square, and three mixing angles and one CP phase which parametrize the PMNS matrix.\footnote{An explicit parametrization can be found in a review~\cite{pdg}.}
Precision measurements regarding neutrino oscillations were conducted in the scenarios involving oscillations among three neutrino flavors. 
From these studies, one mass square difference$\Delta m_{21}^2$ and one mixing angle $\theta_{12}$ have been determined by the Kamioka Liquid Scintillator Antineutrino Detector (KamLAND) experiments~\cite{KamLAND03, KamLAND05, KamLAND08, KamLAND13}
and the solar neutrinos experiments~\cite{Davis:1968cp, CHLORINE, Kamiokande:1996qmi, Super-Kamiokande:1998qwk, Super-Kamiokande:2016yck, SNO:2001kpb, SNO:2002tuh, SNO:2011hxd}.
Another mass square difference $|\Delta m_{32}^2|$ (or equivalently $|\Delta m_{31}^2|$) and another mixing angle $\theta_{23}$ have been determined by the accelerator neutrinos~\cite{K2K:2006yov, MINOS:2014rjg, MINOS:2020llm, T2K:2011ypd, T2K:2013ppw, T2K:2023smv, NOvA:2019cyt, NOvA:2021nfi} and reactor neutrinos~\cite{RENO:2018dro, DayaBay:2022orm}
and also by the atmospheric neutrinos~\cite{Super-Kamiokande:1998kpq, Super-Kamiokande:2004orf, Super-Kamiokande:2017yvm, IceCubeCollaboration:2023wtb}.
%One mixing angle, $\theta_{12}$, was measured by solar neutrinos~\cite{CHLORINE, SK1, SK2, SK3, SNO1, SNO2} and the KamLAND~\cite{KamLAND}, and another, $\theta_{23}$, by atmospheric neutrinos~\cite{SK4, IceCube, NOnA} and the long-baseline accelerator Tokai to Kaminoka (T2K) experiment~\cite{T2K}. 
The third neutrino mixing angle $\theta_{13}$ has been determined by the Double Chooz~\cite{DChooz}, Daya Bay~\cite{DAYABAY1} and RENO~\cite{RENO1} experiments using commercial nuclear reactors at
$\sim1\,\mathrm{km}$ baselines, employing the liquid scintillator (LS) detector technology.

However, several anomalies have been reported in various experiments, such as the Liquid Scintillator Neutrino Detector (LSND)~\cite{LSND}, Mini Booster Neutrino Experiment (MiniBOONE)~\cite{MiniBooNE1, MiniBoone}, gallium anomaly~\cite{BEST1, GALLEX1, GALLEX2, SAGE1, SAGE2}, and reactor antineutrino anomaly (RAA)~\cite{Mention1, ILL, Bugey3, Bugey4, DoubleCh}. 
Despite efforts to fine theoretical contributions~\cite{StNTh1, StNTh2},
the discrepancies persist between and theoretically predicted $\nuebar$ ﬂuxes from nuclear reactors and experimental observations~\cite{StNTh3, StNTh4, StNTh5}.
These anomalies can be explained by introducing an additional type of neutrino, known as the sterile neutrino, which does not participate in weak interactions but can oscillate with other neutrino types~\cite{StNTh6}.

Inclusion of the sterile neutrino leads to the extension of the PMNS matrix containing the admixture of the 4th neutrino flavor,\footnote{An explicit parametrization is shown in recent reviews, {\it e.g.}, Ref.~\cite{Boser:2019rta}.} 
In this 3(active)+1(sterile) neutrino scheme~\cite{Goswami:1995yq,Okada:1996kw,Bilenky:1996rw,Bilenky:1999ny}, The neutrino survival probability in a short baseline experiment is given by%~\cite{SurProb},
\begin{equation}
\begin{aligned}
    P_{\nuebar \to \nuebar} \simeq \left|1 - \sin^2 2\theta_{14} \sin^2 \left(\frac{\Delta m^2_{41}L}{4E}\right)\right|
\end{aligned}
\label{eq:neuosc}
\end{equation}
%where $|U_{ei}|$ and $|U_{ej}|$ denote the elements of the neutrino mixing matrix, 
%$L$ represents the baseline distance between the reactor and detector, $E$ denotes $\nuebar$ energy, and $\Delta{}
%m_{ij}^2$ represents the mass splitting between the $i$-th and $j$-th mass eigenstates. 
%Further, the term 
The mass square difference $\Delta m_{41}^2=m_4^2-m_1^2$ is nearly the same as the sterile neutrino mass square, $m_s^2\sim 1\,{\text{eV}}^2$ because the expected sterile neutrino is much heavier than the other active neutrinos.
%The mixing angle $\theta_{14}$ denote 
%A nonzero $\Delta m^{2}_{41}$ suggests the existence of a sterile neutrino.
Therefore, if the sterile neutrino mixes with active neutrinos with a sizable mixing angle, $\theta_{14}$,
$P_{\nuebar \to \nuebar}<1$ can be observed in the short baseline experiments.
%}
%%% edited by KJB
%({\color{red} edited by KJB up to this point})

\begin{figure}[h]
\centering
\includegraphics[width=0.75\textwidth]{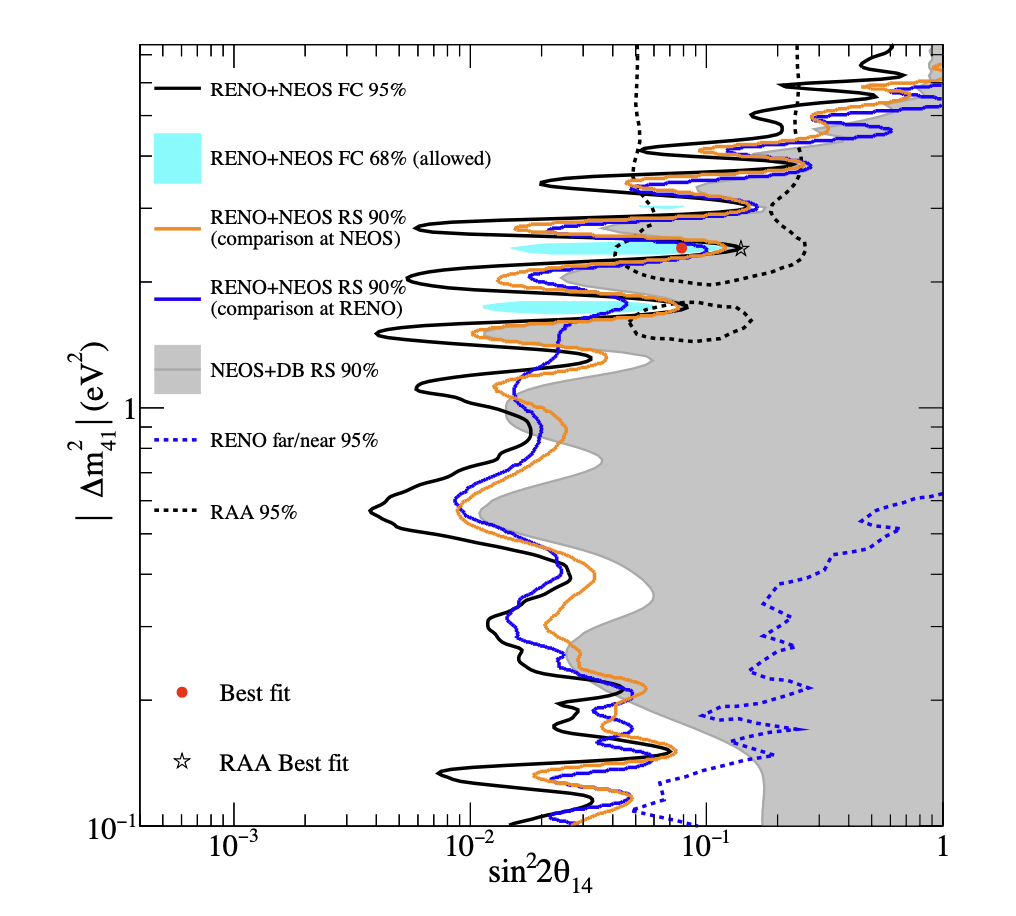}
\caption{\setlength{\baselineskip}{4mm} Exclusion limits and allowed regions for sterile neutrino oscillations. 
        The black curve represents the 95\% confidence level (CL) exclusion contour obtained from the Reactor Experiment for Neutrino Oscillation (RENO) 
        and Neutrino Oscillation and Neutrino Experiment for Oscillation at Short Bassline (NEOS) combined search using the Feldman-Cousins method, 
        while the cyan-filled area denotes to the 68\% CL allowed region. 
        The best-fit point (red dot) is found at $|\Delta{}m^2_{41}| = 2.41\,\rm{eV}^2$ and $\sin^2{}2\theta_{14} = 0.08$ ~\cite{NEOSRENO1}. 
}
\label{fig:exclusion_limit}
\end{figure}

%{\color{red}
The NEOS experiment~\cite{NEOS1} has searched for sterile neutrinos by measuring neutrino oscillations at a short baseline of approximately 24 m from one of the reactors used in the RENO experiment~\cite{RENO3}. The RENO and NEOS collaborations performed a joint analysis~\cite{NEOSRENO1}, which significantly constrained the allowed parameter space for sterile neutrinos, 
particularly in the region of $|\Delta m^2_{41}| \sim 2\,\rm{eV}^{2}$ and $0.01 < \sin^{2}(2\theta_{14}) < 0.1$, as indicated by the light blue region in Fig~\ref{fig:exclusion_limit}.
However, there are also results that disfavor the sterile neutrino hypothesis. The STEREO and PROSPECT collaborations have recently excluded the RAA-allowed region with a 95\% confidence level (C.L.) 
~\cite{STEREO, PROSPECT}. 
%[Nature 613 (2023) 257-261, Phys. Rev. Lett. 134 (2025) 151802]. 
This highlights the importance of conducting alternative meticulous investigations using independent different experimental setup such as detector configurations, baselines, and energy ranges, as
well as extending the search to the parameter space around $|\Delta{}m^2_{41}| = 2.41\,\rm{eV}^2$ and $1.7\,\rm{eV}^2$ with $\sin^2{}2\theta_{14} <$ 0.01. To enable such exploration, systematic improvements in the detector energy resolution are essential for enhancing sensitivity to these small mixing angle regions.
%}

%However, the results do not yet exclude the possibility of sterile neutrino survival. Further reduction of systematic uncertainties is needed to narrow down the remaining parameter space.

%\begin{figure}[!hbpt]\centering
%\includegraphics[width=0.75\textwidth]{Figures/Comparison of prompt-energy.png}
%\caption{\setlength{\baselineskip}{4mm} Comparison of the prompt-energy spectra at RENO and NEOS experiments. Upper: Ratio of the NEOS observed prompt spectrum relative to the 3$\nu$ best-fit prediction at NEOS from the RENO measurement. The error bars represent the statistical uncertainty only. The error bands represent the systematic and prediction uncertainties. The areas of the two spectra are normalized for shape comparison. The gray band indicates systematic uncertainty. Lower: Ratio of the NEOS 3$\nu$ best-fit prediction at RENO relative to the RENO observed prompt spectrum. The red and magenta curves represent the best fit for the data. The blue curves represent spectral ratios expected with one of the sterile neutrino oscillation parameters excluded by this analysis.} 
%\label{fig:prompt-energy}
%\end{figure}

\subsection{Motivation for developing a new detector}

\begin{figure}[h]
\centering
\includegraphics[width=1.0\textwidth]{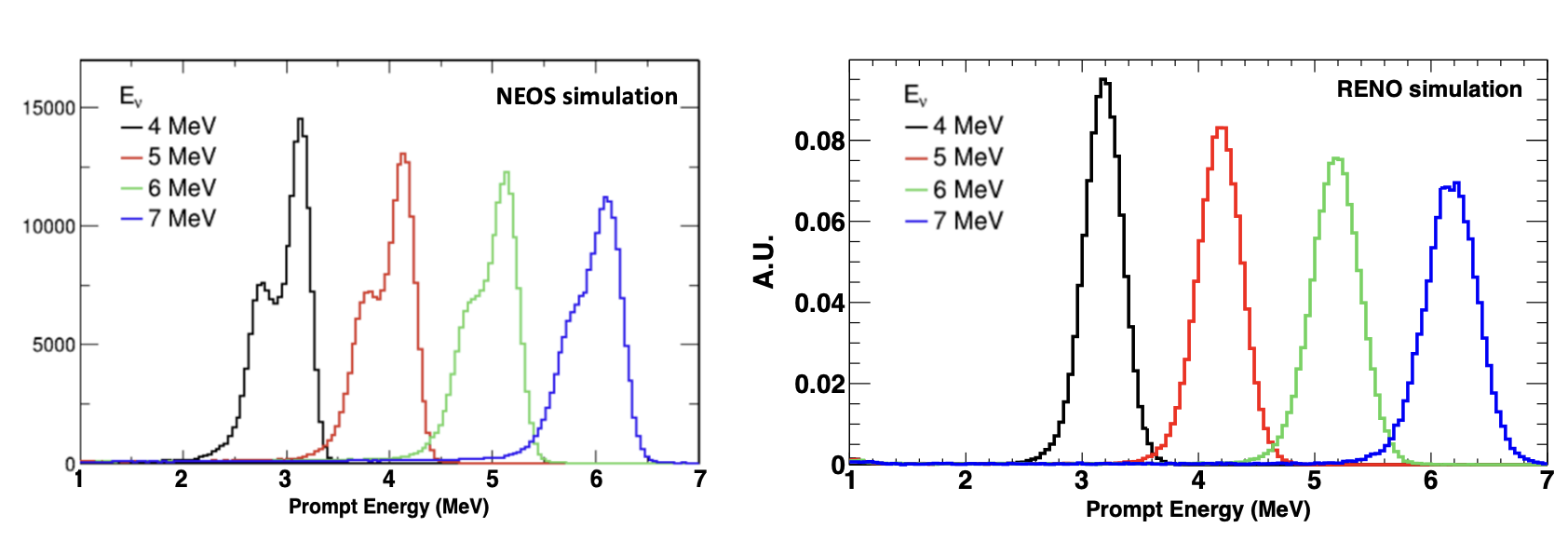}
\caption{\setlength{\baselineskip}{4mm} Prompt energy spectra corresponding to the neutrino energy of 4, 5, 6, and 7 MeV for NEOS design (left) and RENO design (right). 
}
\label{fig:2nd_peak_RENO_NEOS}
\end{figure}

In the NEOS experiment, the detector features a neutrino-target-only design to maximize the target volume, which allows some $\gamma$ rays to escape due to its compact size requirement.
The energy resolution of the detector could be improved by incorporating an additional active detector layer, known as the $\gamma$-catcher, around the target volume to recover the energy of escaping $\gamma$-rays. However, this would lead to a reduction in the target volume.
Figure~\ref{fig:2nd_peak_RENO_NEOS} shows prompt energy distributions obtained from the \GEANTfour-based Monte-Carlo (MC) simulations~\cite{Geant4Pap} of positrons corresponding to neutrino energies
ranging from 4 MeV to 7 MeV. 
The low-energy secondary peaks and tails can be observed in the NEOS detector due to the absence of a $\gamma$-catcher layer in contrast to the RENO detector. This suggests a possible opportunity to reduce systematic uncertainties.

\begin{figure}[h]
\centering
\includegraphics[scale=0.6]{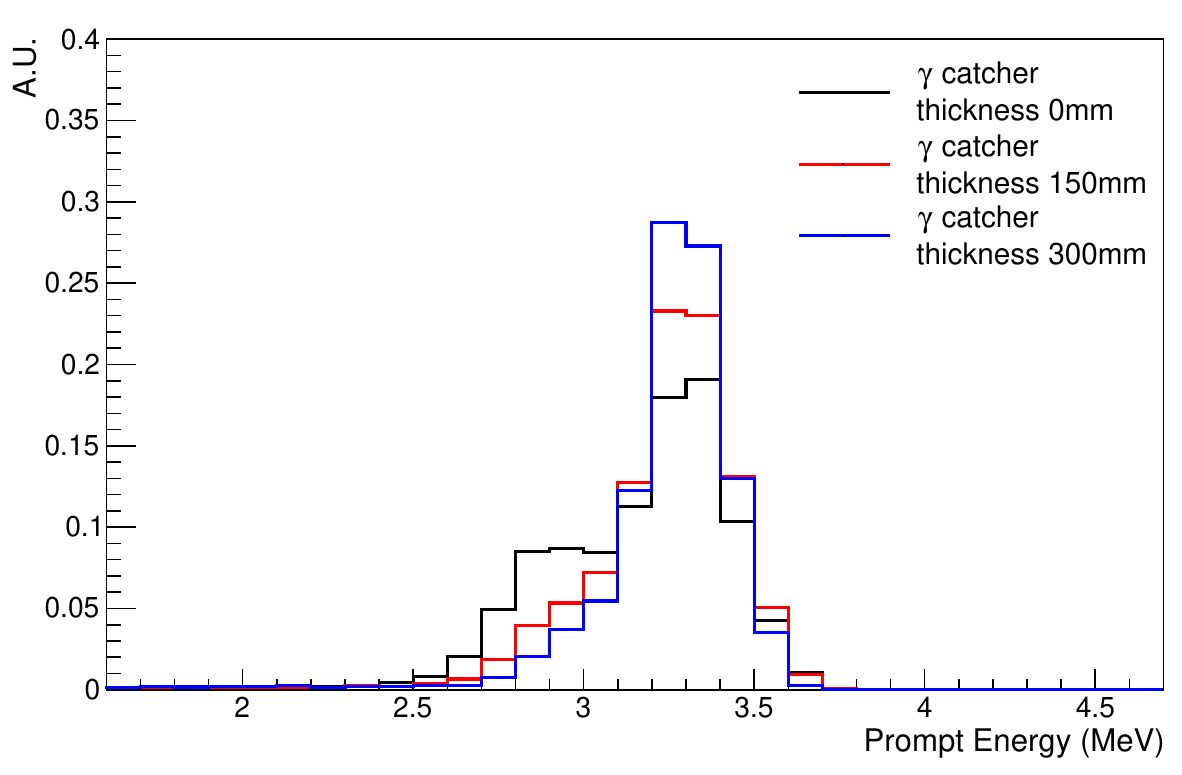}
\caption{\setlength{\baselineskip}{4mm}Prompt energy distributions for NEOS-sized detectors with various $\gamma$-catcher configurations. 
        A low-energy secondary peak is observed in the absence of a $\gamma$-catcher. 
        This peak diminishes with the addition of the $\gamma$-catcher, indicating a notable improvement in energy resolution.} 
\label{fig:positron_MC}
\end{figure}

Here, we propose a new reactor neutrino experiment named Reactor Experiment for Neutrino and Exotics (RENE). 
The initiative targets to explore the $|\Delta{}m_{41}^2| \sim 2\,\rm{eV}^2$ region corresponding to the parameter space allowed by the combined RENO and NEOS joint analysis. 
The key design feature of the RENE detector is improving energy resolution by increasing the thickness of the $\gamma$-catcher while maintaining an acceptable reduction in the available target volume.
Additionally, improved precision enables a better understanding of the neutrino energy spectrum contributed by the reactor fuel components of reactors helping clarify the origin of the 5 MeV
excess~\cite{RENO2, NEOSDeCompose} and establishing better theoretical constraints. To determine the optimal $\gamma$-catcher thickness for our physics goal, the MC simulation is performed as shown
in Fig.~\ref{fig:positron_MC}, which presents the prompt energy distributions of 4 MeV positrons for NEOS-sized generic detectors varying $\gamma$-catcher volumes. 
The low-energy secondary peak around 3 MeV diminishes notably with increasing $\gamma$-catcher thickness.
%The key design feature of the RENE detector is improving energy resolution by increasing the thickness of the $\gamma$-catcher, while maintaining an acceptable reduction in the available target volume. 
A thickness of 150 mm is deemed optimal for achieving the scientific goals of the RENE experiment.
More detailed study on energy resolution as a function of energy will be discussed in Section~\ref{sec::expected_results}.

\begin{figure}[h]
\centering
\includegraphics[width=.9\textwidth]{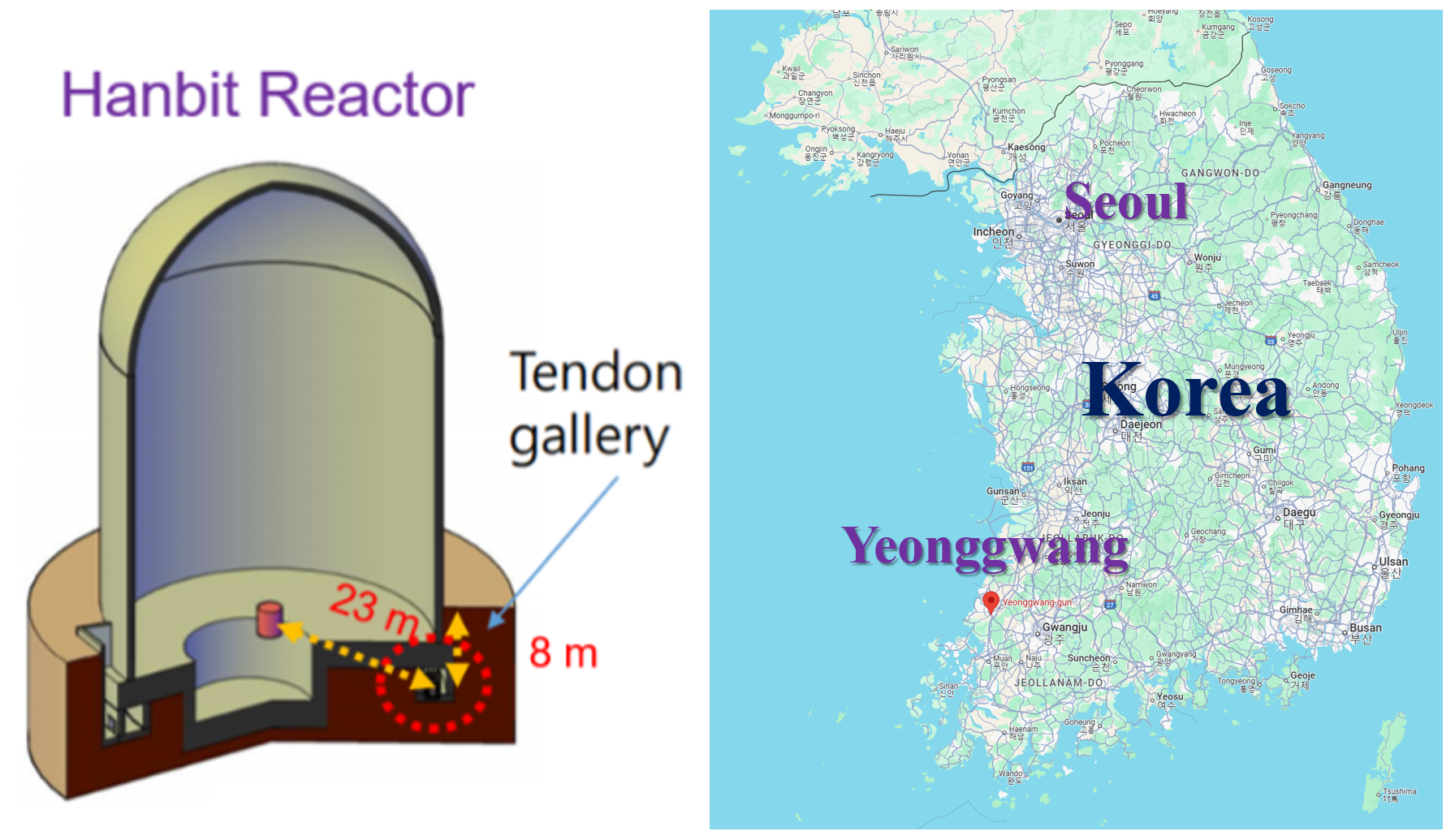}
\caption{\setlength{\baselineskip}{4mm} A schematic representation of a Hanbit reactor building (left) and the geographical location of the Hanbit reactor in Yeonggwang (right). The RENE detector will be installed in the tendon gallery.} 
\label{fig:site_location}
\end{figure}

The RENE experiment will be conducted at the same experimental site as the NEOS experiment specifically in the tendon gallery in one of the reactors in the Hanbit nuclear power plant in Yoeonggwang, Jeollanam-do, the Republic of Korea, as depicted in Fig.~\ref{fig:site_location}. %, approximately 250{\km} southwest of Seoul, on Korea's western coast. 
The power plant houses six Pressurized Water Reactors (PWR) fueled with low-enriched uranium dioxide (UO$_{2}$). 
The reactors generate an average thermal power output of 16.4$\,{GW_{th}}$ in total, with an average operational capacity factor exceeding over 90\% by the Korea Hydro \& Nuclear Power Co. Ltd.~\cite{KHNP}. 
The reactor is expected to serve as a stable and high-flux of antineutrinos during the data-taking periods.

The remainder of this paper is structured as follows: 
Section~\ref{sec::detector} offers a comprehensive overview of the proposed detector, including detailed descriptions of its components.
Section~\ref{sec::daq} outlines the data acquisition (DAQ) strategy and the trigger logic. 
Section~\ref{sec:detector_HVSCM} describes the Slow Control Monitoring system (SCM) and High Voltages (HV). 
%{\color{red}
Section~\ref{sec::expected_results} outlines our expectations regarding detector performance and sensitivity analysis results, including the detector resolution study.
%}
Finally, Section~\ref{sec::summary_plan} details the project plan.

\newpage
\section{Detector Description}
\label{sec::detector}
\subsection{Overall Design}
%The RENE detector consists of a liquid scintillator (LS), an acrylic vessel, a \GC, and two 20-inch photomultiplier tubes (PMTs) for the detection of electron antineutrinos ($\nuebar$) produced by nuclear fission within the reactor at the closest accessible baseline. 
The RENE detector is a neutrino detector based on the LS technology.
A gadolinium-loaded (Gd) liquid scintillator (Gd-LS) is selected as the target material and housed in the cylindrical-shaped acrylic target vessel.
Two 20-inch photomultiplier tubes (PMTs) are positioned at both ends of the target vessel.
The target vessel and the PMTs are installed in the box-shaped chamber built with stainless steel, which is the $\gamma$-catcher chamber,which is filled with the LS.
Thus, the target vessel and PMTs are immersed in the liquid.
Reflector cones are attached to each PMT to enhance overall detection efficiency.

\begin{figure}[h]
\centering
\includegraphics[width=\textwidth]{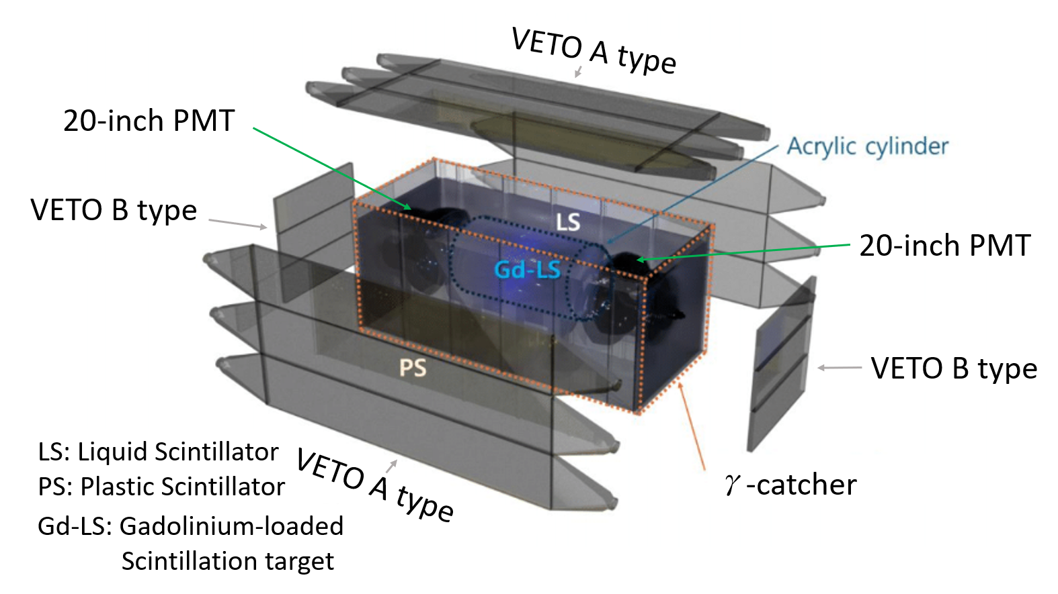}
\caption{\setlength{\baselineskip}{4mm} Conceptual illustration of the RENE detector. Details such as passive shielding, and supporting structures are omitted for clarity.}
\label{fig:detector_rene_isometric}
\end{figure}

The VETO system forms the outermost part of the RENE detector to reject the cosmic muon-induced events surrounding the $\gamma$-catcher chamber with the plastic scintillators.
%Two types of plastic scintillator (PS) panels, (A, B), are used to maximize the coverage.
To reduce environmental backgrounds, borated polyethylene (Boron-PE) and lead bricks are incorporated between the $\gamma$-catcher chamber and the VETO system as passive shielding components.
A conceptual drawing of the major components of the RENE detector is presented in Fig.~\ref{fig:detector_rene_isometric} 
%\textcolor{red}{
and summarized in Tab.~\ref{tab:detector_rene_isometric}
%}. %%% jspark added.

%Schematic drawing of fully assembled RENE detector. The order of assembly from the outside is the VETO

\begin{table}[h]
    \centering
    \begin{tabular}{ccc} \hline
         & Material & Amount \\
        \hline \hline
        Target  & Gd-LS & 270 L \\
        $\gamma$-catcher  & LS  & around 3000 L \\
        Veto  & Plastic scintillator  & Type A: 9 EA   \\
        Veto  & Plastic scintillator  & Type B: 6 EA\\
        \hline
    \end{tabular}
    \caption{Summary table with key features of the detector.}
    \label{tab:detector_rene_isometric}
\end{table}

The RENE detector is designed to detect electron antineutrinos ($\nuebar$) produced through nuclear fission within the reactor at the closest accessible baseline. 
When an electron antineutrino with an energy level above approximately 1.806 MeV interacts with a proton, it undergoes the inverse beta decay (IBD), producing positron and neutron ($\nuebar+p\to e^{+}+n$). 
The positron promptly generates $\gamma$-rays within $\sim20\,rm{ns}$ through electromagnetic interactions and pair annihilation (prompt signal), providing a primary measurement of the $\nuebar$ energy. 
The $\gamma$-rays are converted to optical photons in LS through the scintillation process and detected by photon sensors, such as the PMT.

\begin{figure}[h]
\centering
\includegraphics[width=.8\textwidth]{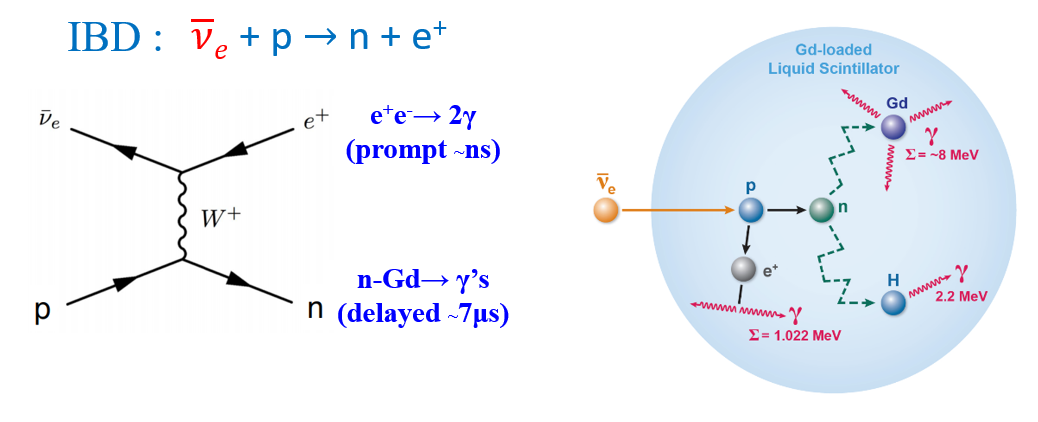}
\caption{\setlength{\baselineskip}{5mm} 
    Feynman diagram (left) and corresponding illustration (right)~\cite{IBDfigure} of electron antineutrino detection based on IBD events.} 
\label{fig:ibd}
\end{figure}

Meanwhile, after undergoing thermalization, various nuclei in LS capture the neutron, and subsequently, the nucleus emits $\gamma$-rays. 
The thermal neutron capture and subsequent decay occur with a mean time delay of 30 to 200$\,\mu s$ (delayed signal), depending on the type of capturing nucleus. 
Therefore, IBD events can be identified as two signals with a characteristic time window, as illustrated in Fig.~\ref{fig:ibd}.

Gd-LS has advantages over unloaded LS in neutron detection. 
The thermal neutron capture cross sections of Gd isotopes are significantly higher than that of free protons, which is the primary target of thermalized neutrons in unloaded LS. 
Additionally, the total energy released through $\gamma$-rays by Gd from neutron capture is approximately 8 MeV, compared to the 2.2 MeV released by proton capture. 
This higher energy output provides a clearer signal, enhancing discrimination against background noise from natural radioactive materials.

\begin{figure}[h]
\centering
\includegraphics[width=.6\textwidth,trim={5cm 6cm 9.5cm 4.5cm},clip]{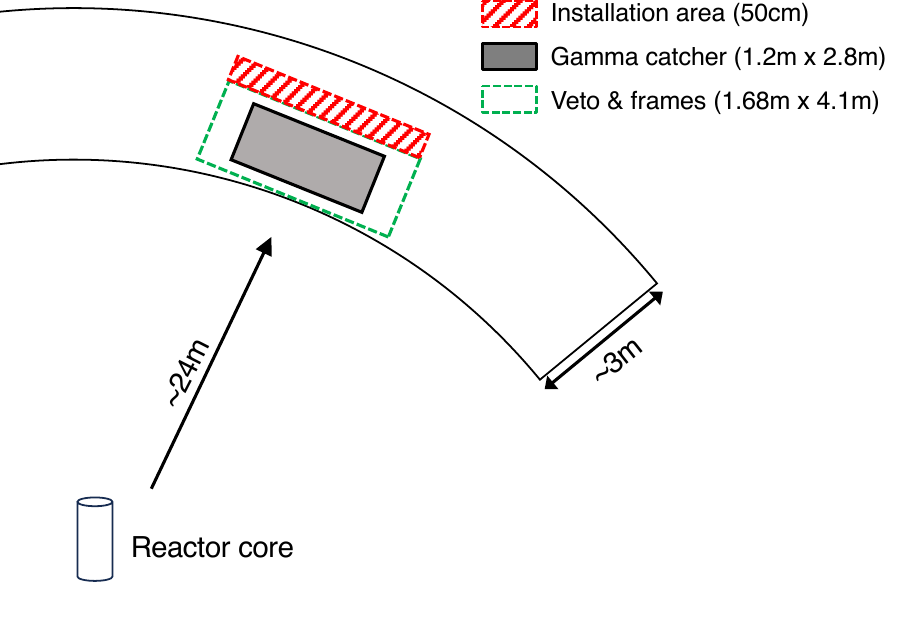}
\caption{\setlength{\baselineskip}{4mm} 
    Top-view layout of the experimental site. 
    The RENE detector will be installed in the underground tendon gallery of the nuclear reactor. 
    The detector system is designed to have a width of less than 1.9$\,$m and a height of less than 4.1$\,$m to ensure sufficient space for personnel and maintenance access. 
    } 
\label{fig:site_schematic}
\end{figure}

The experimental site is located in the tendon gallery of the Hanbit Nuclear Power Plant. 
It is a narrow underground corridor reserved for maintenance of the nuclear power plant building as depicted in Fig.~\ref{fig:site_schematic}. Owing to space limitations, the dimensions of the detector system are strictly constrained.
Access to this area requires special authorization with remote access being strictly prohibited due to the national security regulations.
Consequently, the detector system must be compact, safe, and capable of standalone operation.

\begin{figure}[hbt!]
\centering
\includegraphics[width=0.9\textwidth]{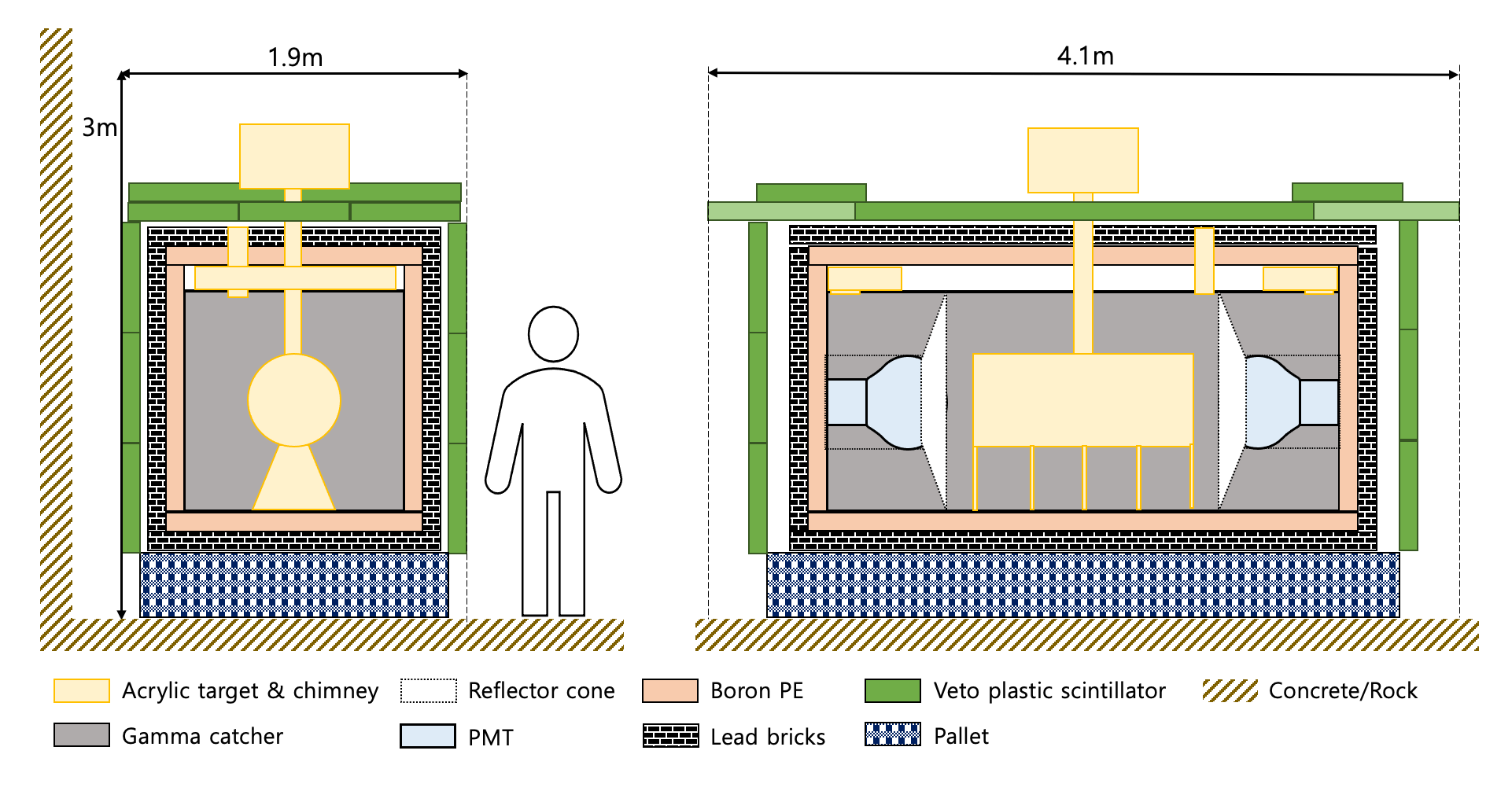}
\caption{\setlength{\baselineskip}{4mm} 
    Schematic representation of the RENE detector shown from a side (left) view and front view (right). 
    The detector comprises the target vessel (yellow) and the $\gamma$-catcher chamber (gray), along with reflector cones (white), and PMTs (light blue). 
    The chimney and buffer (yellow) are used for filling the LS and Gd-LS, or for source injection. 
    The target vessel and $\gamma$-catcher chamber are enclosed by layers of Boron-PE (light orange) and lead bricks (black) for passive shielding. 
    The outermost layer houses the VETO detector system (green). 
    The entire detector system is mounted on the pallet (blue).}
\label{fig:detector_rene_geometry}
\end{figure}

%% Describe about the overall structure again, but focus on the mechanical structures
%% Additionally, we keep description on the passive shielding

Figure~\ref{fig:detector_rene_geometry} presents a schematic diagram of the RENE detector system.
The RENE detector is installed near a wall in the tendon gallery.
To isolate this system from the ground, it is mounted onto a stable base structure called a pallet.
Multiple pillars and additional supporting frames are attached to the pallet to protect the detector system and support the VETO panels. 
The pallet and all supporting frames are made of stainless steel known as Steal Use Stainless (SUS) to enable the structure to bear the weight of the detector 
and remain unaffected by the surrounding magnetic field.
When installed on the pallet, the overall dimensions of the detector system are expected to be 3 m in height, 1.9 m in depth, and 4.1 m in length.

On the pallet, the lead bricks and Boron-PE will be placed as the passive shielding, with a thickness of 100 mm for each component. %sykim add 
Each of these lead bricks measures 50 mm in width, 200 mm in length, and 100 mm in height. 
The lead shielding at the bottom of the detector includes 264 bricks per layer, stacked in two layers, covering an area of 2.6 square meters. 
Similarly, the top shielding will include two layers, each containing 262 bricks, covering an area of 2.6 square meters. 
The side sections of the detector are shielded with a total of 1,470 bricks, covering an area of 14.7 square meters. 
Boron-PE material used in the shielding is a high-density polyethylene (HDPE) polymer containing more than 56 grams of boron per liter. 
The side and top sections of the detector are shielded using a total of nine cut panels, each measuring 100 mm in thickness, 1000 mm in width, 
and 2000 mm in length. Meanwhile, the bottom section is shielded using high-density polyethylene panels without boron, 
covering an area of 2.35 square meters. 
The combination of lead bricks and Boron-PE are expected to reduce environmental $\gamma$-rays and fast neutrons to below 1\% and 10\%, respectively~\cite{GammaShield,NuetronShield,NEOS2}.
The $\gamma$-catcher chamber is installed inside of the passive shielding.
It is covered by the $\gamma$-catcher lid, drilled with multiple holes for access into the chamber through chimneys.
The chimneys will serve as a passage to fill LS and Gd-LS, insert radioactive sources, and establish connections with PMTs and sensors.
To prevent leakage caused by the thermal expansion of the LS and Gd-LS, buffer tanks are installed on the lid.
The target vessel, PMTs, and reflectors are installed inside the $\gamma$-catcher chamber, with
various mechanical structures securing each component in place.

\begin{figure}[hpt]
\centering
\includegraphics[width=0.85\textwidth]{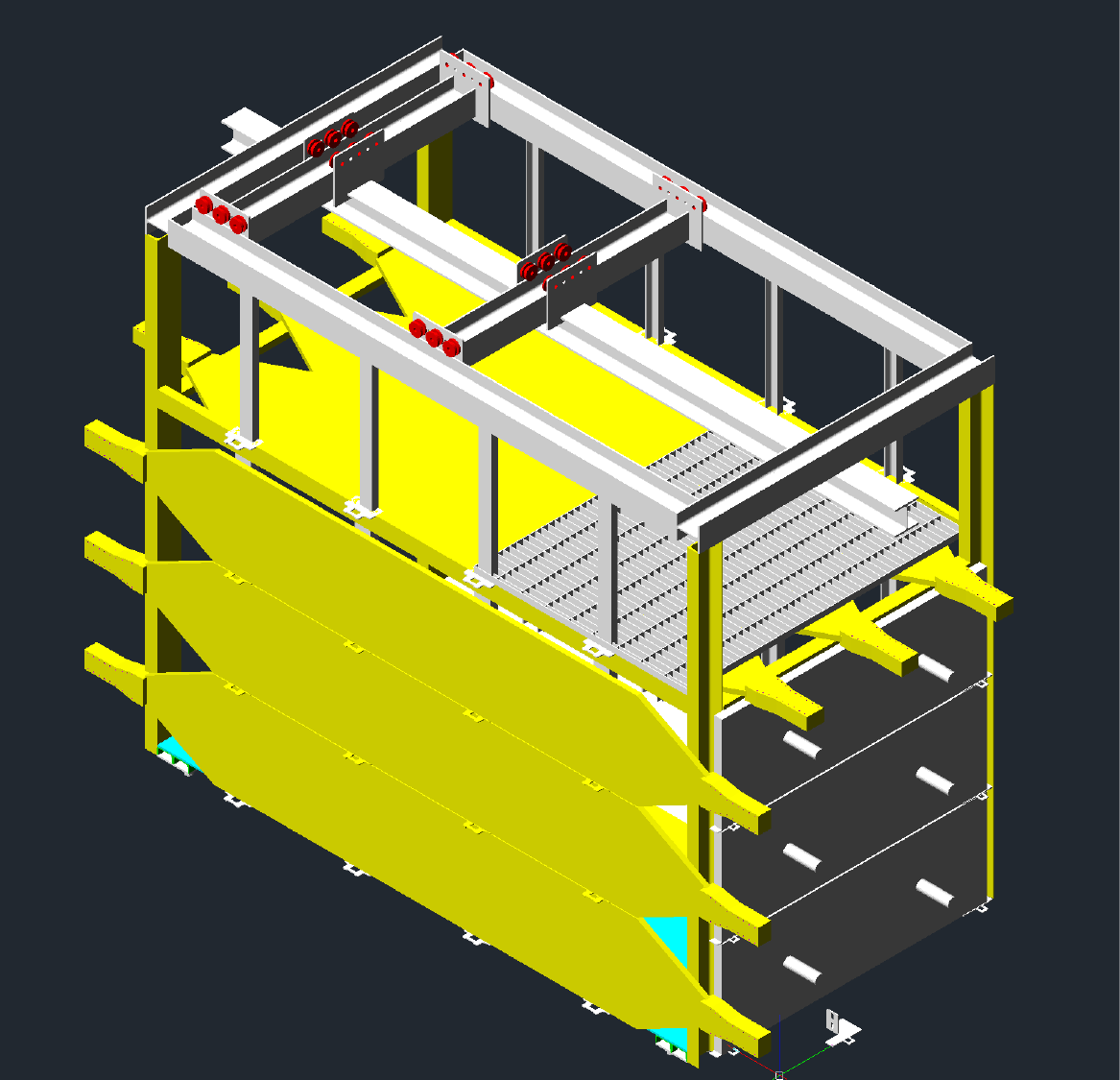}
\caption{\setlength{\baselineskip}{5mm}Isometric view of the RENE detector system.} 
\label{fig:shielding}
\end{figure}

\subsection{Target Vessel}
\label{sec:detector_at}
The target vessel is a cylindrical container filled with Gd-LS as the target material.
The material of the vessel is made from polymethyl methacrylate (PMMA), a type of acrylic, to ensure transparency for the optical photons 
and its long-term chemical compatibility with the LS and Gd-LS. 
The inner dimensions of the target vessel are 1200 mm in length and 534 mm in diameter, allowing it to hold 270 liter of Gd-LS. The vessel wall thickness of the target vessel of 8 mm, optimized to minimize $\gamma$-ray escape
while withstanding the internal pressure of the Gd-LS and LS.

\begin{figure}[h]
\centering
\includegraphics[width=0.8\textwidth]{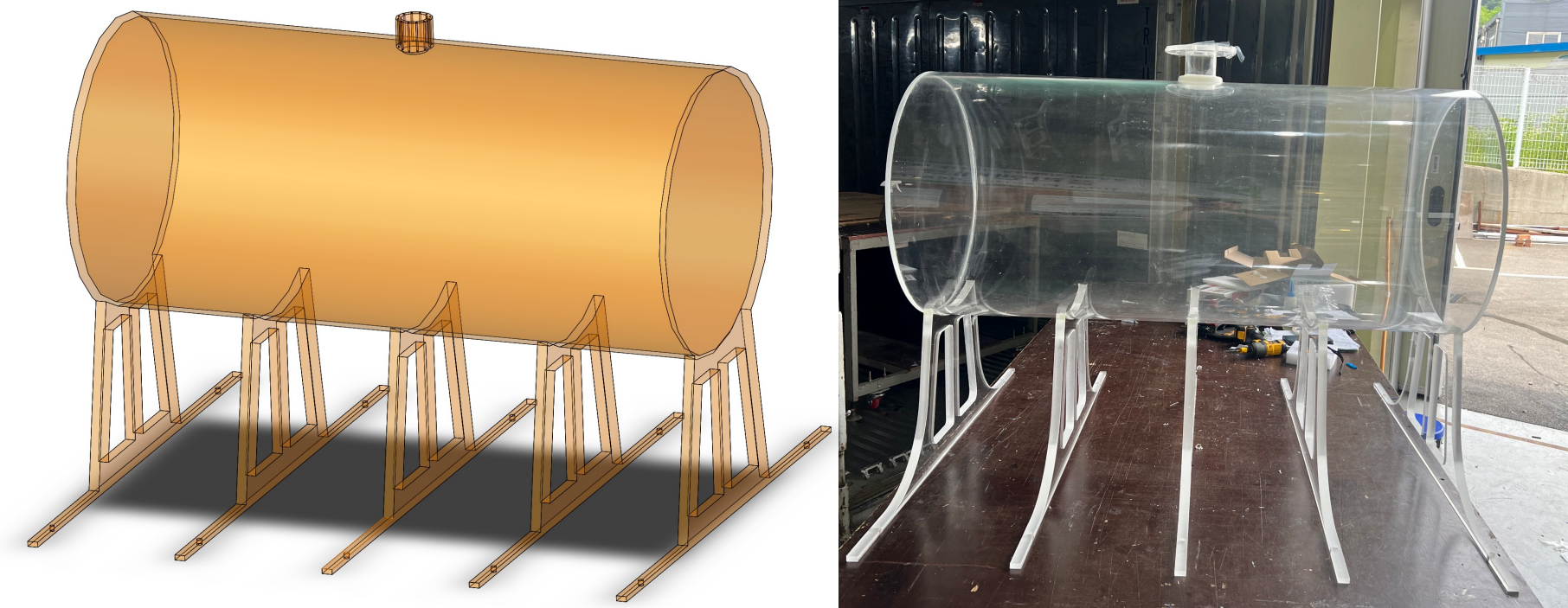}
\caption{\setlength{\baselineskip}{5mm} Isometric depiction of the acrylic target chamber (left) and the target chamber (right). %The main volume of the target chamber is covered with protection vinyl sheets before integration.
} 
\label{fig:detector_acylicTarget}
\end{figure}

\begin{figure}[h]
\centering
\includegraphics[width=0.85\textwidth]{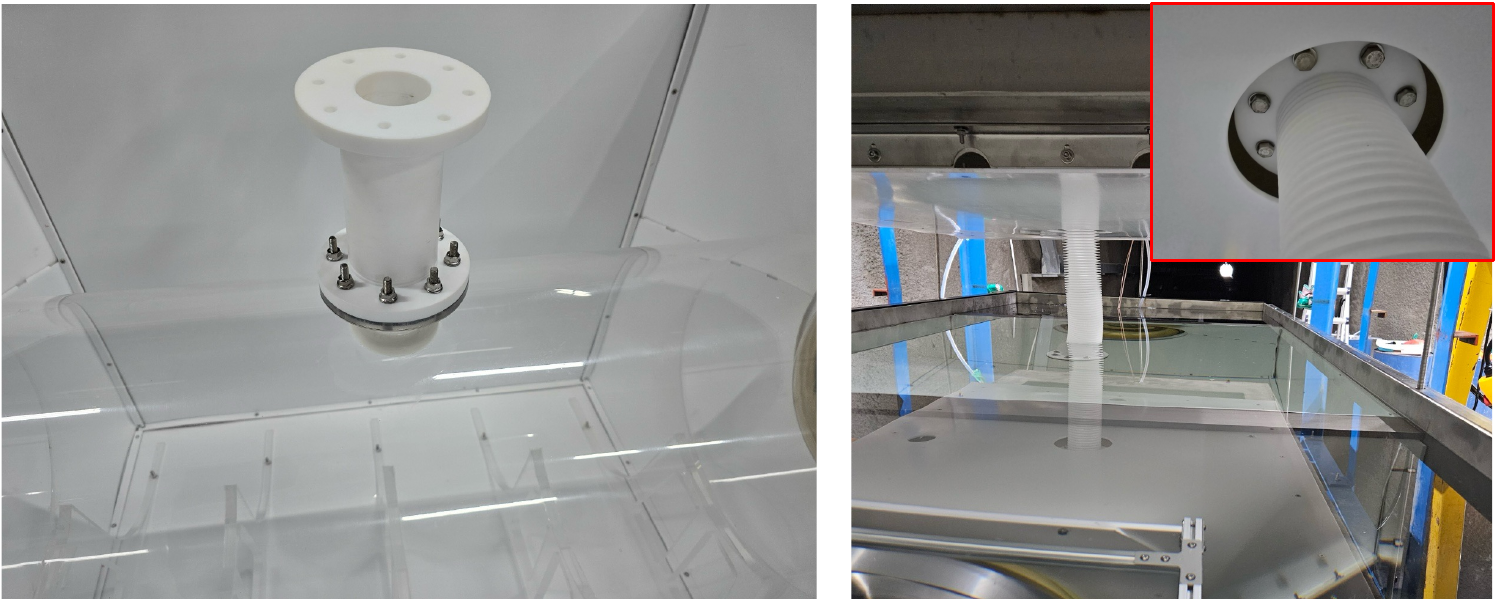}
\caption{\setlength{\baselineskip}{5mm} 
The lower end of the bellows is fastened to the flange at the central top of the acrylic target (left). Similarly, the upper end of the bellows is secured to the central flange of the detector lid, enabling flexible extension and contraction as the lid opens and closes (right). The inset image within the red box in the right figure shows a detailed view of the bellows assembly affixed to the lid.} 
\label{fig:bellows}
\end{figure}

%The vessel will be horizontally placed at the center of the detector. 
%The inner dimensions of the vessel are $1200\,{\mm}$ in length and $534\,{\mm}$ in diameter. 
A chimney is incorporated at the top of the vessel to facilitate Gd-LS filling and radiation source calibration, as depicted in Fig.~\ref{fig:detector_acylicTarget}.
The chimney has an inner diameter of 60 mm and an outer diameter of 80 mm. A teflon bellows serves as a conduit 
connecting the acrylic target vessel to the $\gamma$-catcher lid. When compressed, the bellows contracts to a length of 240 mm, 
and when fully extended, it reaches a length of 840 mm. As illustrated in Fig.~\ref{fig:bellows}, this flexibility allows the bellows 
to function as a chimney for the acrylic target vessel, enabling the controlled introduction of a radiation source from outside the detector 
for calibration purposes.

At the bottom of the vessel, five trapezoidal support structures with a wall thickness of 20 mm are attached to distribute the weight of the vessel.
Each support structure will include Windows designed to maximize scintillation probability and light yield.

\begin{figure}[h]
\centering
\includegraphics[width=.8\textwidth]{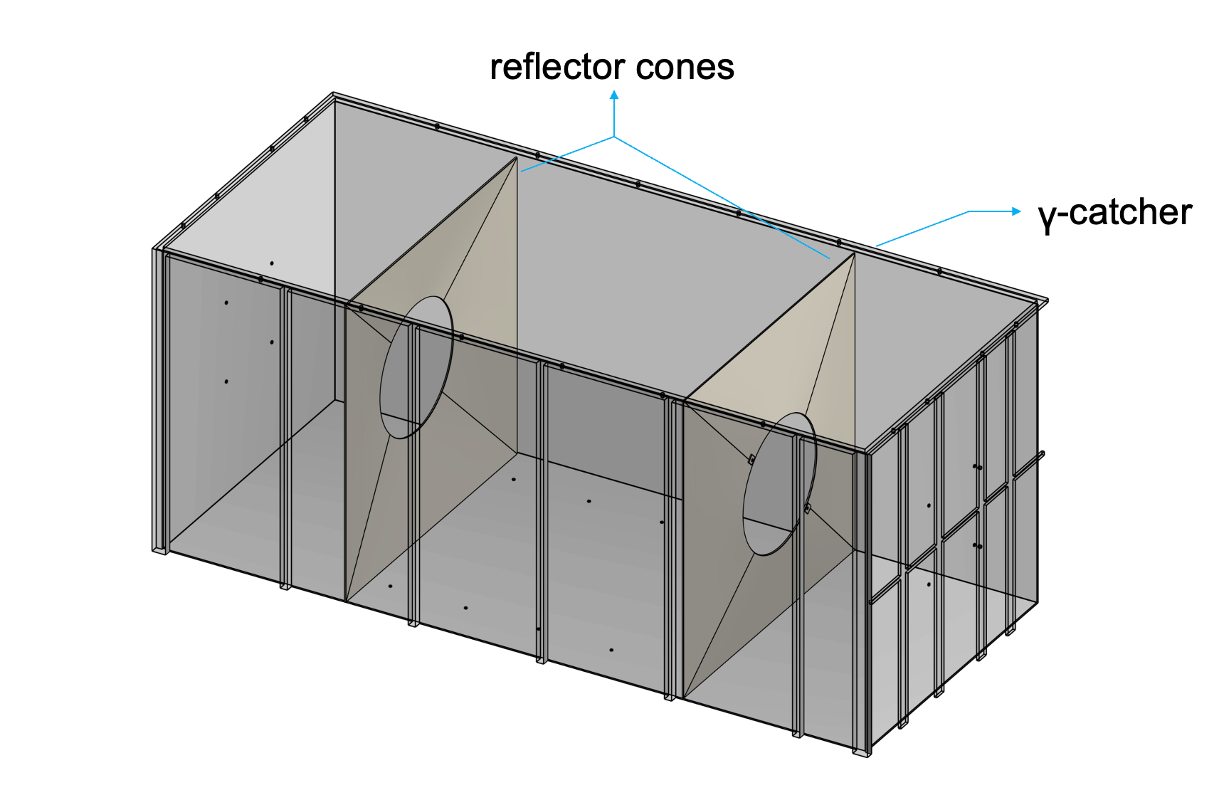}
\includegraphics[width=.6\textwidth]{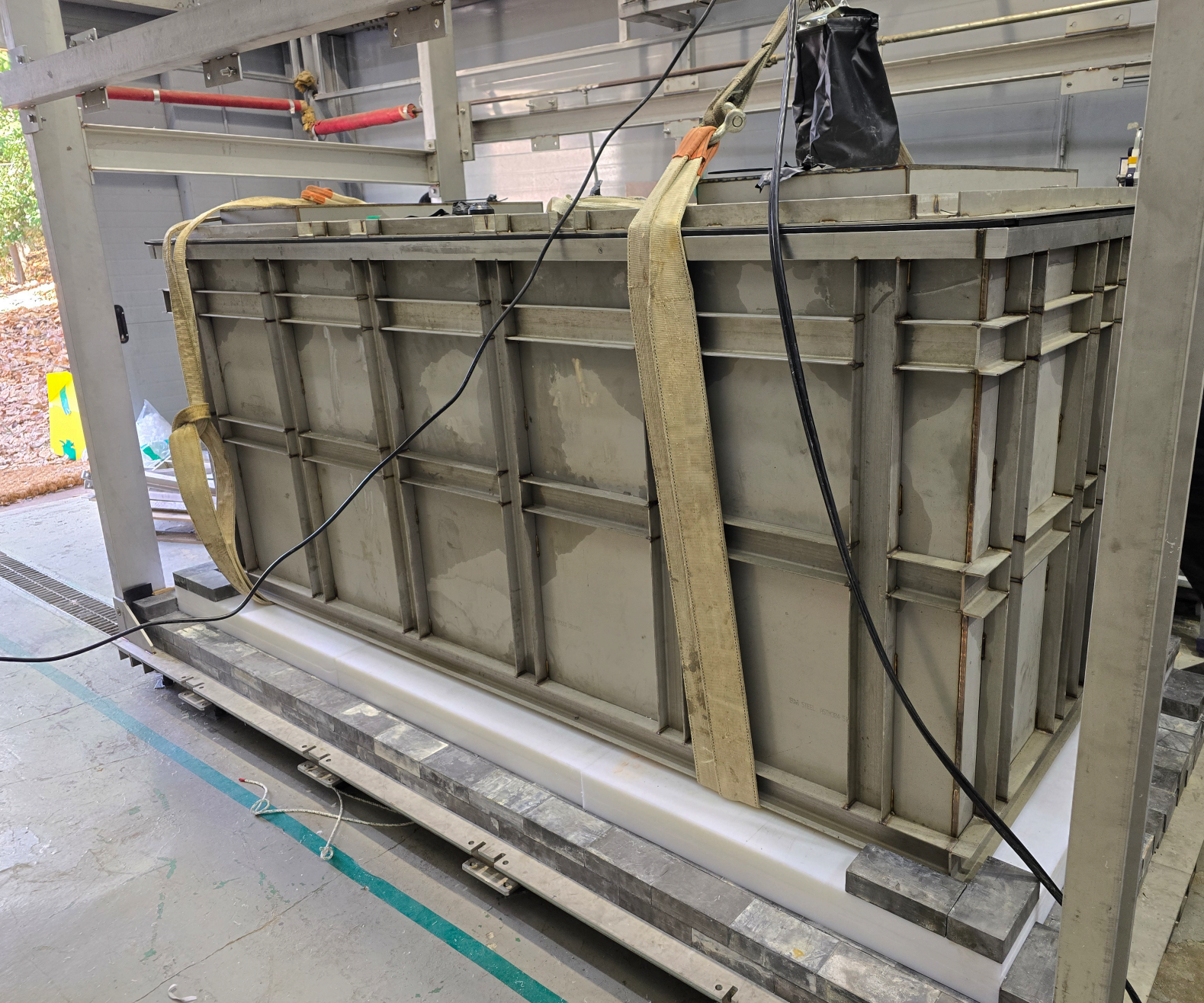}
\caption{\setlength{\baselineskip}{5mm} Isometric view of the $\gamma$-catcher chamber and the reflector cones (top). Fully assembled $\gamma$-catcher chamber (bottom).} 
\label{fig:detector_gc_and_cone}
\end{figure}

\subsection{$\gamma$-catcher Chamber and Reflector Cones}
\label{sec:detector_GC}
The $\gamma$-catcher chamber is a rectangular stainless steel structure housing the target vessel, LS, reflector cones, and two 20-inch PMTs. 
The chamber's inner dimension of the $\gamma$-catcher chamber is $2800\times1250\times1250 \rm{mm}^{3}$ ensuring sufficient thickness for the LS surrounding the target vessel to catch the escaping one $\gamma$-ray from the election-positron annihilation. 
This design is expected to enhance the energy resolution of events occurring in the target.

\begin{figure}[h]
\centering
\includegraphics[width=1.0\textwidth]{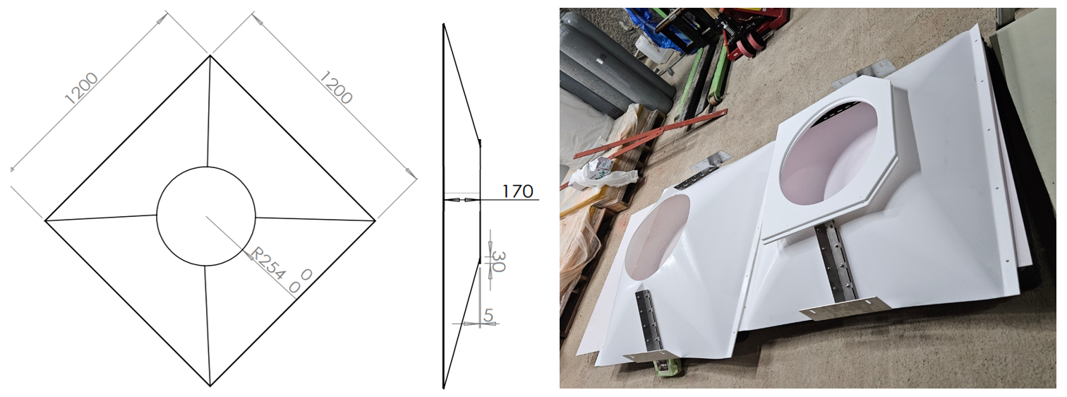}
%\caption{\setlength{\baselineskip}{5mm} Front of reflector cone made by Teflon (left). Back of reflector cone (right). } 
\caption{\setlength{\baselineskip}{5mm} Blueprint of reflector cone and fully assembled reflector cone. } 
\label{fig:relector_cone}
\end{figure}

\begin{figure}[h]
\centering
\includegraphics[width=.8\textwidth]{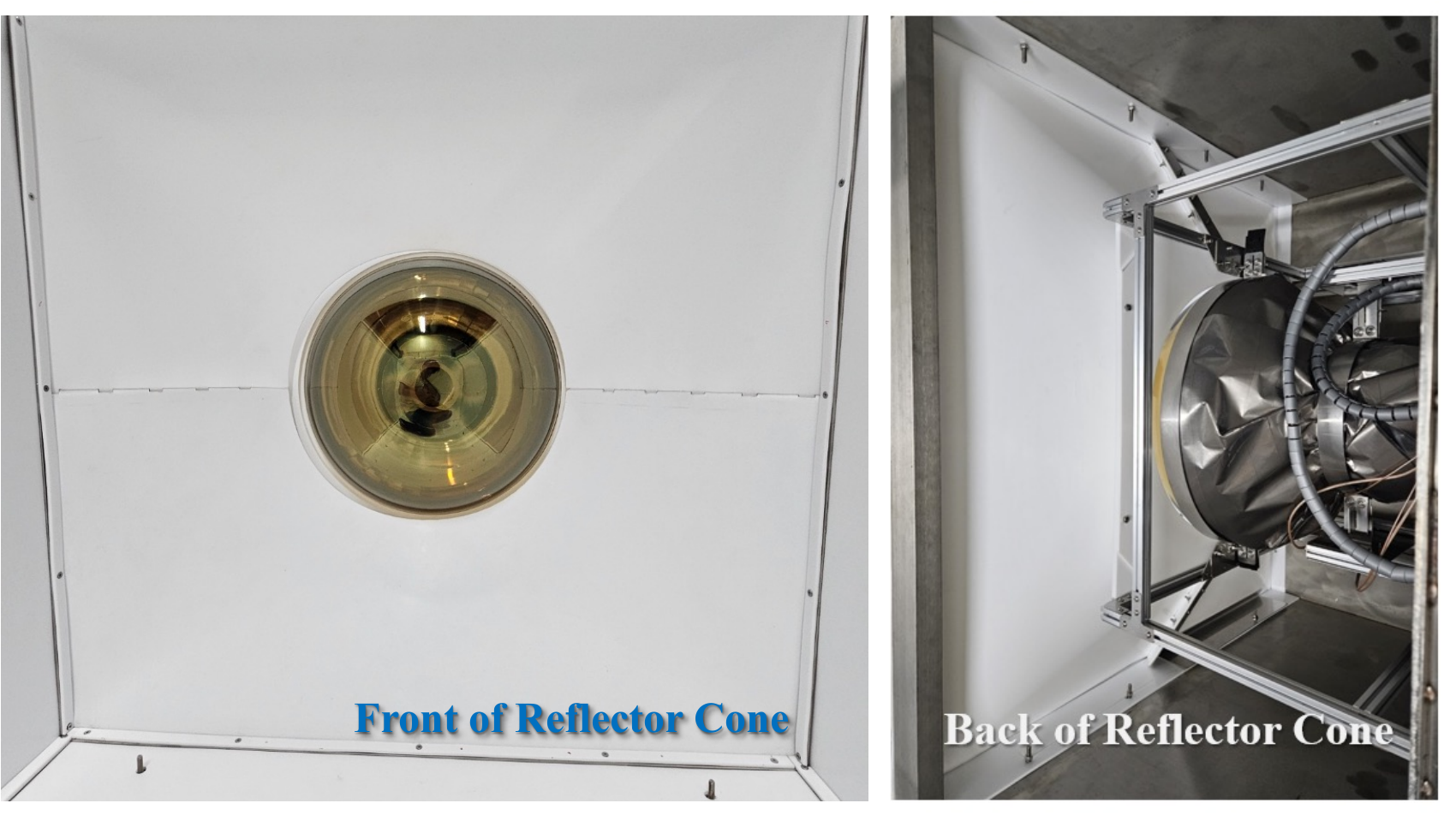}
\caption{\setlength{\baselineskip}{5mm} Front view of a reflector cone made from Teflon (left) and back view of reflector cone (right). } 
\label{fig:relector_cone2}
\end{figure}

Figure~\ref{fig:detector_gc_and_cone} illustrates the $\gamma$-catcher chamber equipped with the reflector cones. 
The reflector cone shape is designed to fit with the $\gamma$-catcher chamber wall on the near side with a circular hole on the far side
to accommodate the 20-inch PMTs. They are made from stainless steel, and installed facing each other with a \textsc{Dupond Tyvek}\texttrademark~ covering
to material to maximize the reflectance. The Tyvek will be applied to the inner surface of the $\gamma$-catcher chamber and reflector cones 
in the region between them direct optical photons toward the PMTs.
%%% dhmoon will explain reflector one %%%
%which can contain about $\num{3000}l$ of LS, as shown in Fig.~\ref{fig:detector_design}. 

The $\gamma$-catcher chamber is covered with a lid featuring five holes. Among these, three central holes are designated for the radioactive source
calibration, while the remaining two accommodate the signal and high voltage (HV) lines of the 20-inch PMT connected to the DAQ system and power supply. 

%%%% SUS frame change and magnetic effect

\begin{figure}[hbt!]
\centering
\includegraphics[width=.9\textwidth]{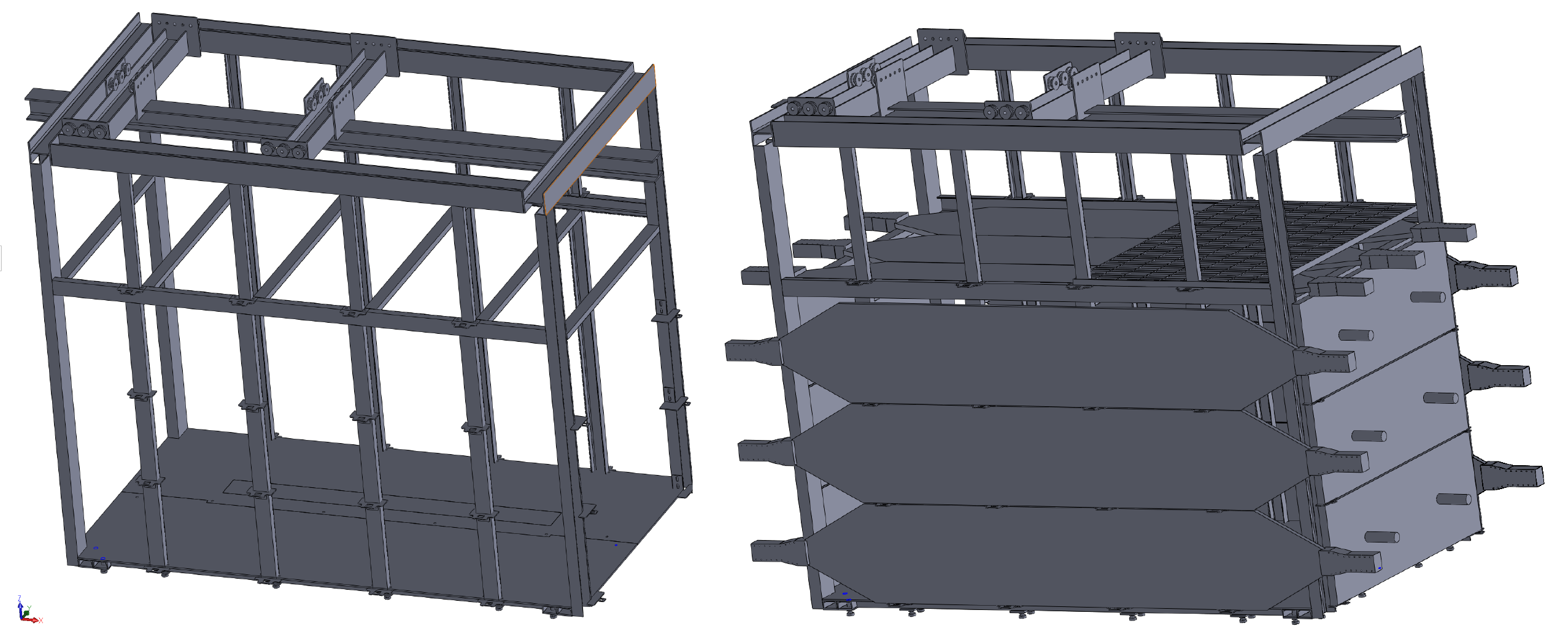}

\caption{\setlength{\baselineskip}{5mm} Surrounding support structure of the VETO system (left) and isometric view of VETO detector system (right)} 
\label{fig:detector_veto_outside}
\end{figure}

\subsection{VETO System}
\label{sec:detector_VETO}
%%% dhmoon writing %%%
The VETO system is placed in the outermost layer of the RENE detector to reject the background events primarily attributed to cosmic muons.
Each VETO detector consists of a plastic scintillator panel (EJ-200~\cite{eljen}) and two 2-inch PMTs (\textsc{Hamamatsu} R7195~\cite{Hamamatsu}).
Figure~\ref{fig:detector_veto_outside} illustrates the corresponding schematic design. Cosmic muons serve as the primary source of background noise in the experiment. 
Detecting incoming muons will be essential, as they can induce neutron production through muon–nucleus interactions inside the detector. Additionally, correlated background events can arise from muon interactions with target materials, as well as $\gamma$-rays generated by muons interacting with the $\gamma$-catcher.

\begin{figure}[h]
\centering
\includegraphics[width=.8\textwidth]{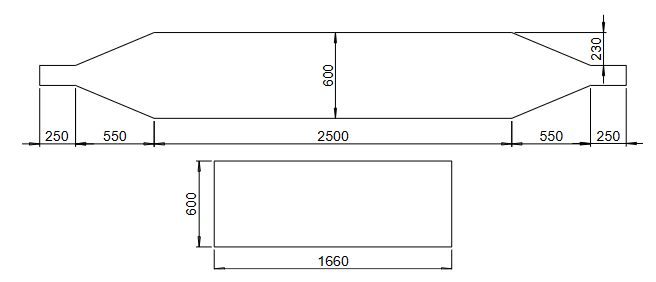}
\caption{\setlength{\baselineskip}{5mm} Detailed geometry of the veto panel Type A (top) and Type B (bottom).} 
\label{fig:detector_veto_geometry}
\end{figure}

Muon signal detected in the veto system will help identify muon-related background events for each candidate event. Notably,
an oil-based LS has an IBD neutrino event rate of approximately 
1 event$\cdot$ km$^2/\rm{GW/ton/day}$. 
Fast neutrons produced in the surrounding rock and subsequently entering the detector volume
are expected to serve as an additional background source in the RENE experiment.
To ensure adequate background suppression, the fast neutron event rate and the accidental coincidence rate from independent positron and neutron signals  
($1\,\rm{MeV} < E <8\,\rm{MeV}$ for the positrons and $6\,\rm{MeV} < E < 10\,\rm{MeV}$ for the neutrons) must remain below 1\% of the IBD event rate~\cite{RENO1}.

The accidental coincidence rate from two unrelated signals, with rates $R_1$ and $R_2$, is calculated as $R_{\text{accidental}} = R_1 \cdot R_2 \cdot \Delta T$, where $\Delta T$ denotes the coincidence time window. 
The veto system is designed to provide shielding against ambient $\gamma$-rays and reduce muon-related background events by applying an offline veto timing cut after each muon passes through the detector system.

\begin{figure}[h]
\centering
\includegraphics[width=.9\textwidth]{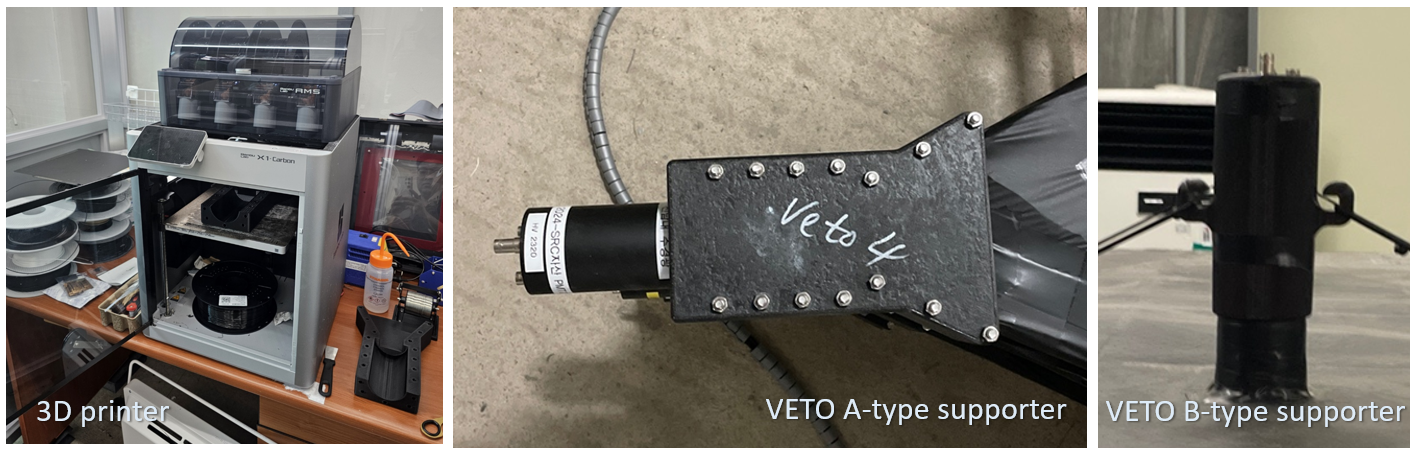}
\caption{\setlength{\baselineskip}{5mm} Manufacturing PMT mounter for A and B type. 3-D printer (left). Type A supporter (middle) and Type B supporter (right).
} 
\label{fig:PMT_supporter}
\end{figure}

\begin{figure}[h]
\centering
\includegraphics[width=.9\textwidth]{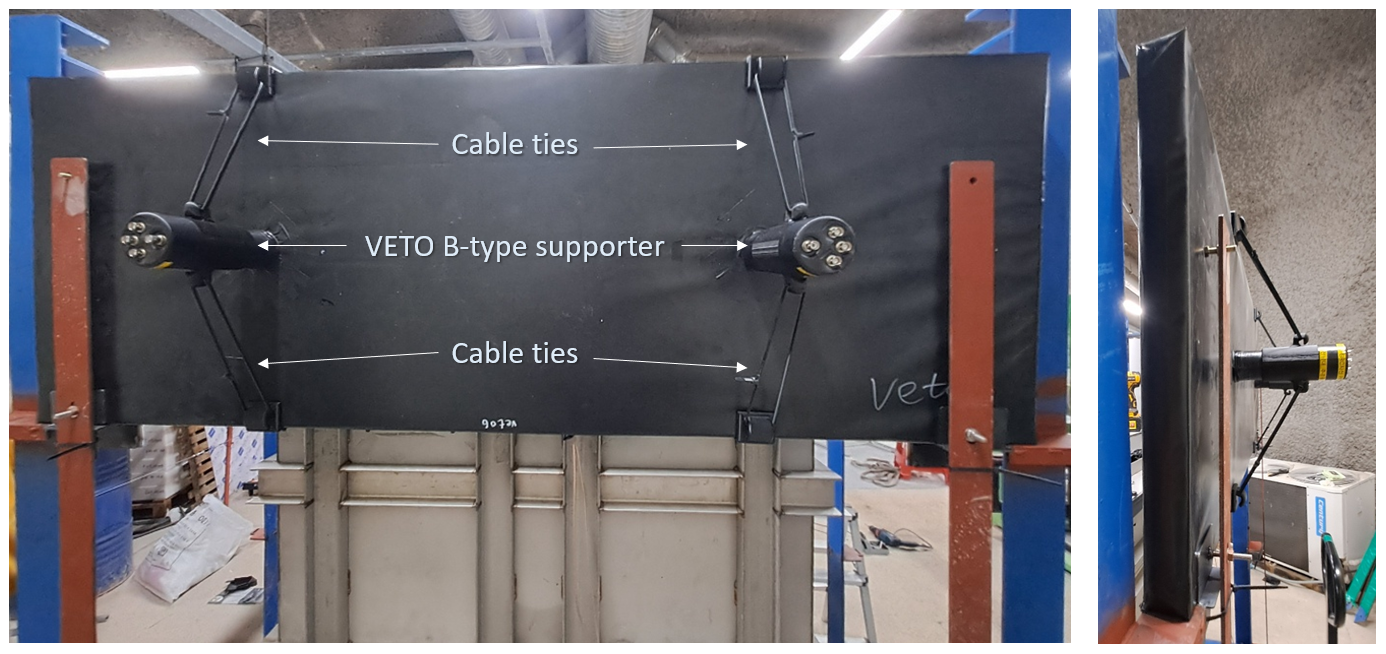}
\caption{\setlength{\baselineskip}{5mm} Long-term mounting test of the Type B PMT mounter.
} 
\label{fig:PMT_mounting_test}
\end{figure}

\begin{figure}[h]
\centering
\includegraphics[width=.9\textwidth]{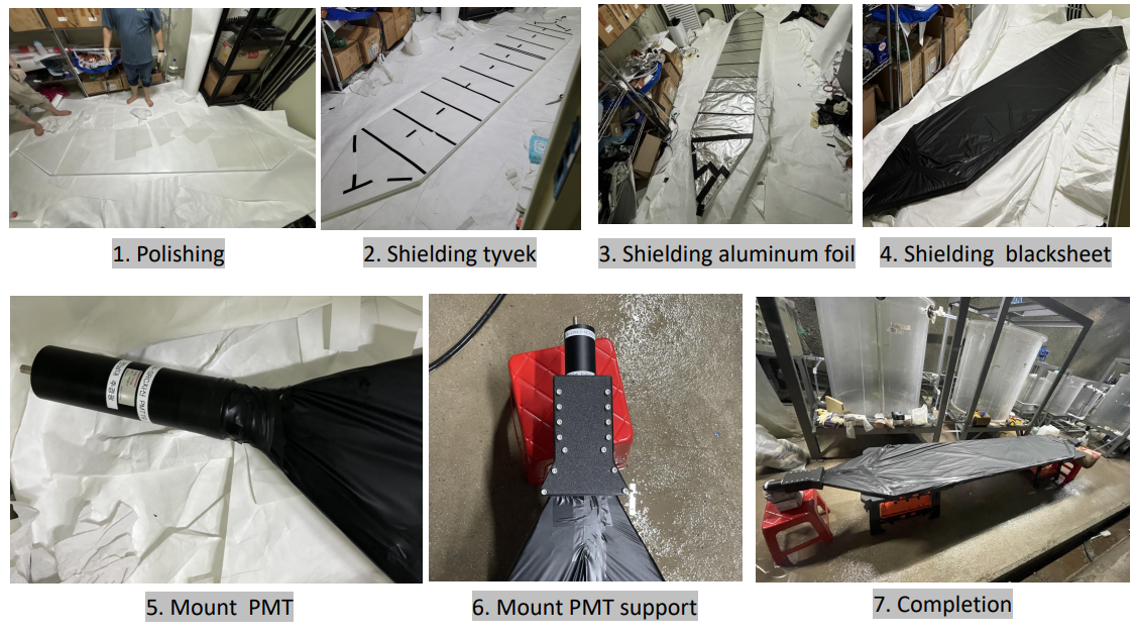}
\caption{\setlength{\baselineskip}{4mm} Manufacturing process of veto panels.} 
\label{fig:veto_process}
\end{figure}

Two types of scintillator panels are used, as depicted in Fig.~\ref{fig:detector_veto_geometry}. Type A, measuring $2500\times 600\times 30\,\rm{mm}$, is connected to light guides on each end to PMTs.
Type B,  measuring $1660\times 600\times 50\,\rm{mm}$, is directly connected to two PMTs.
Figure~\ref{fig:detector_veto_outside} illustrates nine Type A VETO panels surrounding the top and long sides, while Type B VETOs cover the front and back sides. The plastic scintillator panels are covered with Tyvek, aluminum foil, and black sheets to enhance reflectivity and ensure that the generated photons effectively reach the PMT photocathodes. 

Each VETO detector was fabricated through a sequential process: surface polishing, followed by layering with Tyvek, aluminum foil, and black sheet for shielding. Subsequently, the detectors were coupled to 2-inch PMTs and mounted onto a support holder provided by a 3-D printer with the unique design, as illustrated in Fig.~\ref{fig:PMT_supporter}. 
The Type A support was securely fastened using 14 bolts and nuts, while the Type B supporter
was connected to PMT vertically. Given that the PMT connected to the PS could shift under the downward force of gravity, 
cable ties were fixed at the top and bottom to create tension, ensuring that the Type B PMT support remained securely fastened to the scintillator panel.

\begin{figure}[h]
\centering
\includegraphics[scale=0.8]{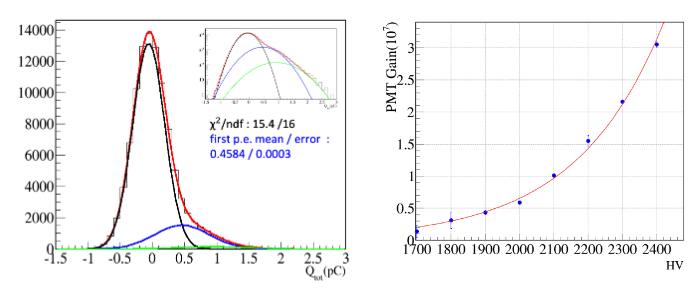}
\caption{\setlength{\baselineskip}{5mm} Charge distribution obtained using a light-emitting diode (LED, left) and gain curve (right)} 
\label{fig:veto_spe}
\end{figure}

To evaluate whether each PMT support could securely hold the PMT for an extended period, it was attached to a vertically standing Type B 
plastic scintillator, secured tightly at the top and bottom with cable ties and tested for approximately two weeks.
Observations revealed that the support remained securely in place, just as it was at the beginning, 
confirming that it could continue to support the PMT effectively over a long-term period. Figure~\ref{fig:PMT_mounting_test} illustrates the setup used for the long-term supporter PMT support mounting test.

\begin{figure}[h]
\centering
\includegraphics[scale=0.8]{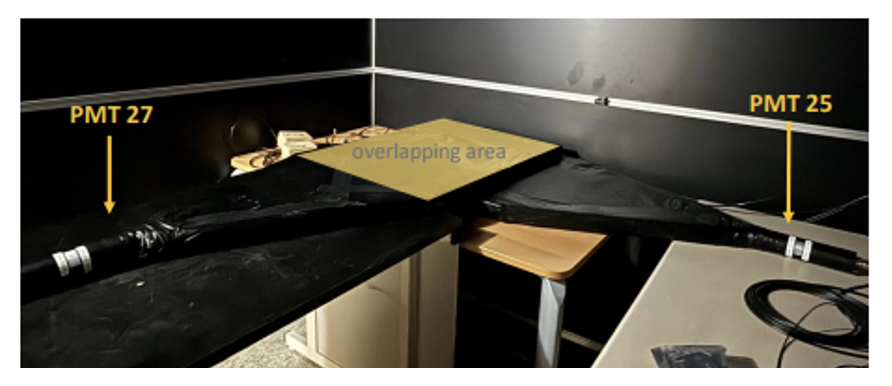}
\includegraphics[scale=0.9]{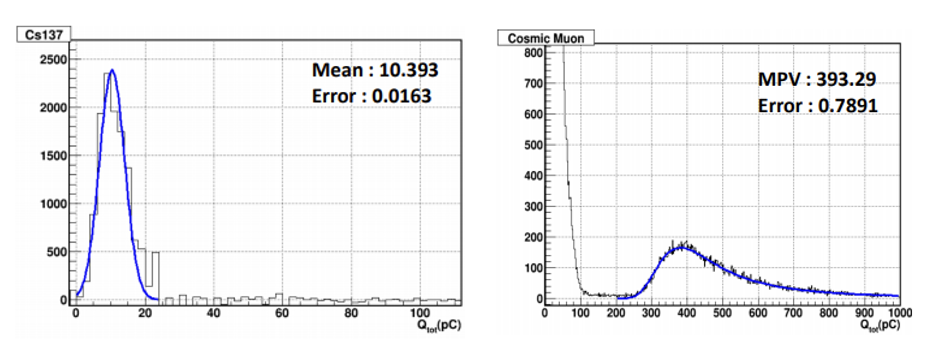}
\caption{\setlength{\baselineskip}{5mm} Coincidence test of the experimental setup (top). Charge distributions of $^{137}$Cs (bottom left) and cosmic muons (bottom right) } 
\label{fig:veto_coin}
\end{figure}

The following tests were conducted to evaluate the Type A and Type B VETO detectors and confirm their detection performance: cosmic muon, $^{137}$Cs, and LED tests. 
The working voltage was set to approximately 2000 V during testing.
The left panel of Fig.~\ref{fig:veto_spe} presents the single photoelectron (SPE) peak obtained from the LED test, 
and the right panel of Fig.~\ref{fig:veto_spe} presents the gain curve fitted against an exponential function. 
These results were found to be consistent with the official data sheet provided by Hamamatsu company, confirming that all 2-inch PMTs functioned properly. 
Figure~\ref{fig:veto_coin} represents the coincidence test setup and the energy distributions of $^{137}$Cs and the cosmic muon.
Then, fit the charge signals of cosmic muons with a Landau distribution.

The most probable value (MPV) of the energy deposited by cosmic muons in the plastic scintillator was approximately 10 MeV, 
which matched the expected value based on an energy loss of 2 MeV per cm over a scintillator thickness of 5 cm.
The M64ADC module will control the 32 channels of 2-inch PMTs installed in the VETO system.
A more detailed description is provided in the section~\ref{sec::daq}, which discusses the DAQ system.

\begin{figure}[h]
\centering
\includegraphics[scale=0.6]{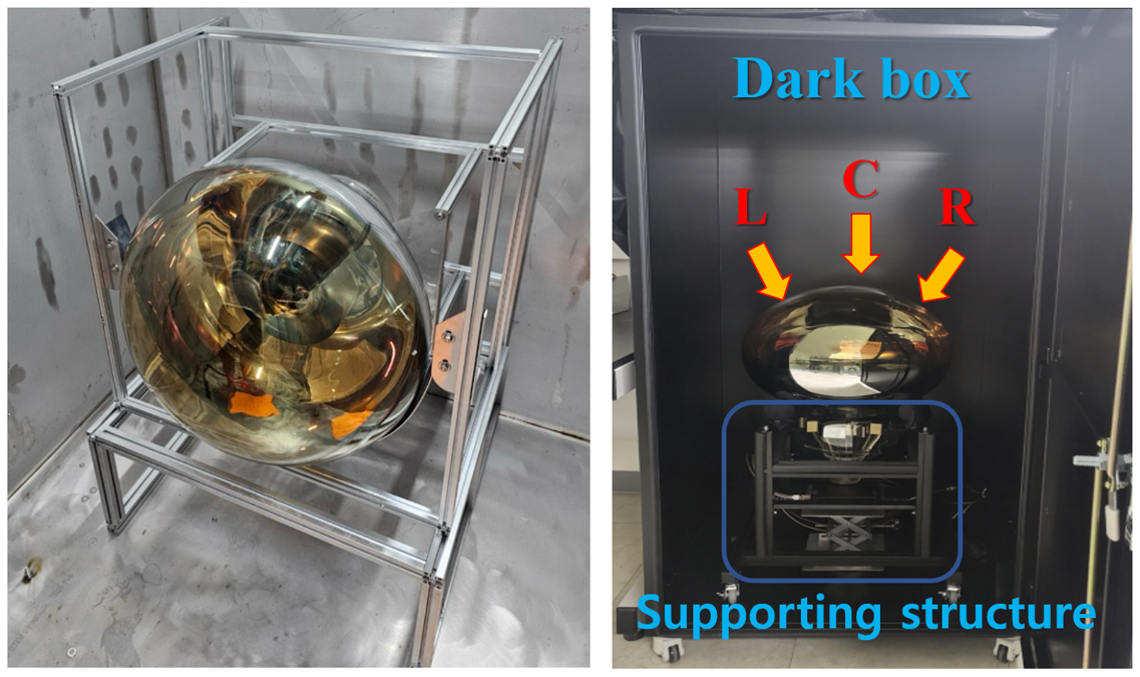}
\caption{\setlength{\baselineskip}{5mm} 20-inch PMT installed on the side surface with an aluminum holder frame (left). The dark box for the PMT test (right). Arrows indicate the positions of the radioactivity sources. } 
\label{fig:20PMT}
\end{figure}

\begin{figure}[h]
\centering
\includegraphics[scale=0.3]{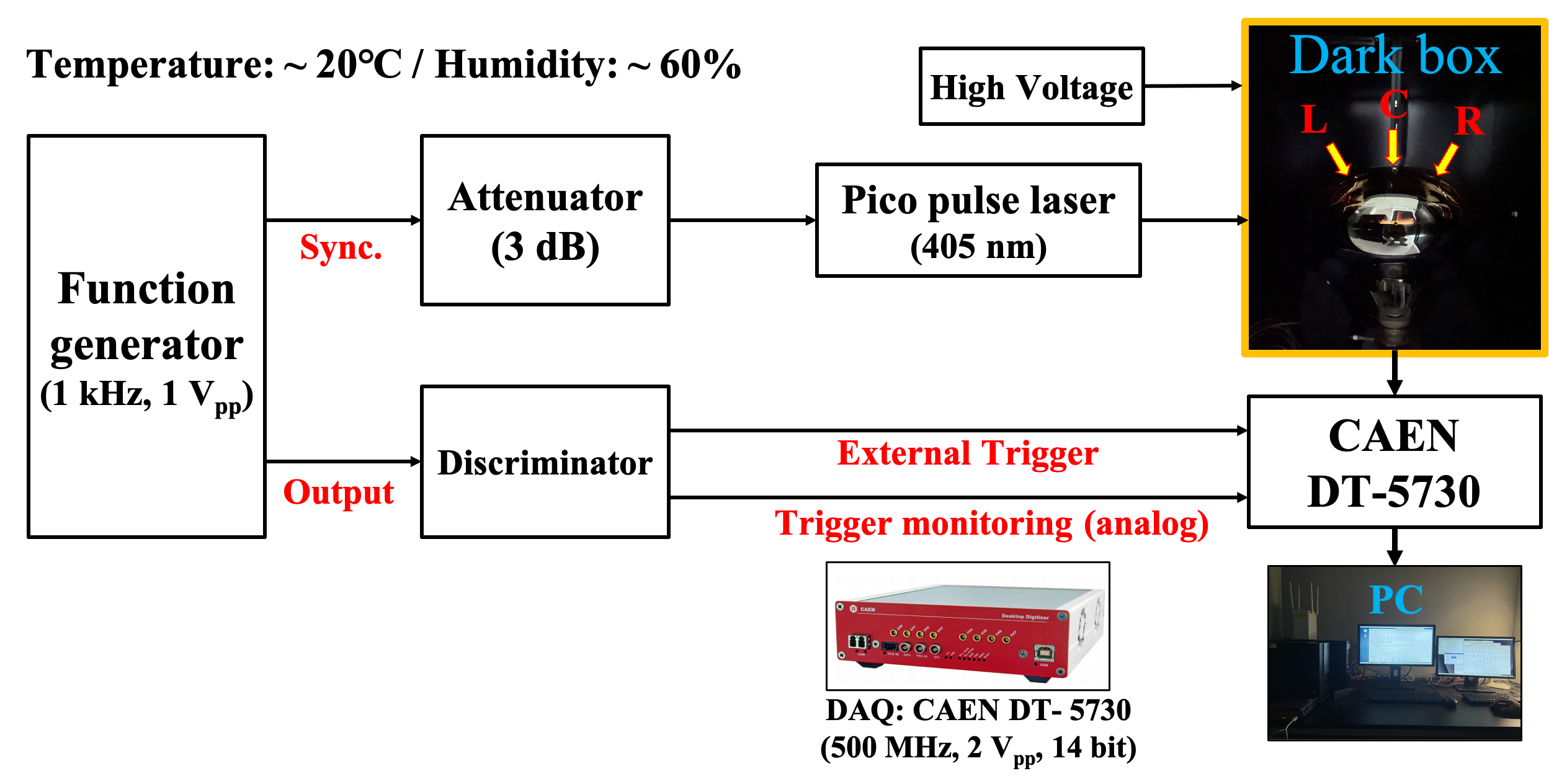}
\caption{\setlength{\baselineskip}{5mm} Setup for assessing PMT characteristics.} 
\label{fig:PMT_measurement_setup}
\end{figure}

\begin{figure}[h]
\centering
\includegraphics[scale=0.45]{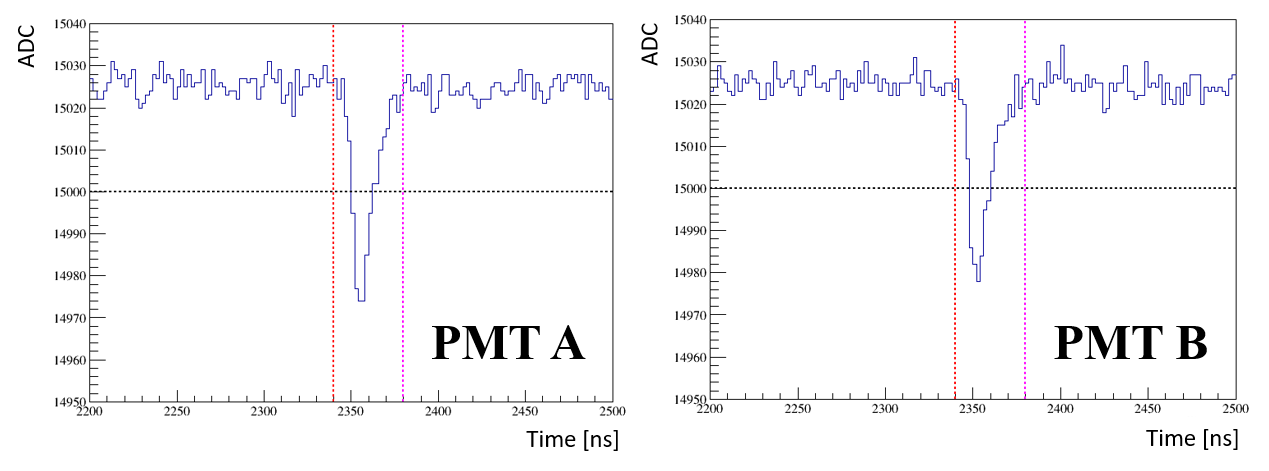}
\caption{\setlength{\baselineskip}{5mm} Example waveforms of PMT A (left) and B (right). The horizontal bars indicate the threshold values.} 
\label{fig:dark_current}
\end{figure}

\begin{figure}[h]
\centering
\includegraphics[scale=0.5]{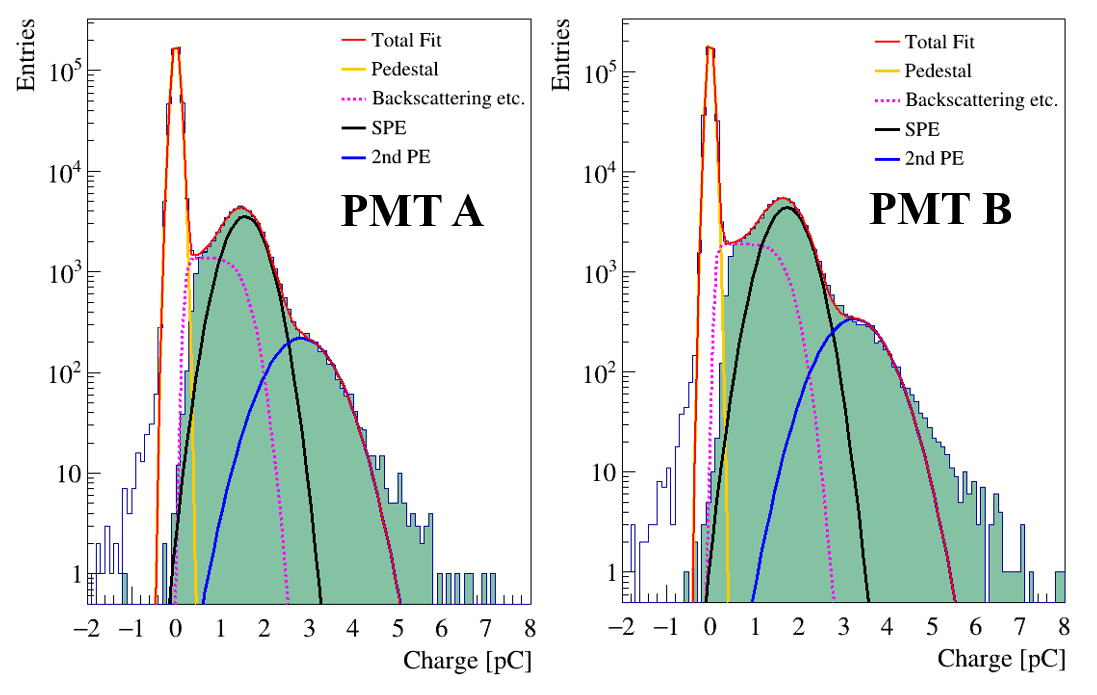}
\caption{\setlength{\baselineskip}{5mm} Charge distributions of PMT A (left) and B (right). The yellow line marks the pedestal peak, 
while the dark and light blue lines represent the photoelectron peaks.} 
\label{fig:PMT_spe}
\end{figure}

\subsection{20-inch PMT}
\label{sec:detector_PMT}
%\subsubsection{PMT Requirements and Specification}

In the detection system of the RENE experiment, 
scintillation light emitted by the active target materials, composed of Gd-LS and LS, is collected by two 20-inch \textsc{Hamamatsu} R12860 PMTs 
mounted onto the inner surfaces of the $\gamma$-catcher chamber, as depicted in the top panel of Fig.~\ref{fig:detector_gc_and_cone}. 
These PMTs are essential for the precise measurement of anti-electron neutrino energy. Key performance factors are the charge response and timing characteristics of the PMTs. 
Since the PMTs will be immersed in mineral oil, the entire PMT assembly should be chemically resistant to the oil, 
which must remain stable throughout the experiment. 
PMTs positioned in peripheral areas will not be included in the detector setup~\cite{RENO1}.
Figure \ref{fig:20PMT} presents a 20-inch PMT installed on the side surface of the inner $\gamma$-catcher chamber, along with the aluminum holder frame. 
To evaluate the performance of the 20-inch PMT A and B, dark condition and pico-pulse laser tests were conducted at room temperature of $\sim 20\,^{\circ}\mathrm{C}$ and $\sim60\%$ humidity, environmental conditions under which the experiment will be performed, as indicated in Fig.~\ref{fig:PMT_measurement_setup}. 
Figure~\ref{fig:dark_current} presents a sample waveform digitized by the \textsc{CAEN} DT-5730 for both PMTs. 
The detection threshold was set at a pulse height of 3$\,$mV. 
Data acquisition was conducted using a self-trigger mechanism, yielding in an observed dark rate of approximately 3 kHz. 

%%%%%%

A pico-pulse laser module operating at a wavelength of 405 nm, combined with an external trigger, was utilized to record the charge distribution of SPE, as depicted in Fig.~\ref{fig:PMT_spe}. 
The majority of recorded events corresponded to pedestals with no discernible signals, while a smaller subset exhibited clear SPE signals. 
We fit the charge distribution with a formula in eq.~\ref{eq:PMT_gain}

\begin{equation}
%\begin{align*}
N_{0} \exp \left(-{(x-\mu_0)^2 \over 2\sigma_0^2} \right) + N_1 \exp \left(-{(x-\mu_1)^2 \over 2\sigma_1^2} \right) + N_2\left( erf\left({x-\mu_0 \over \sigma_0} \right) -erf\left({x-\mu_1 \over \sigma_1} \right) \right)
%\end{align*}
\label{eq:PMT_gain}
\end{equation}

where, the first term represents the pedestal, the second term represents the single photoelectron distribution, and the third term represents the flat inelastic backscattering distribution.

\begin{figure}[h]
\centering
\includegraphics[scale=0.4]{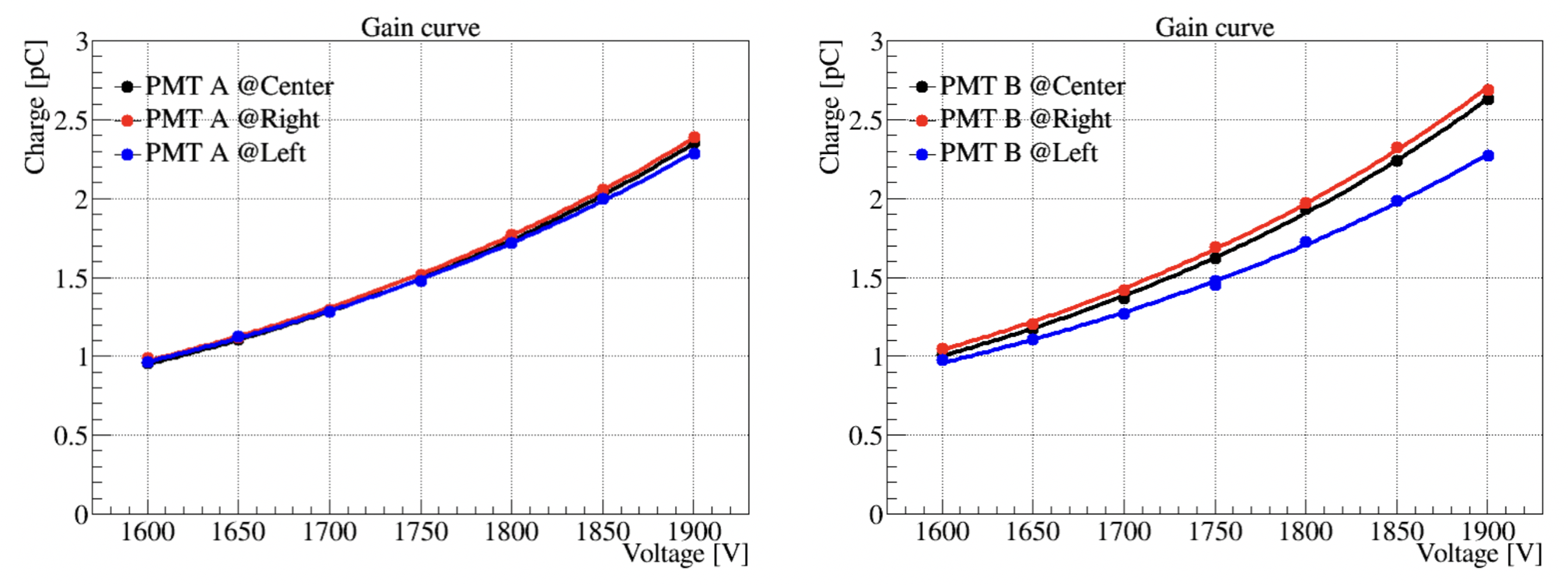}
\caption{\setlength{\baselineskip}{5mm} Mean values of the SPE charge distributions as a function of HV for PMT A (left) and PMT B (right). The black circles indicate the cases where the laser light was introduced from the center, while 
the red and blue circles denote the introduction from the left and the right, respectively.} 
\label{fig:PMT_gains}
\end{figure}

In the charge distribution plot, the yellow line represents the pedestal level of the PMT, and the black lines denote the SPE peaks, and the blue lines represent the two-photoelectron (2PE) peaks, which provide a more precise determination of the SPE. 
Consequently, the gain of PMTs can be calculated from the mean values of the first peak, which corresponds to the SPE. 
Figure~\ref{fig:PMT_gains} presents the mean values of charge distributions as a function of HV for PMT A shown on the left 
and B on the right. 
The measured charge values for the SPE at 1750 V was 1.6 pC, which is consistent with
values reported by the \textsc{Hamamatsu}.

\begin{figure}[h]
\centering
\includegraphics[scale=0.5]{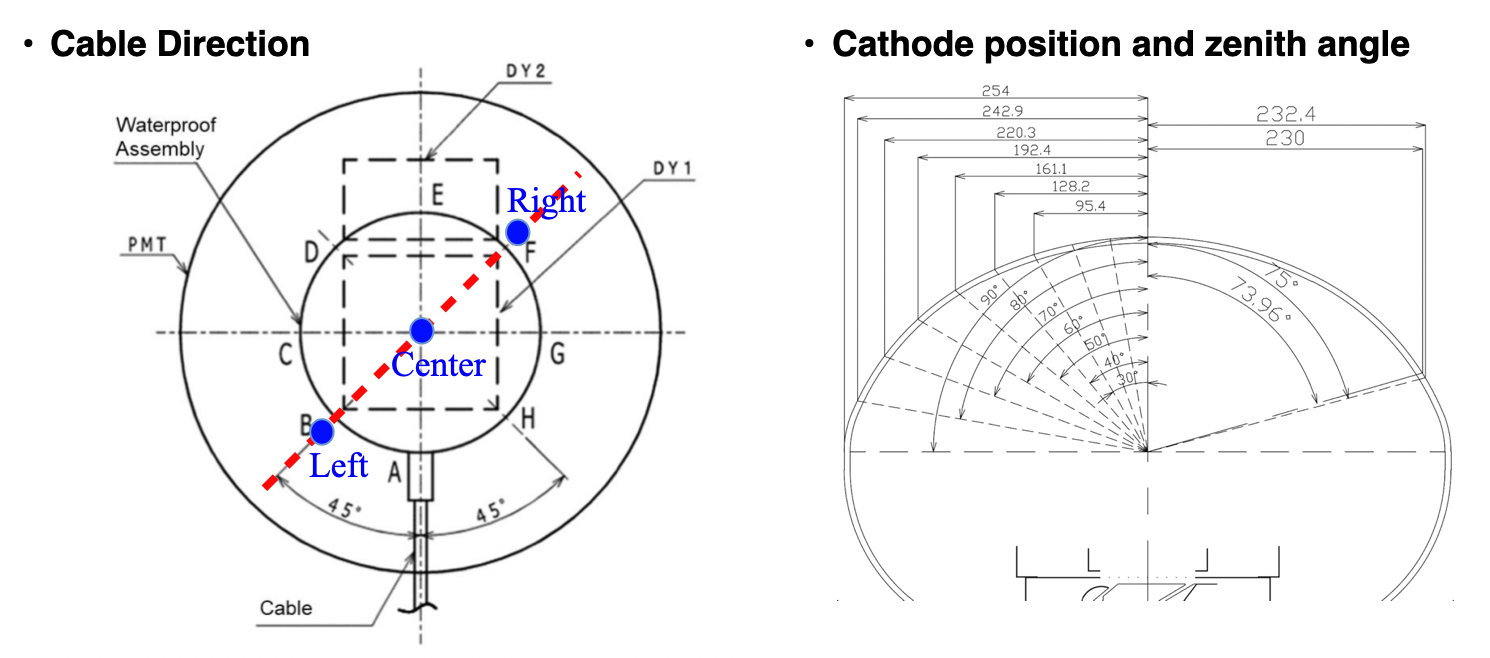}
\caption{\setlength{\baselineskip}{5mm} Various directions of cable outlet for 20-inch \textsc{Hamamatsu} PMT (Left) and the zenith angle corresponding to cathode position (Right). The cable outlet of both PMTs in this test is in the B direction, and the test positions are on the PMT cathode.} 
\label{fig:cable direction}
\end{figure}

%%%%%%%%
Among the manufactured PMTs, given that the 20-inch PMT demonstrate the widest coverage, evaluating the position dependence of their
charge response is essential. The yellow arrows in Fig.~\ref{fig:PMT_measurement_setup} indicate the laser injection directions, while
Fig.~\ref{fig:cable direction} represent the laser injection positions based on the dynode orientation.
To assess position dependence laser light was injected at three locations on the PMT surface: the center (black dot),
left (blue dot), and right (red dot) as depicted in Fig.~\ref{fig:PMT_gains}. 

The observed deviations in charge response were within $\pm$3\%, which will be considered as systematic uncertainties. 
Position-dependent variations in PMT performance, particularly in terms of charge response and gain, 
are influenced by dynode orientation and cathode uniformity in large-diameter PMTs. 
However, the extent of performance variation with respect to the cathode position differs for each PMT. 
For instance, the two PMTs in the RENE detector exhibited different gain variation patterns 
depending on the cathode position as illustrated in Fig.~\ref{fig:PMT_gains}).

\begin{figure}[h]
\centering
\includegraphics[scale=0.6]{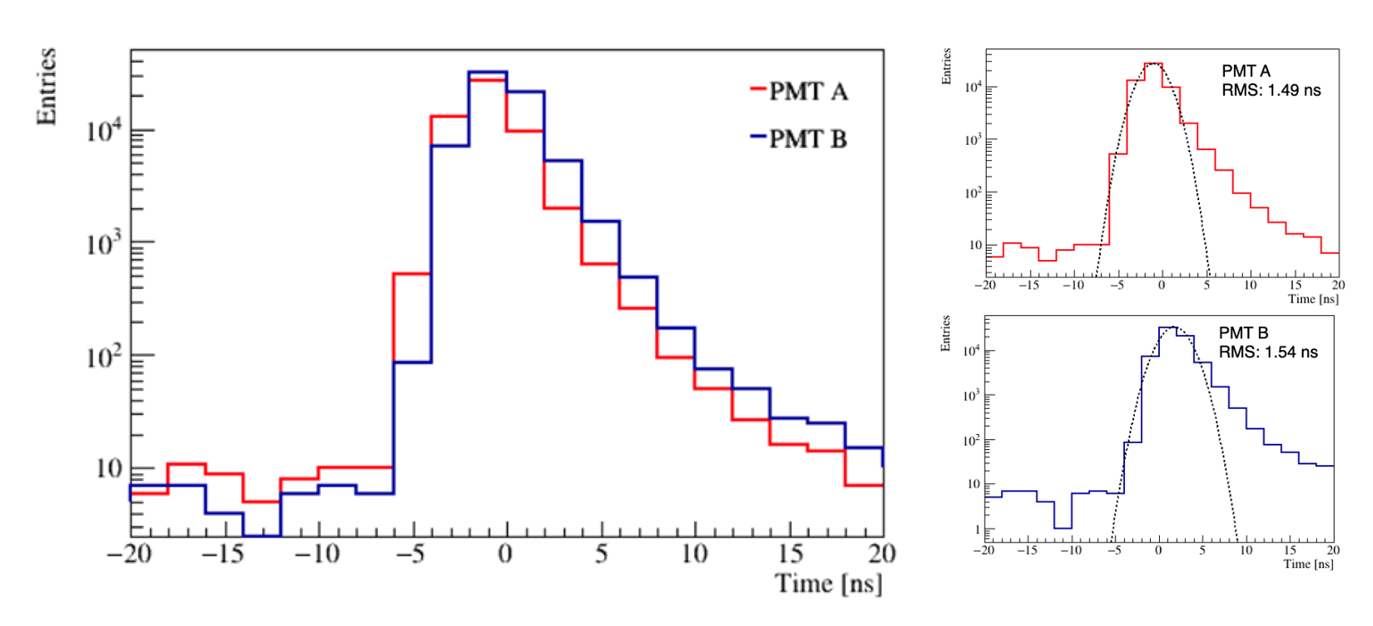}
\caption{\setlength{\baselineskip}{5mm}  Measured time response. The red line represents the time response of PMT A, while the blue line denotes the time response of PMT B (left). The Gaussian fits are shown around the main peaks, with the fitted results for the main peak interpreted as transit time spread (TTS, right).} 
\label{fig:PMT_time_response}
\end{figure}

To evaluate the time response and TTS of the PMTs, the laser-trigger time was subtracted from the PMT falling time, which is defined as the instant when the PMT waveform surpasses the 3$\,$mV pulse
height threshold. Figure~\ref{fig:PMT_time_response} presents the time response distribution, constructed from cumulative recorded events. The measured TTS was found to be less than 2$\,$ns, with both PMTs exhibiting consistent results, indicating similar performance characteristics.

\begin{figure}[h]
\centering
\includegraphics[scale=0.46]{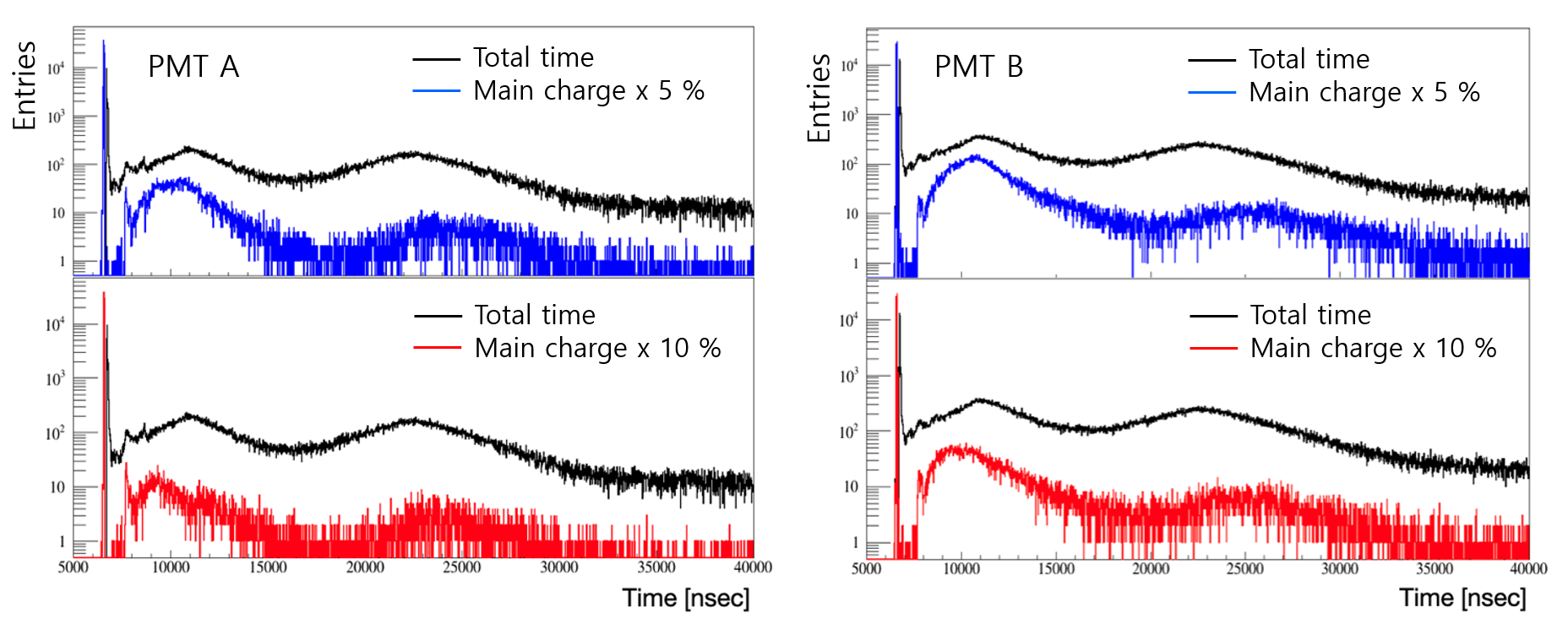}
\caption{\setlength{\baselineskip}{5mm} Time distributions of PMT A (left) and PMT B (right). The black lines indicate the total time, while the blue and red lines represent the time measured for events exceeding 5 \% (blue) and 10 \% (red) of the charge for the main pulse}
\label{fig:PMT_afterpulse_time}
\end{figure}

\begin{figure}[h]
\centering
\includegraphics[scale=0.45]{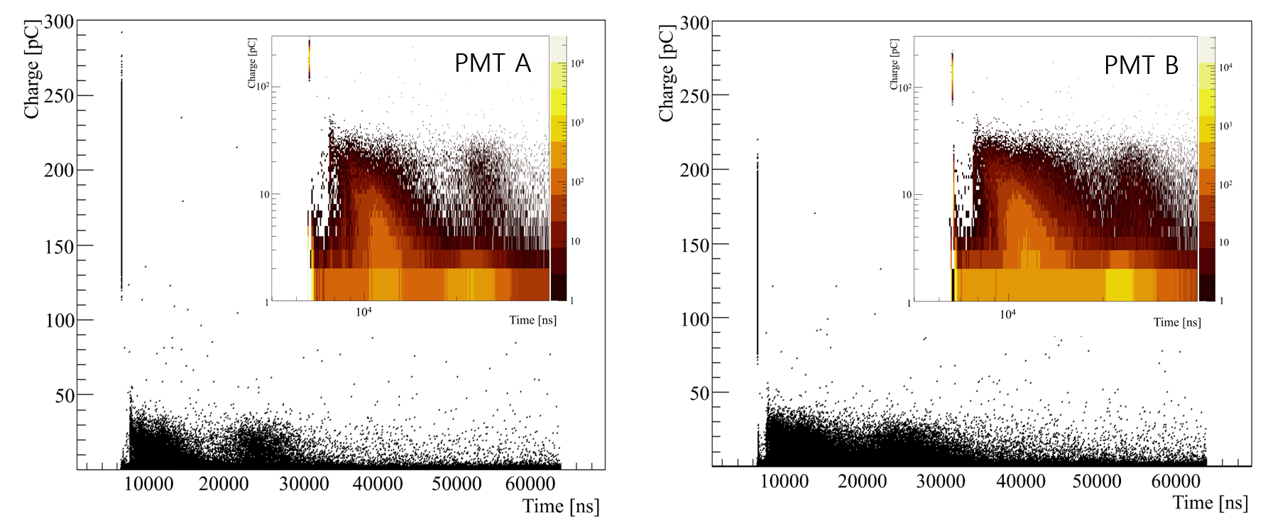}
\caption{\setlength{\baselineskip}{5mm} Two-dimensional histogram of time versus the charge for PMT A (left) and PMT B (right). The plots within the figure are drawn with a logarithmic scale on the y-axis.}
\label{fig:PMT_afterpulse_charge}
\end{figure}

Afterpulsing, a well-documented phenomenon in PMTs~\cite{JUNOAfter}, was measured using a laser coupler that bifurcates the input laser into two output beams. Measurements were conducted under
high-intensity light conditions exceeding 100 photoelectrons. Figure~\ref{fig:PMT_afterpulse_time} represents time distributions of events with charge exceeding 5\% (blue) and 10\% (red) of the
primary charge pulse for PMTs A and B. In the blue histogram of Fig.~\ref{fig:PMT_afterpulse_time}, the first peak before 7000 ns corresponds to the main pulse, while afterpulses begin to appear
approximately 1$\,\mu$s after the main pulse and persist for up to 20${\,\mu}$s. The two-dimensional distributions of charge and time are shown in Fig~\ref{fig:PMT_afterpulse_charge}. 
The similarity between the distributions suggests that both PMTs exhibit consistent performance.

Afterpulses are undesirable signals that appear after the primary light-induced photoelectron pulse in PMTs. This phenomenon has been extensively studied and is known to negatively impact PMT timing accuracy~\cite{PMTAfter8, PMTAfter9, PMTAfter10, PMTAfter12}. The primary cause of afterpulses is the emission of positive ions generated by the ionization of residual gases inside the PMT~\cite{PMTAfter7}, making it difficult to distinguish afterpulses from actual light-induced signals. These afterpulse signals closely resemble the primary signal and degrade both the timing and charge resolution of the PMTs, thereby adversely impacting particle identification and event reconstruction performance.

Tests on the 20-inch PMTs were conducted to evaluate afterpulse rates, revealing two distinct afterpulses occurring approximately 10$\,{\mu}$s and 
20$\,{\mu}$s after the main pulse, consistent with observations from the Jiangmen Underground Neutrino Observatory (JUNO) ~\cite{JUNOAfter}. 
Two-dimensional charge and time histograms offer a clear distinction between the main pulse and afterpulses. 
Although the charge of the main pulse varies substantially, the detected timing of afterpulses remains consistent between both PMTs, as shown in Fig~\ref{fig:PMT_afterpulse_charge}.

Another critical factor to be considered when using PMTs is the influence of the surrounding magnetic field. 
A PMT detects photons incident on the photocathode, where they generate photoelectrons via the photoelectric effect. 
These photoelectrons are multiplied by more than a million times through multiple dynodes under HV conditions, producing an electrical signal. For optimal performance, the generated photoelectrons must travel through all dynodes under the electric field created by the HV condition. 
However, the external magnetic fields can alter the electron trajectories, reducing the multiplication efficiency and affecting the PMT’s detection timing, energy resolution, and photon detection efficiency. Since the external magnetic fields can impact PMT performance, effective magnetic shielding is essential. In the Republic of Korea, the Earth’s magnetic field averages approximately 500 mG, but it can be reduced to a level that does not significantly affect PMT operation using Permalloy shielding. 
This issue is particularly relevant for the 20-inch PMTs used in the RENE
detector, where the longer electron paths inside the PMT and the wider photocathode
surface make photon detection efficiency more susceptible to external magnetic field
disturbances. Given these factors, minimizing the use of ferromagnetic materials,
which generate additional magnetic fields around the detector,
is crucial. Additionally, shielding measures should be implemented to 
maximally reduce external magnetic field interference for the detector.

\begin{figure}[h]
\centering
\includegraphics[scale=0.55]{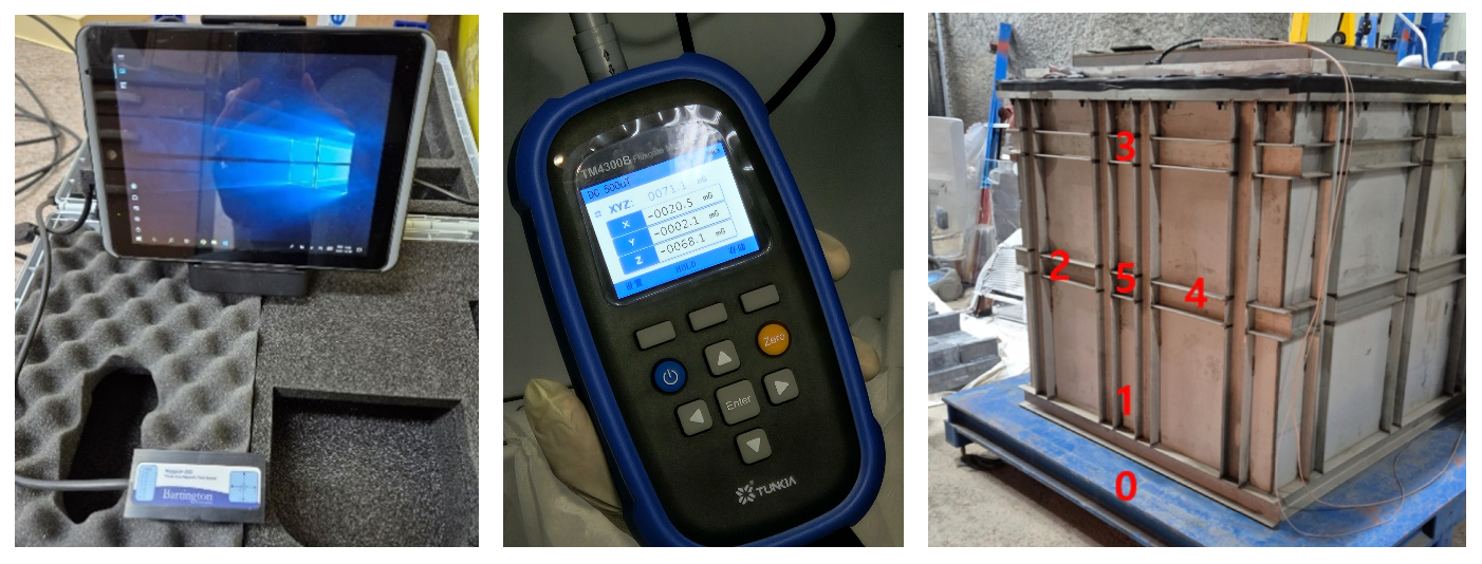}
\caption{\setlength{\baselineskip}{5mm} Magnetic field measuring devices: TFM1186 (left), TM4300B (middle). Measurement positions of the magnetic field (right).} 
\label{fig:PMT_magnet2}
\end{figure}

\begin{table}[h]
    \centering
    \begin{tabular}{ccc} \hline
        Position & Magnetic field (mG) with Iron pallet & Magnetic field (mG) with SUS pallet \\
        \hline \hline
        0  & 1898.56 & 389.60  \\
        1  & 431.83  & 392.94  \\
        2  & 332.10  & 369.18  \\
        3  & 308.27  & 415.86  \\
        4  & 302.29  & 441.51  \\
        5  & 338.90  & 394.44  \\ \hline
        Variation & 636.90 & 24.95 \\
        \hline
    \end{tabular}
    \caption{Magnetic fields at various positions for two pallet configurations.}
    \label{tab:magnetic_field}
\end{table}

To explore shielding materials, two pallet configurations were considered: one made from ferromagnetic iron and another made from non-magnetic SUS. 
During the test, the magnetic field surrounding the detector was measured, and comparisons were made between the two configurations. 
Additionally, the magnetic fields at the positions where PMTs would be installed inside the detector were measured to determine methods for minimizing magnetic field interference.
Figure~\ref{fig:PMT_magnet2} presents the magnetic field measuring devices (TFM1186, TM4300B) and the six measurement positions. The results, summarized in Table ~\ref{tab:magnetic_field}, indicate that when using the iron pallet, 
the magnetic field varied considerably across the measurement locations. In contrast, when using the SUS pallet, the magnetic field exhibited minimal variations across the measurement locations.

\begin{figure}[h]
\centering
\includegraphics[scale=0.7]{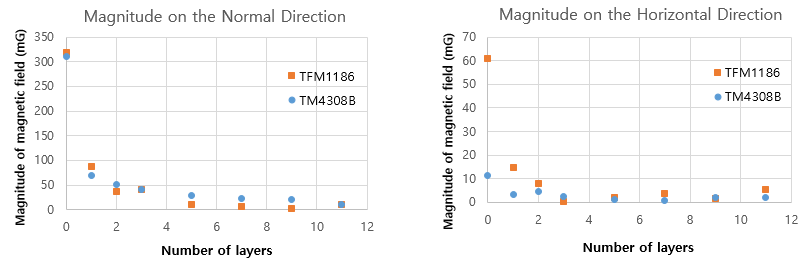}
\caption{\setlength{\baselineskip}{5mm} Magnitude of the magnetic field in the normal direction to the ground (left) and the magnitude of the magnetic field in the horizontal direction to the ground (right) as a function of the number of permalloy sheet layers. Results from two different sensors exhibit a similar trend.} 
\label{fig:PMT_magnet}
\end{figure}

Notably, the magnetic field surrounding the PMTs inside the detector was passively compensated using Permalloy shielding. The compensation rate was measured by incrementally adding Permalloy sheets, and the results are summarized in Fig.~\ref{fig:PMT_magnet}. 
Even a single layer of Permalloy substantially reduced the magnetic field. Given that the target compensation level was less than 100 mG around the PMTs, two layers of Permalloy sheets were used for the RENE detector.

\newpage

\subsection{LS Production}

The detection of reactor electron antineutrinos relies on the IBD process, followed by neutron capture. 
This interaction produces positrons, which immediately release energy in the range of 1 MeV to 8 MeV.
When a hydrogen nucleus captures a neutron, deuteron formation occurs, accompanied by
the emission of photons with an energy of approximately 2.2 MeV. 
However, when the LS is doped with Gd, which has a considerably higher capture cross section for thermal neutrons than for free protons~\cite{GdCross}, 
the delayed neutron capture signal is substantially amplified. This amplification results 
in the emission of photons with a total energy of about 8 MeV, which considerably
improving detection sensitivity by reducing background interference.

In the RENE detector, both the target vessel and $\gamma$-catcher contain the LS.
Effective detection of low deposited energies is possible only if the scintillator exhibits high light yield and optical transparency. 
Additionally, the scintillator must be easy to handle, cost-efficient, and possess favorable physical and chemical stability.

The LS comprises an aromatic organic solvent, a fluor, and a wavelength shifter, as described in Table~\ref{tb:liguid_scint}~\cite{RENOTDR}. 
Generally, aromatic compounds such as benzene ($\rm{C_{6}H_{6}}$) or its derivatives are commonly used as solvents due to their excellent light transmission properties.

\begin{table}[h]
\centering
\begin{tabular}{c|c|c|c}
\hline
\textbf{Aromatic Solvent} & \textbf{Diluent Oil} & \textbf{Fluor} & \textbf{WLS}         \\ \hline
PC, PXE, LAB  & Mineral Oil, Dodecane  & PPO, BPO, PTF      & bis-MSB, POPOP     \\ \hline  
\end{tabular}
\caption{Commonly used chemicals for LS admixtures.~\cite{RENOTDR}}
\label{tb:liguid_scint}
\end{table}

% book-keeping
%Pseudocumene (PC or TMB, \ch{C9H12}, 1,2,4-trimethylbenzene) \sout{is} \textcolor{red}{was}the most frequently utilized solvent in Gd-LS. Among the commonly employed scintillators, PC delivers the highest light output. However, it is known to degrade acrylic materials and has a relatively low hydrogen-to-carbon (H/C) ratio of 1.66. Additionally, it is highly flammable, with a flash point of \SI{48}{\celsius}, and emits harmful vapors. For this reason, PC is typically blended with other solvents to reduce these risks. The concentration of PC is optimized based on factors such as flash point, light yield, and transparency. It has limited solubility in water but dissolves well in ethanol and benzene.

Pseudocumene (PC, $\rm{C_{9}H_{12}}$, 1,2,4-trimethylbenzene) is the most frequently utilized solvent in Gd-LS. 
Among the commonly employed scintillators, PC-based scintillators provide the highest light output. 
However, PC degrades acrylic material properties and has a relatively low hydrogen-to-carbon (H/C) ratio of 1.66. 

%\begin{figure}[hpt]
%\centering
%\includegraphics[scale=0.7]{Figures/rene_LS_component.png}
%\caption{\setlength{\baselineskip}{4mm}Chemical components used in LS~\cite{RENOTDR}.} 
%\label{fig:LS_comp}
%\end{figure}

An alternative aromatic solvent, 1,2-dimethyl-4-(1-phenylethyl)-benzene (phenyl-o-xylythane, PXE, $\rm{C_{16}H_{18}}$), has been considered as a viable substitute. 
PXE has a flash point temperature of 145$\,^{\circ}\mathrm{C}$ and a density ranging from 
0.980$\,\rm{g}/\rm{cm}^3$ to 1.000$\,{\rm{g}/\rm{cm}^3}$
at 15$\,^{\circ}\mathrm{C}$. Despite having a lower H/C ratio
($\sim1.37$) compared to PC, PXE was selected as the LS in the Borexino experiment~\cite{borexino} owing to its higher density and flash point temperature. 
The Double Chooz experiment has also evaluated PXE for similar reasons~\cite{PXE}.

The organic solvent used in LS must be compatible with the acrylic vessels. 
Although PC provides high light yield and favorable optical properties, 
it is not highly compatible with acrylic materials. To mitigate this issue, dilution components were added to PC to enhance its compatibility.
Mineral oil (MO, C$_n$H$_{2n+2}$, where $n = 11$ to $44$) is commonly used as a diluent. 
Its density ranges from 0.7$\,{\rm{g}/\rm{cm}^3}$ to 0.9$\,{\rm{g}/\rm{cm}^3}$ depending on the manufacturer. 
A frequently used LS formation consists of 40\% PC and 60\% MO, with an H/C ratio of approximately 1.87.

Normal dodecane also has a higher flash point temperature (83$\,^{\circ}\mathrm{C}$) than PC. 
Consequently, adding normal dodecane to aromatic solvents with lower flash points enhances the safety of LS. 
Dodecane has a high H/C ratio of 2.17. It is typically produced via distillation from paraffin within a narrow temperature range, ensuring high purity.

An alternative to PC is linear alkyl benzene (LAB, C$_n$H$_{2n+1}$-C$_6$H$_5$, where $n = 10$ to $13$), which was recently developed for next-generation neutrino detectors. 
LAB offers several advantages: It does not require a diluent solvent, provides a relatively high light yield, is environmentally safe, 
and is available domestically in high quality from Isu Company. Consequently, LAB was selected as the scintillator solvent, 2,5-diphenyloxazole (PPO) was chosen 
as the primary fluor, and 1,4-bis(2-methylstyryl)benzene (bis-MSB) was selected as the wavelength shifter for the RENE detector.

\begin{figure}[hpt]
\centering
\includegraphics[scale=0.45]{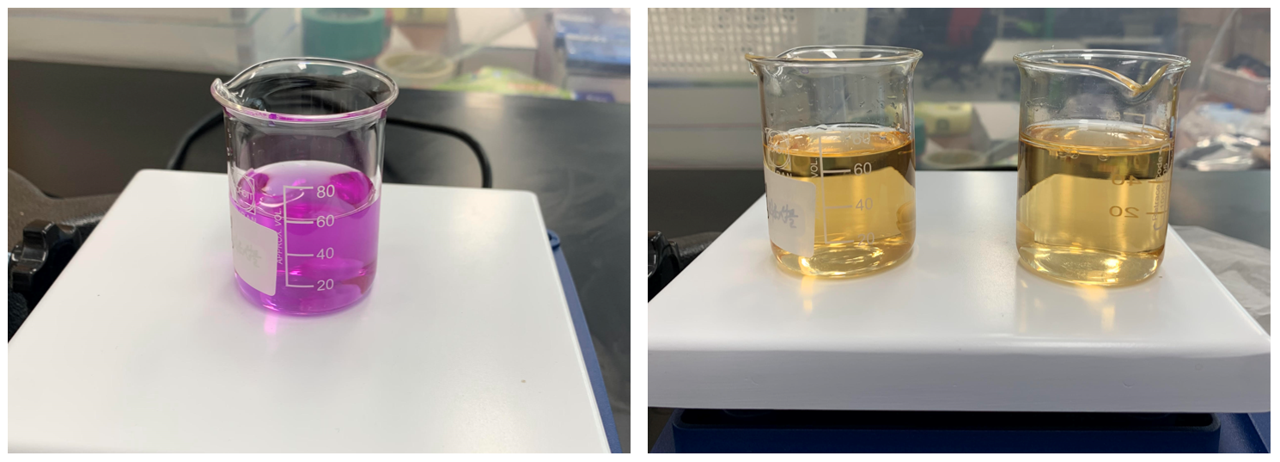}
\caption{\setlength{\baselineskip}{4mm} Ethylenediaminetetraacetic acid (EDTA) titration method. Before adding Gd to LAB (left), after completing the addition of Gd to LAB at 0.6\% concentration.} 
\label{fig:gd_lab-edta}
\end{figure}

The target vessel of the RENE detector is filled with a Gd-LS, synthesized following the method used in the RENO experiment, as described in~\cite{LSpaper}. 
To reduce accidental background noise, a 0.5\% Gd-LS mixture was used for the RENE detector. This mixture consists of Gd-doped LAB (Gd-LAB) and an LS master solution. 
A total of 240 L of a 0.6\% Gd-LAB solution was produced at the refurbished RENO LS production facility. 
The LS master solution, which is an unloaded LS formulation enriched by a factor of ten, contains 30$\,{\rm{g}/\textit{l}}$ of the primary fluor (PPO for the RENE) and 
0.3$\,{\rm{g}/\textit{l}}$ of the secondary wavelength shifter (bis-MSB for the RENE), both used specifically in the RENE detector. 
To adjust the Gd concentration, the EDTA titration method was adopted. 
As depicted in Fig.~\ref{fig:gd_lab-edta}, when the Gd concentration reaches the desired level, the solution transitions from purple to yellow.

\begin{figure}[hpt]
\centering
\includegraphics[scale=0.7]{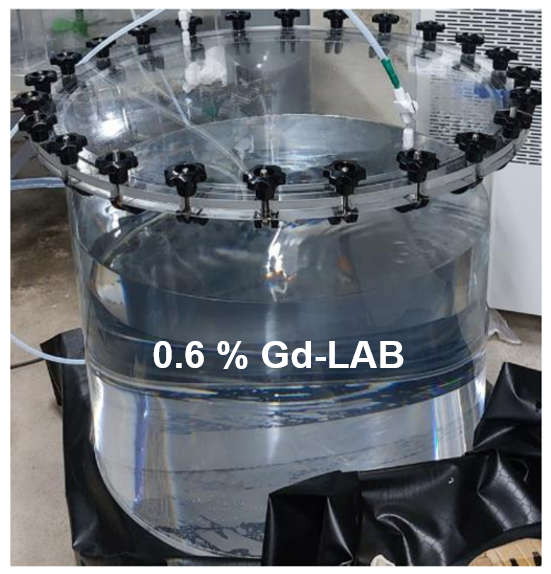}
\caption{\setlength{\baselineskip}{4mm} The produced 0.6 \% of Gd-LAB solution.} 
\label{fig:Gd-LAB-stability}
\end{figure}

RENE is designed as an above-ground experiment, where fast neutron background levels are expected to be one or two orders of magnitude higher than those in typical underground experiments. 
Effective pulse shape discrimination (PSD) is crucial for mitigating this background. 
To enhance PSD performance, 10\% by weight of a di-isopropyl-naphthalene-based commercial LS (EJ-309) is incorporated into the Gd-LS. 
The unloaded LS is used in the $\gamma$-catcher. Figure~\ref{fig:Gd-LAB-stability} presents the measured Gd concentration in the Gd-LAB solution, 
determined using the EDTA titration method. The concentration remained stable for at least 15 months after production, as independently verified by two institutions: 
Kyungpook National University (KNU) and Chonnam National University (CNU).

\begin{figure}[hpt]
\centering
\includegraphics[scale=0.3]{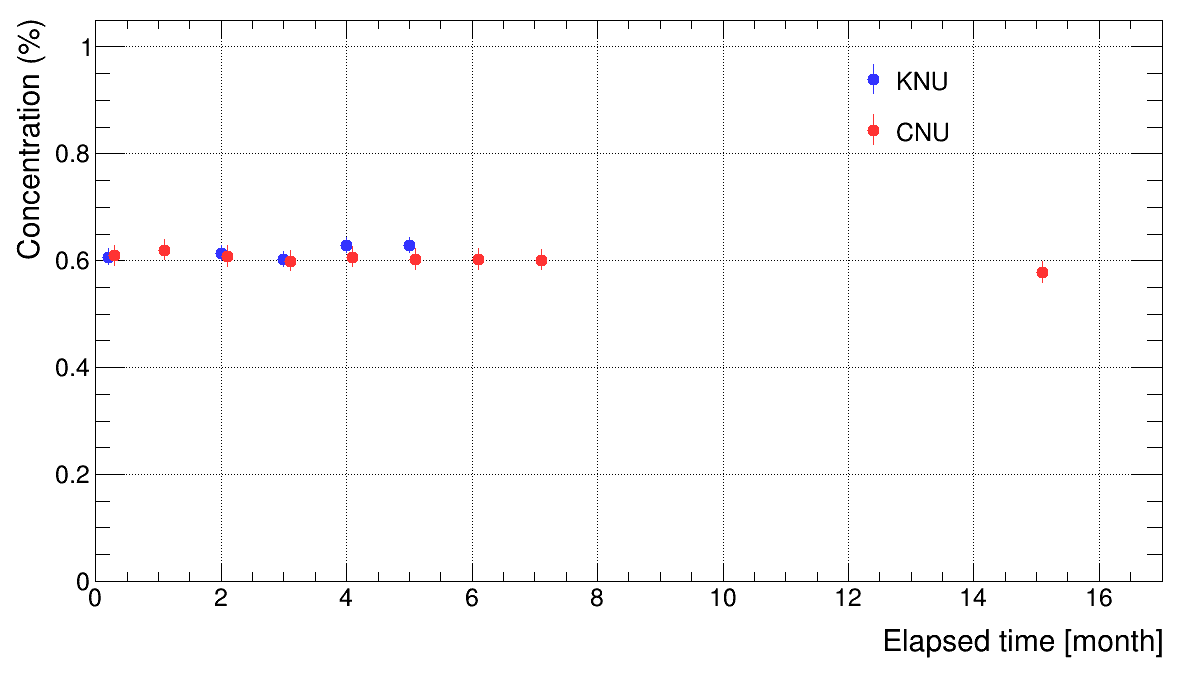}
\caption{\setlength{\baselineskip}{4mm} Long-term measurement of Gd concentration in the Gd-LAB solution, conducted independently at KNU and CNU.} 
\label{fig:Gd-LAB-stability}
\end{figure}

\newpage
\section{Data Acquisition}
\label{sec::daq}
\label{sec:detector_DAQ}

%%%%% 여기까지 잠시만 20250304 1:26

The DAQ system included in the RENE setup processes signals from the 20-inch PMTs and the PMTs in the VETO system. It consists of an analog-to-digital converter (ADC), a flash ADC (FADC), a trigger control board (TCB), and a DAQ computer. Signals from the PMTs are digitized by the FADC and ADCs before being sent to the TCB. The TCB determines whether to record data based on the digitized signals according to predefined criteria.

\begin{figure}[htb!]
    \centering
    \includegraphics[width=0.9\linewidth]{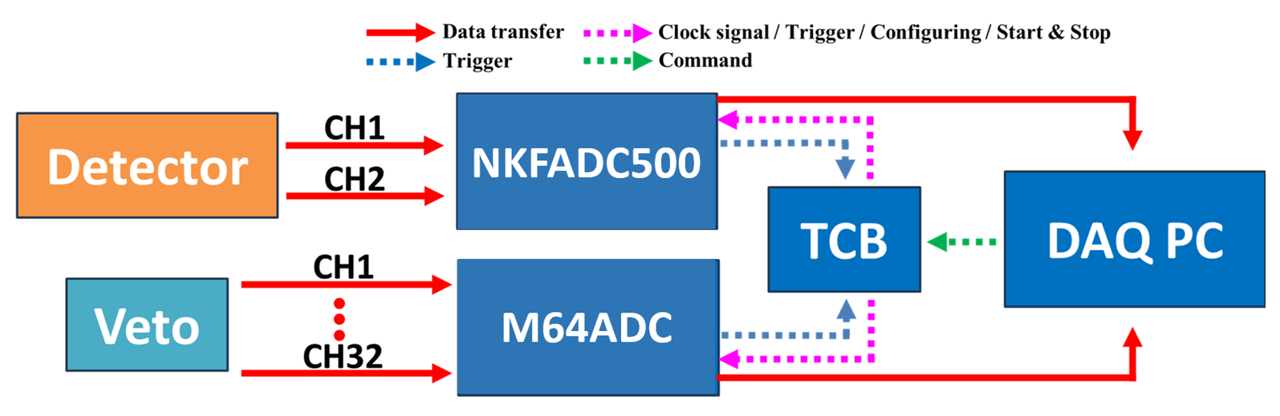}
    \caption{DAQ workflow of RENE.}
    \label{fig:daq_scheme}
\end{figure}

Figure ~\ref{fig:daq_scheme} illustrates the data flow and signal interactions among these devices. The system operates as follows: First, the DAQ computer initializes the TCB, FADC, and ADC simultaneously. Then, the FADC and ADC are configured according to predefined settings. When triggered, the FADC or M64ADC generates a hardware signal indicating how many channels exceed the predefined pulse height threshold (i.e., multiplicity) and transmits this information to the TCB. The TCB evaluates whether all ADCs should record data. Once a decision is made, the relevant events are stored in the data storage system.

\begin{figure}[htb!]
    \centering
    \includegraphics[width=0.9\linewidth]{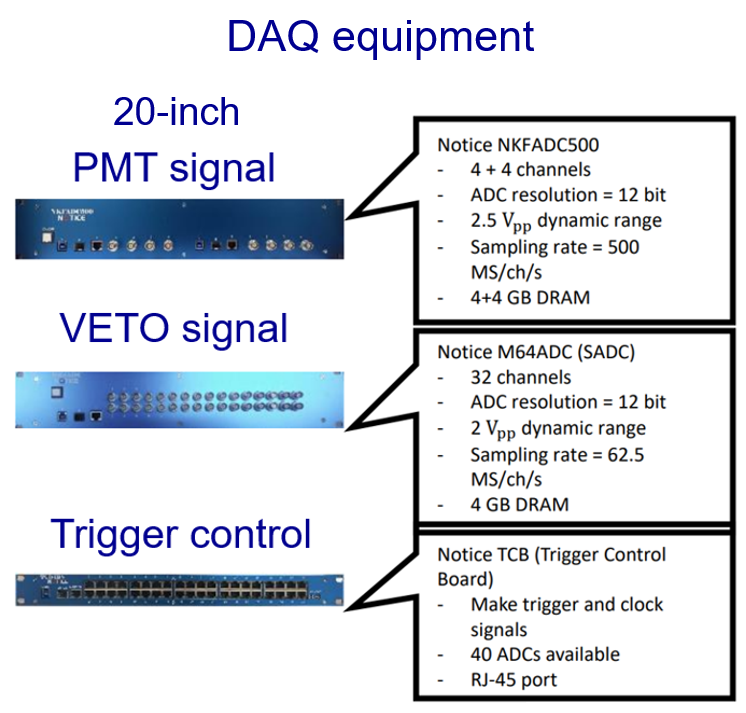}
    \caption{Electronics of the DAQ system. }
    \label{fig:electronics}
\end{figure}

The FADC (NKFADC500) digitizes signal waveforms from the 20-inch PMTs at a sampling rate of  500 MHz per channel. 
This high sampling rate is essential for performing pulse shape discrimination (PSD), a technique used to differentiate neutron and positron signals when reconstructing IBD events, 
which involve both prompt (ns) and delayed ($\mu$s) signals. The 12-bit sampling resolution provides a dynamic peak-to-peak voltage range of 2.5$\,$V $_{\rm{pp}}$. 
Each digitized unit corresponds to one FADC count, with the NKFADC500 featuring 8 GB of dynamic random-access memory (DRAM), equivalent to $0.61\,$mV$ \simeq 2.5\,$rm{V}/$2^{12}$ per count. 
The power module CAEN SY4517~\cite{CAEN} supplies the operating voltage (1750$\,\rm{V}$) for the 20-inch PMTs.

\begin{table}[ht]
\centering
\caption{DAQ system speculations}
\begin{tabular}[t]{lcc}
\hline
                        & NKFADC500   & M64ADC \\
\hline \hline
Number of Channels      & 8           &  32   \\
ADC resolution          & 12 bit      &  12 bit \\
Dynamic range           & 2.5$\,\rm{V}_{\rm{pp}}$ &  2$\,rm{V}_{\rm{pp}}$ \\
Sampling rate           & 500$\,${MS/ch/s} & 62.5$\,${MS/ch/s} \\
DRAM                    & 8$\,$GB & 4$\,$GB\\
\hline
\end{tabular}
\label{tb:daq}
\end{table}%

VETO signals are processed by the M64ADC, an ADC equipped with 32 channels to connect to the 2-inch PMTs mounted on the VETO PS panels. The M64ADC system samples signals based on the charge of the
analog input at a rate of 62.5$\,$MS/s per channel. Although this sampling rate is lower than that of the FADC, it is sufficient for rejecting incident cosmic muon background based on 
timing information The dynamic range of the M64ADC is 2$\,{\rm{V}_{\rm{pp}}}$, with a 12-bit charge resolution, and data are temporarily stored in 4$\,$GB of DRAM. 
Figure~\ref{fig:electronics} presents images of the electronics, showing the FADC, M64ADC, and TCB from top to bottom, respectively.

The DAQ workflow is illustrated in Fig.~\ref{fig:daq_scheme}. 
Signals from the two 20-inch PMTs installed within the $\gamma$-catcher are transmitted to the FADC module, where they are digitized and temporarily stored in 8$\,$GB of DRAM. 
When a digitized signal exceeds the threshold voltage of approximately 5$\,$mV, the FADC module opens a coincidence time window of 256$\,$ns. 
Once the window closes, the module records the number of triggered channels and sends the trigger bit to the TCB.

%\begin{figure}[htb!]
%    \centering
%    \includegraphics[width=0.8\linewidth]{Figures/daq_trg_id.png}
%    \includegraphics[width=0.8\linewidth]{Figures/daq_trg_veto.png}
%    \caption{Trigger identification of signals from 20-inch PMTs (top) and VETO system (bottom)}
%    \label{fig:Trigger timing ID}
%\end{figure}

The TCB performs essential functions, including trigger signal generation, logic processing, and clock distribution. 
Equipped with 40 RJ-45 ports, it can interface with up to 40 compatible FADCs and M64ADCs. The TCB distributes the clock signal across these devices, collects their multiplicity or trigger bit information, and applies predefined logic to determine the trigger. Additionally, it counts event numbers and transmits both the trigger signal and event number to the connected FADCs and M64ADCs.

\begin{table}[ht]
\centering
\caption{DAQ computing system speculations}
\begin{tabular}[t]{ll}
\hline
                        & Speculations    \\
\hline
\hline
CPU                     & Intel Xeon W3-2423 (15 MB cache, 6 core, 12 threads, 2.1 GHz to 4.2 GHz)   \\
OS                      & Rocky Linux 8 \\
RAM                     & 32 GB, $2\times16$ GB, DDR5, 4800 MHz, RDIMM ECC \\
GPU                     & NVIDIA\textsuperscript{\textregistered} T400, 4 GB GDDR6 \\
SSD                     & 512 GB, M.2, PCIe NVMe, SSD, Class 40 \\ 
HDD                     & $4\times12\,$TB, 7200$\,$RPM, 3.5 inch, SATA\\
Power                   & Precision 5860 Tower 750 W \\
\hline
\end{tabular}
\label{tb:daqpc}
\end{table}%

When an FADC module sends a trigger bit to the TCB, the TCB evaluates whether the received trigger bit meets the predefined threshold conditions. 
If the conditions are satisfied, the TCB sends the trigger signal to both the FADC and M64ADC. Once the window closes, the TCB transmits the start time for waveform recording to the FADC, which is
set to 996$\,$ns before the end of the coincidence window. The FADC records waveforms with a duration of 996$\,$ns for the active channels. Simultaneously, 
the M64ADC records waveforms within the same event window as the FADC to facilitate the rejection of external particles during data analysis. 
Both the FADC and M64ADC then transmit the temporarily recorded waveforms to the DAQ server for storage. The details are listed in Table~\ref{tb:daq}.
Finally, the events that satisfy the trigger requirement will be recorded, obtaining a muon signal in the veto system and detecting a signal from the 20-inch PMTs within the IBD time window, which is 150 $\mu$s, between the prompt and delayed signals. During the IBD analysis, events triggered by the veto system will be excluded, as they are clearly induced by cosmic muons.

\begin{figure}
    \centering
    \includegraphics[width=0.9\linewidth]{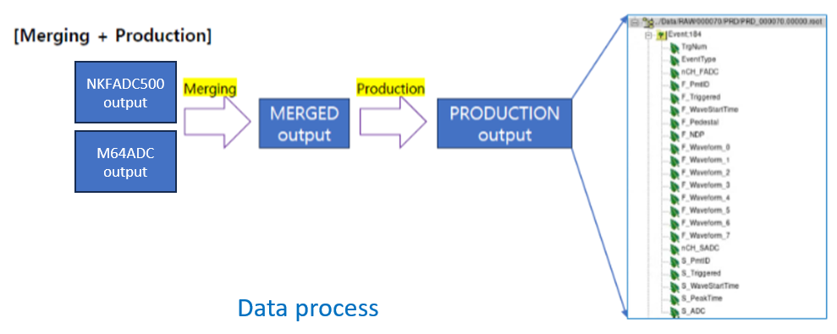}
    \caption{Data processing sequence}
    \label{fig:Event build}
\end{figure}

The stored raw data are analyzed using the ROOT toolkit~\cite{root}. After merging the FADC and M64ADC data, the pre-processed data are recorded in TTree format. The left panel of Fig.~\ref{fig:Event build} illustrates the data processing scheme, while the right panel shows the run control system. The run control system issues commands to the DAQ components and establishes the operating conditions. The shift crew operates through an integrated graphical user interface (GUI), which allows the selection of the run mode, trigger type, and detector parameters.

Available run modes in the system include data recording and calibration, while the trigger type can be selected from predefined trigger sets. Detector parameters, such as high voltage settings for the PMTs, can also be configured through the interface. The DAQ process is managed by a dedicated computing system, which must operate stably during data taking and provide sufficient memory and storage. The detailed specifications are listed in Table~\ref{tb:daqpc}.

Regarding the expected data volume, each event generates 2$\,$kB of data from the FADC and 0.4$\,$kB from the M64ADC. 
This results in an estimated daily data output of approximately 100$\,$GB. 
Over a 300-day period, the total data accumulation is projected to reach 30$\,$TB.

\begin{figure}
    \centering
    \includegraphics[width=1\linewidth]{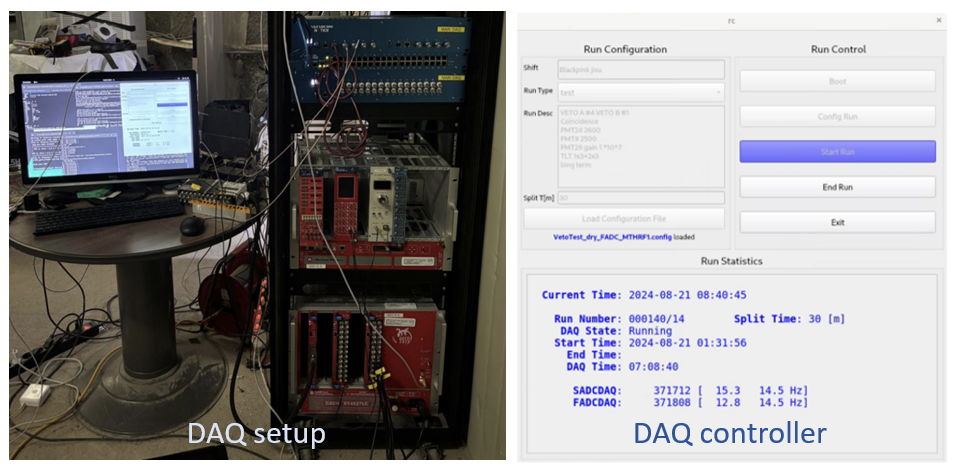}
    \caption{DAQ setup (left) and run controller (right).}
    \label{fig:DAQ_setup}
\end{figure}

\newpage
\section{HV and Detector Control Monitoring}
\label{sec:detector_HVSCM}

%%%%  여기까지 20250304 4:49

The RENE detector must be reliably controlled to meet the safety requirements of the nuclear power plant. Variables related to the detector and its environment must be continuously monitored to ensure optimal performance and prevent safety issues. The detector system must be designed to remain robust against anomalies and environmental changes.

To meet the above requirements, the RENE setup includes an SCM, which is based on the Supervisory Control and Data Acquisition (SCADA) concept widely used for system management. 
The SCM records various environmental variables, including ambient temperature, oxygen concentration, liquid level, and radioactivity. It also monitors and controls the HV applied to the 20-inch PMTs and the VETO system. A supervisory computer processes sensor data and manages the detector. Additionally, the sensor-recorded data can be visualized on the computer display during scheduled on-site operations.

The ADC module used for acquiring detector variables is obtained from National Instruments (NI)~\cite{NI}. Accordingly, an NI cDAQ USB/Ethernet crate capable of accommodating NI slot modules was selected for the RENE setup. 
The Laboratory Virtual Instrument Engineering Workbench (LabVIEW) a graphical programming environment compatible with NI systems, supports SCADA-format frameworks and is used for system control. The NI cDAQ system is managed through the Data Acquisition Measurement and Automation Explorer (DAQmx) module, while serial communication is handled via the Virtual Instrument Software Architecture (VISA) module adopting the Universal asynchronous receiver/transmitter (UART) format. In the RENE system, both GUI-based programming in LabVIEW and text-based programming in NI-DAQmx have been implemented to enable monitoring and recording functions.

The primary variables monitored by the SCM can be categorized as follows:
\begin{itemize}
    \item \textbf{LS status}: LS temperature and level to prevent potential chemicals.
    \item \textbf{Environment}: Temperature and humidity at the experimental site.
    \item \textbf{Radioactivity}: Concentration of radioactive elements.
    \item \textbf{HV supply}: HV supply for the PMTs.
    %\item \textbf{Oxygen level (optional)}: Oxygen level for operator's safety
    %\item \textbf{Geomagnetism (optional)}: External magnetic field for correction
    %\item \textbf{Video surveillance (optional)}: Photographic records of the detector system and analogue meters.
\end{itemize}

\begin{figure}[h]
\centering
\includegraphics[width=0.8\textwidth]{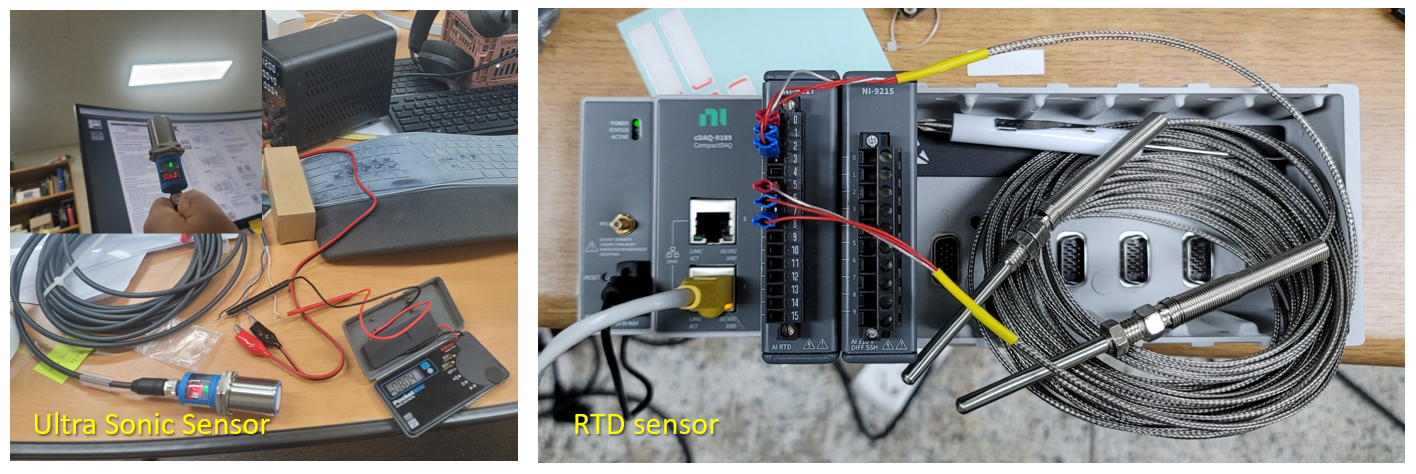}
\caption{\setlength{\baselineskip}{5mm} US sensor for monitoring the LS level (left) and RTD sensor for
measuring the LS temperature (right).}
\label{fig:SCM_sensor1}
\end{figure}

Although the LS is inflammable, the LS level and temperature are among the most critical variables to be monitored for ensuring general safety. In the RENE system, the primary sensors for monitoring
the LS level and temperature are installed on the lid of the $\gamma$-catcher chamber, as depicted in Fig.~\ref{fig:SCM_sensor1}. An ultrasonic (US) sensor provides a precise, non-contact measurement
of the LS level in the $\gamma$-catcher chamber. The required accuracy of the level sensor is less than 0.1$\,^{\circ}\mathrm{C}$.  Notably, the LS temperature must be monitored concurrently with the LS
level as the LS undergoes thermal expansion. LS temperature is a crucial variable that affects detector performance and stability. Resistance temperature detector (RTD) sensors are included in the
RENE setup to measure LS temperature at two to three positions, with an accuracy requirement of $< 0.1\,^{\circ}\mathrm{C}$.

\begin{figure}[h]
\centering
\includegraphics[scale=0.7]{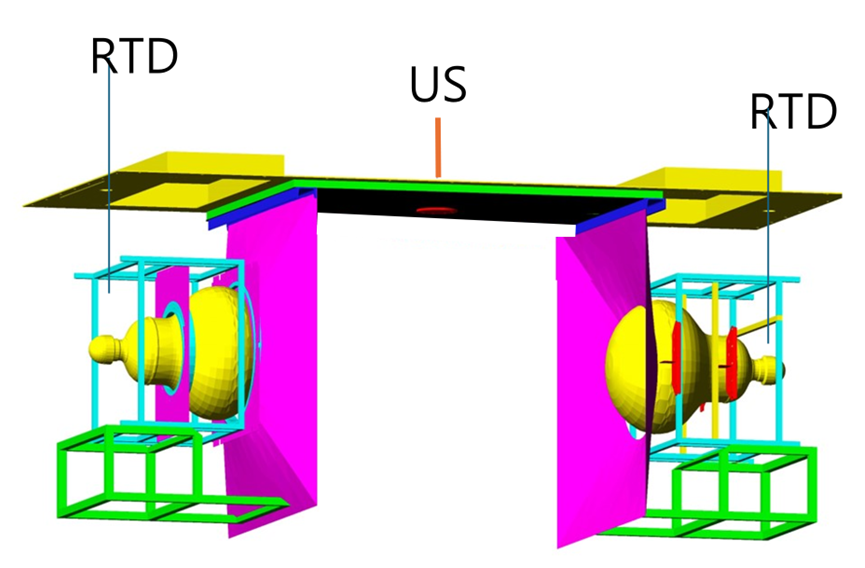}
\caption{\setlength{\baselineskip}{5mm} RTD and US sensors attached to the lid of the $\gamma$-catcher chamber}
\label{fig:HVSCM}
\end{figure}

The US sensor functions as both a transmitter and a receiver. In the transmitting mode, the sensor emits sound waves toward a target surface, and in the receiving mode, it detects the reflected echoes.   
The SICK UM model outputs an analog voltage when supplied with 24$\,$V and an analog current when the supply voltage is lower.
Voltage output is typically preferred for short monitoring distances with minimal electrical noise, whereas current output is better suited for harsher environments.

The PT-100 RTD device is made from platinum with a resistance of 100$\,{\rm{\Omega}}$ and adopts a three-wire configuration. 
The NI 9217 module, specifically designed for RTDs, supplies power to the RTD elements. The probe is constructed from corrosion-resistant stainless steel.

\begin{figure}[!h]
\centering
\includegraphics[width=0.8\textwidth]{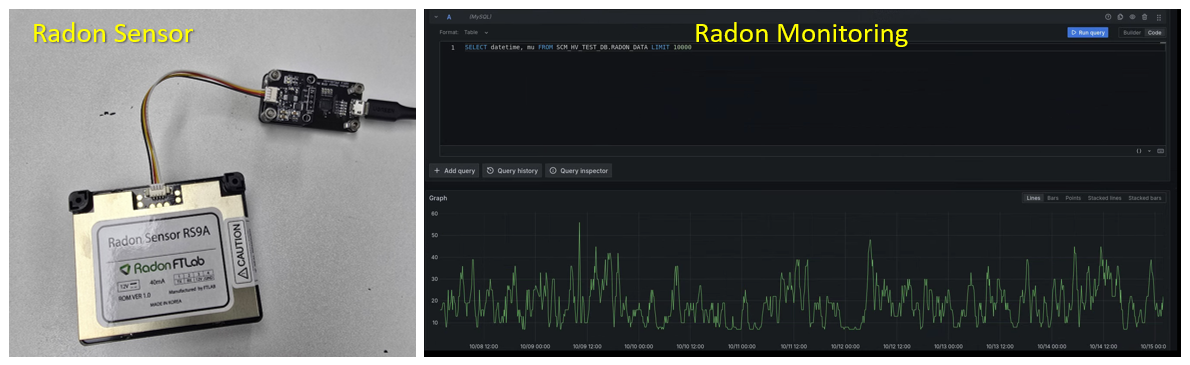}
\caption{\setlength{\baselineskip}{5mm} Radon sensor (left) and radon monitoring display (right).}
\label{fig:SCM_sensor2}
\end{figure}

Radon detectors monitor variations in environmental radioactivity. These detectors must be capable of measuring radioactivity levels ranging 
from 1 pCi/L 100 pCi/L with an accuracy of $< 2\,$pCi/L and 
a compact size of 10$\,$cm$\times$10$\,$cm.
The radon detector selected for the RENE setup is illustrated in Fig.~\ref{fig:SCM_sensor2}. 
It adopts a miniaturized double differential amplified pulse ion chamber design for detection and signal processing~\cite{RADON}. 
A double differential amplifier circuit is used to address the low signal-to-noise ratio (SNR) issue caused by 
the high input impedance of the reference detection circuit and ionization chamber. 
This configuration effectively eliminates common-mode noise during signal amplification, improving the SNR. 
The system’s setup time is limited to within 60 min, and the minimum measurement cycle is set to 10 min. 
The right panel of Fig.~\ref{fig:SCM_sensor2} illustrates radon monitoring data recorded at 10 min intervals.

\begin{figure}[h]
\centering
\includegraphics[scale=0.7]{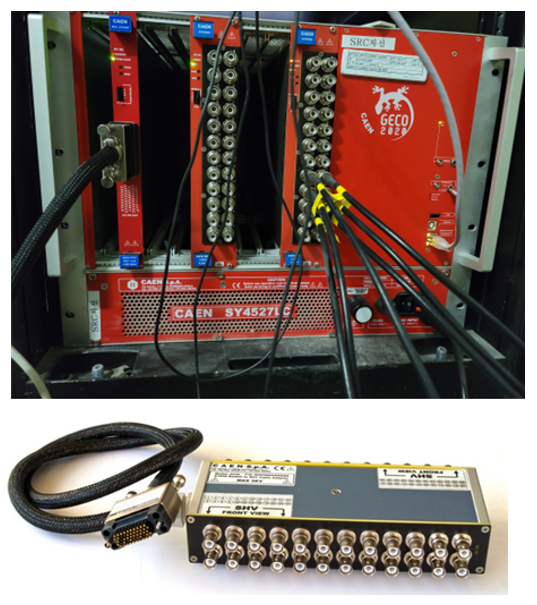}
\caption{\setlength{\baselineskip}{5mm}HV supply system for the RENE detector.}
\label{fig:SCM_HVS}
\end{figure}

Maintaining the amplification ratio of dynode-type PMTs within the target range requires precise adjustment of the supply voltage. 
The difference between the voltage of the HV supply device and the actual voltage applied to the PMT must remain within an acceptable tolerance. 
The HV module requires an accuracy of $< 1\%$ and a voltage stability of 10$\,$V per week on 1000$\,$V. 
The RENE detector employs the CAEN SY4527LC system to reliably supply and control the HV of 1700$\,$V at 1$\,$mA, as shown in Fig.~\ref{fig:HVSCM}. 
The RENE configuration is planned to include an A7030P board, two A7435SN boards, and one A648 adapter, all mounted onto the SY4527LC crate. 
This setup is anticipated to supply the necessary voltage and current to the PMT via the A7030P.

Table~~\ref{tb:scm} summarizes the monitoring variables and selected sensor models.
\begin{table}[ht]
\centering
\caption{Summary of SCM sensors and speculations}
\label{tb:scm}
\begin{tabular}[t]{llll}
\hline
Monitoring                  & Sensor Model & Communication \\
Variable                    &              & Module \\
\hline
\hline
LS level                    & SICK UM30-213113  & NI 9201/9203  \\ \hline
LS temperature              & PT-100            & NI 9217       \\ \hline
Radon concentration         & RS9A              & UART          \\ \hline
Environmental Temperature   & DHT-22            & UART          \\
and Humidity                &                   &               \\ \hline
Leakage                     & HCSR04+           & UART          \\ \hline
\end{tabular}
\end{table}%

We also developed a GUI for the HV control system to enable intuitive and advanced control of the RENE detector, 
as shown in Fig.~\ref{fig:HVSCM}. The HV control system SY4527LC was implemented using the CAEN GECO 2020 software 
and the \textsc{CAEN} HV Wrapper library for the Linux operating environment. 
Connections between the SY4527LC device and the control computer were configured via transmission control protocol/internet protocol (IP), using a fixed IP address and port number 4527. 
The software includes an additional feature for recording long-term data history, which is interfaced with the\textsc{MariaDB}~\cite{MariaDB} database software.

\begin{figure}[h]
\centering
\includegraphics[scale=0.7]{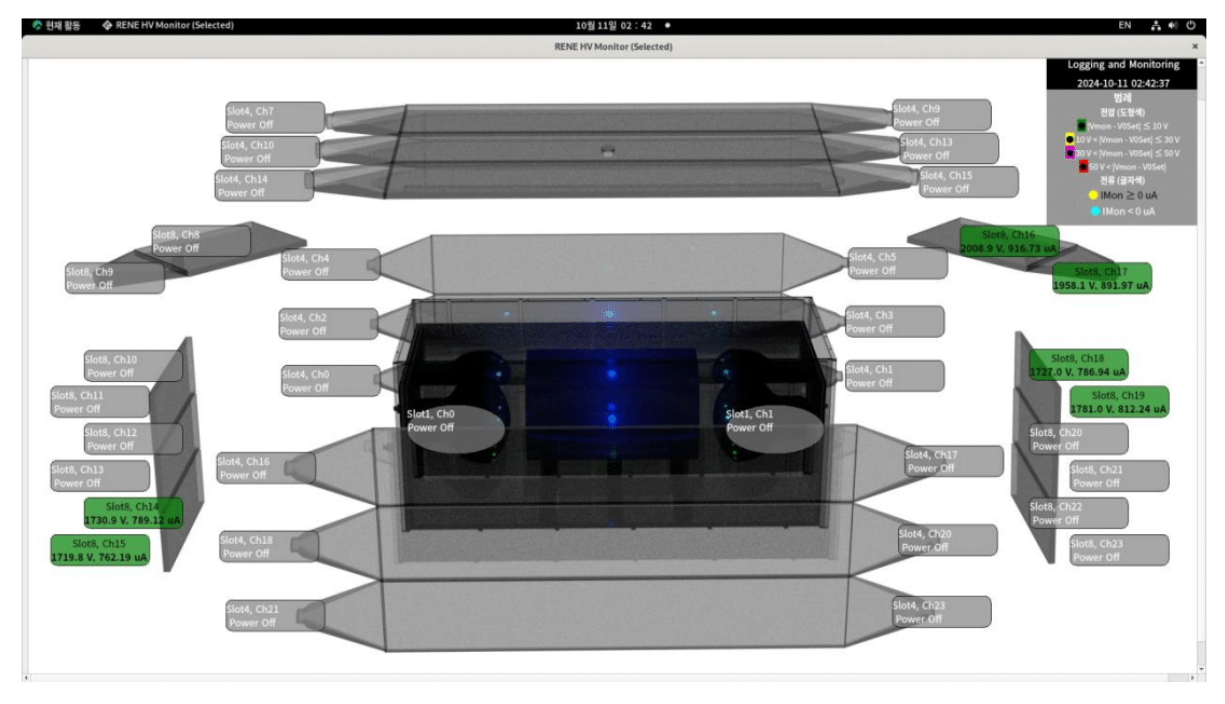}
\caption{\setlength{\baselineskip}{5mm} HV GUI system}
\label{fig:HVSCM}
\end{figure}

The communication and GUI display update cycle between the SY4527LC and the host PC is set to 1 s, 
while data transmission to the host PC and the database server occurs every 1 min. 
Stable communication between the SY4527LC and the host PC is maintained by calling the \verb|get_ch_param()| method, 
instead of the event-based \verb|subscribe_channel_params()| method, at a polling interval of 1 s to retrieve data, along with an implemented keep-alive function. The HV GUI created using this implementation is shown in Fig.~\ref{fig:HVSCM}.

\newpage
\section{Detector Performance}
\label{sec::expected_results}

\subsection{MC Simulation}

The performance of the RENE detector was evaluated using a detailed MC simulation based on GLG4SIM~\cite{GLG4SIM}. 
This simulation included detailed modeling of the detector geometry, encompassing the target vessel, \GC, 
reflector cones, and 20-inch PMTs. It was designed to analyze the detector’s response to IBD events.

This simulation framework was selected owing to its remarkable capabilities in modeling LS detectors, 
enabling accurate simulation of both prompt and delayed signals. The geometry used in the simulation corresponds to the actual detector design, ensuring that the results closely represent the expected detector performance.

\begin{figure}[hpt]
\centering
\includegraphics[scale=1.0]{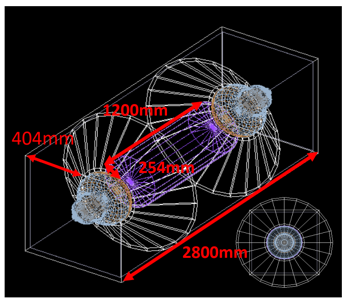}
\caption{\setlength{\baselineskip}{4mm} Schematic view of the RENE detector modeled using the GLG4SIM MC simulation} 
\label{fig:gd_lab-edta}
\end{figure}

\subsection{Energy Resolution}
\label{subsec:EneRes}

The energy resolution critically impacts the detector’s ability to reconstruct neutrino events and to resolve spectral distortions arising from oscillation phenomena with high precision.  
The energy resolution of the RENE detector was evaluated using positrons from IBD events.

For monoenergetic neutrinos undergoing IBD, the momenta of the resulting positrons and neutrons were calculated. Assuming that IBD events occur uniformly throughout the detector volume, each event was assigned a random vertex position. Based on the positron momentum and position, a full detector simulation was performed to estimate the number of photoelectrons (\( N_{\mathrm{PE}} \)) generated in the photodetectors.

The true prompt energy, \( E_{\mathrm{prompt}}^{\mathrm{true}} \), is defined as the kinetic energy of the positron plus the energy of its annihilation gammas. This quantity is directly calculated from the positron kinematics in the simulation and can be expressed as:
\[
E_{\mathrm{prompt}}^{\mathrm{true}} = T_{e^+} + 2m_e,
\]
where \( T_{e^+} \) is the positron kinetic energy and \( m_e \) is the electron mass (511$\,$keV). The two annihilation gammas, each with energy \( m_e \), are assumed to deposit their energy fully within the detector volume.

Under the assumption of negligible neutron recoil, the positron kinetic energy can be approximated by the difference between the incoming neutrino energy, \( E_\nu \), and the reaction threshold:
\[
T_{e^+} \approx E_\nu - \Delta,
\]
where \( \Delta = m_n - m_p = 1.293\,\mathrm{MeV} \) is the mass difference between the neutron and the proton. Substituting this into the expression for \( E_{\mathrm{prompt}}^{\mathrm{true}} \), we obtain:
\[
E_{\mathrm{prompt}}^{\mathrm{true}} \approx E_\nu - \Delta + 2m_e = E_\nu - 0.782\,\mathrm{MeV}.
\]
This linear relationship between the neutrino energy and the true prompt energy is widely used in reactor neutrino experiments to estimate the visible energy deposited in the detector.

To estimate the observed prompt energy, \( E_{\mathrm{prompt}} \), the number of photoelectrons was converted into energy using an energy-conversion function.
This function, \( E_{\mathrm{prompt}}(N_{\mathrm{PE}}) \), was derived from simulations of monoenergetic positrons uniformly generated in the target volume.
For each input energy value, a large number of positron events (\( N_{\mathrm{sim}} \)) were simulated to determine the average photoelectron yield, \( \langle N_{\mathrm{PE}} \rangle \).
This mapping defines the function \( E_{\mathrm{prompt}}(N_{\mathrm{PE}}) \), which was then applied to the full simulation dataset to obtain the prompt energy distribution.

\begin{figure}[b]
\centering
\includegraphics[scale=0.38]{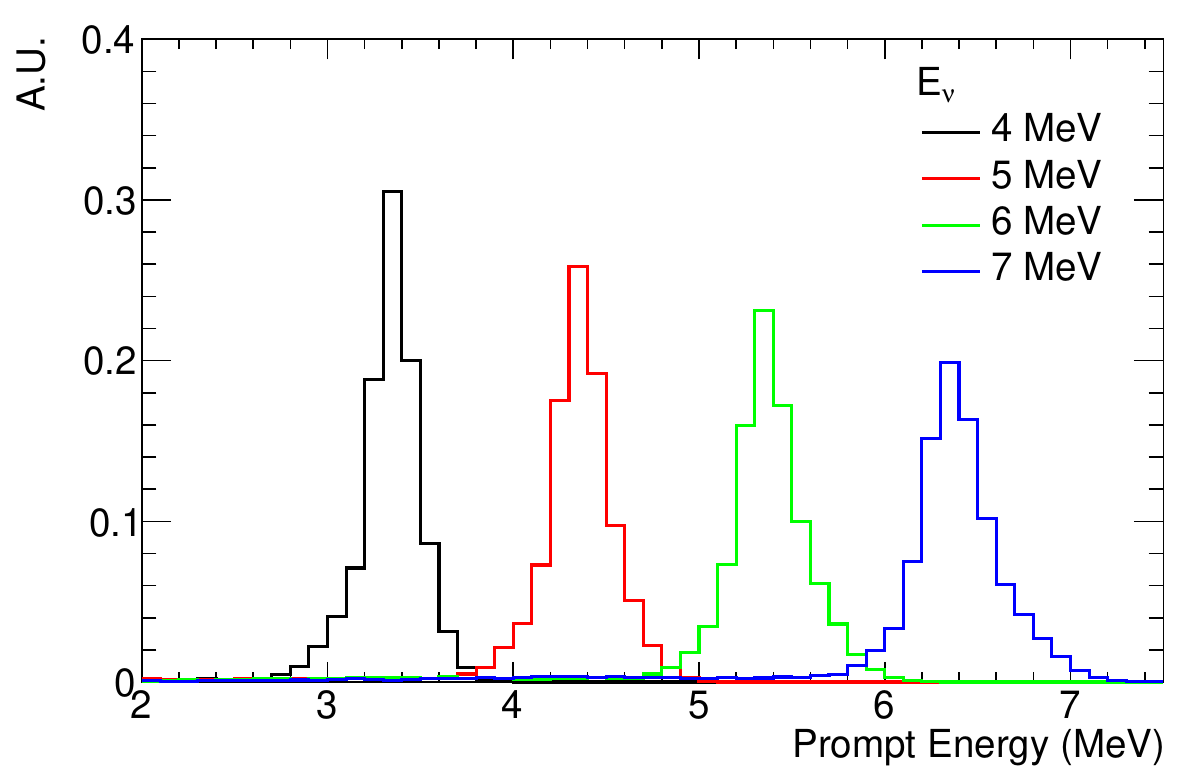}
\includegraphics[scale=0.38]{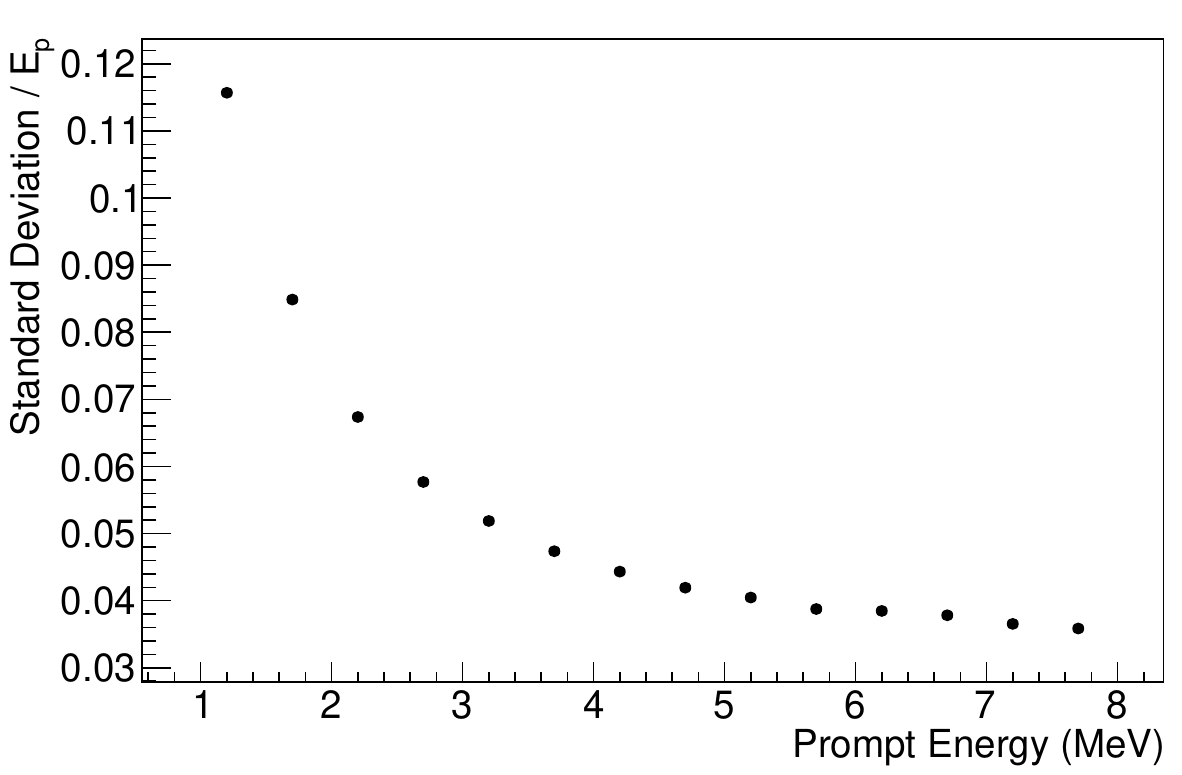}
\caption{Simulated prompt energy spectra for four neutrino energies (left) and energy resolution as a function of true prompt energy (right). Events below $\sim$0.7$\,$MeV were excluded from each spectrum to reduce the influence of low-energy tails on the resolution calculation.}
\label{fig:energy_resol}
\end{figure}

From these simulations, the detector response function \( R(E_{\mathrm{prompt}}; E_\nu) \) was constructed numerically:
\begin{equation}
R(E_{\mathrm{prompt}}; E_\nu) = \frac{1}{N_{\text{sim}} (E_\nu)} \frac{dN_{\text{sim}}}{dE_{\mathrm{prompt}}}(E_{\mathrm{prompt}}; E_\nu),
\end{equation}
where \( dN_{\text{sim}}/dE_{\mathrm{prompt}} \) is the distribution of prompt energies for a fixed neutrino energy \( E_\nu \), and \( N_{\text{sim}} \) is the total number of simulated events at that energy. This function satisfies the normalization condition:
\begin{equation}
\int R(E_{\mathrm{prompt}}; E_\nu) \, dE_{\mathrm{prompt}} = 1.
\end{equation}
This response function, constructed from the simulation, is used directly in the spectrum folding process described in Section~\ref{subsec:ExpectedSpectrum}.

Figure~\ref{fig:energy_resol} shows the simulated prompt energy spectra for four representative neutrino energy values (left) and the resulting energy resolution as a function of the true prompt energy (right).  
To obtain a more representative estimate of the energy spread, events with prompt energy below approximately 0.7$\,$MeV were excluded to mitigate the effect of asymmetric low-energy tails on the standard deviation:
\begin{equation}
\frac{\sigma_E}{E_{\mathrm{prompt}}^{\mathrm{true}}} = \frac{\sqrt{\langle E_{\mathrm{prompt}}^2 \rangle - \langle E_{\mathrm{prompt}} \rangle^2}}{E_{\mathrm{prompt}}^{\mathrm{true}}}.
\end{equation}
Compared to the energy resolution reported by the NEOS experiment (as shown in Fig.~\ref{fig:2nd_peak_RENO_NEOS}), the RENE detector demonstrates improved performance.
In the high-energy region, the energy resolution of the RENE detector reaches approximately 4\,\%, which corresponds to an estimated $\sim$20\,\% enhancement in sensitivity to sterile neutrino oscillations around $\Delta m^2_{41} \sim 2\,\text{eV}^2$,
%compared to the case where the NEOS-like resolution of approximately 10--20\,\% is assumed for the same detector.
compared to the case where a lower-resolution scenario similar to that of the NEOS experiment is assumed for the same detector.

%------------------------------------------------------------

\color{black}

\subsection{Expected Energy Spectrum}
\label{subsec:ExpectedSpectrum}

\begin{figure}[b]
\begin{center}
\includegraphics[scale=0.6]{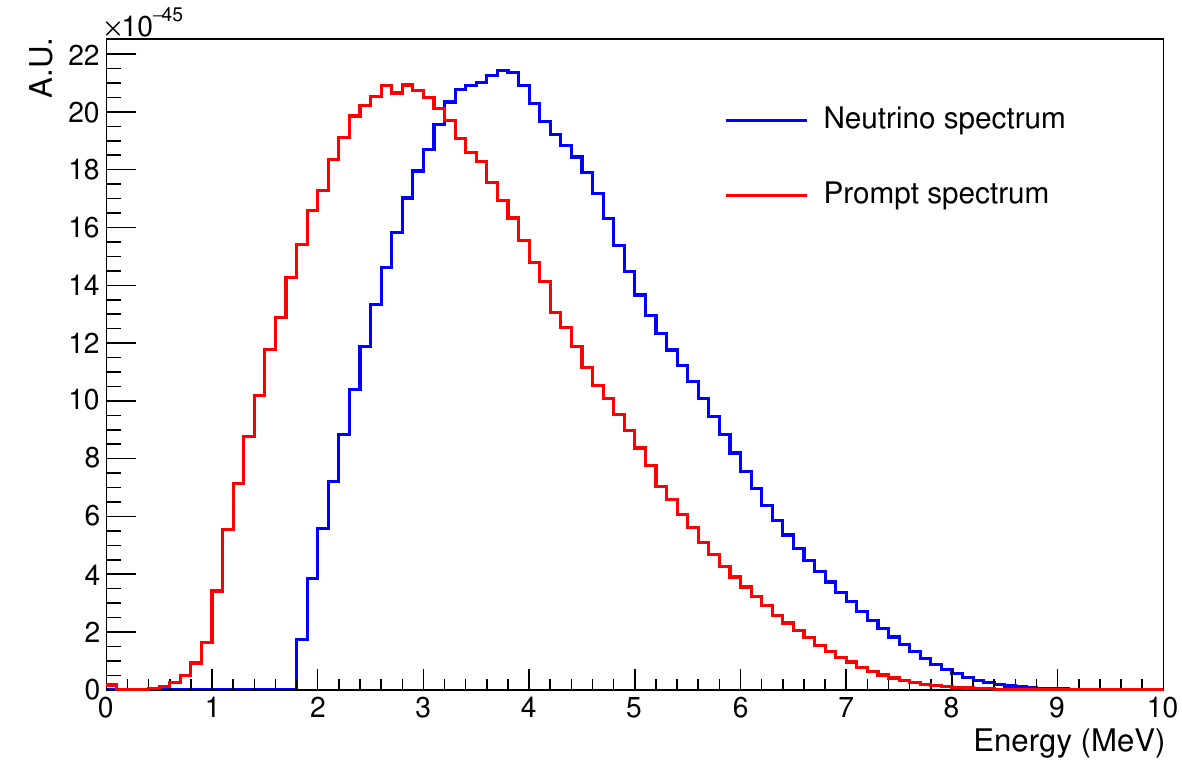}
\end{center}
\caption{\setlength{\baselineskip}{5mm} Theoretical energy spectrum of reactor antineutrinos weighted by the IBD cross section (blue) and the corresponding prompt energy spectrum after detector response folding (red).}
\label{fig:energy_spectrum}
\end{figure}

Figure~\ref{fig:energy_spectrum} shows both the theoretical IBD neutrino energy spectrum and the corresponding prompt energy spectrum after detector response.  
The expected IBD event rate induced by reactor antineutrinos is given by:
\begin{equation}
\frac{dN}{dE_\nu} (E_\nu) = N_p T \int \frac{1}{4\pi L^2} \, \phi(E_\nu) \, \sigma_{\mathrm{IBD}}(E_\nu) \, \epsilon \, P(L) \, dL,
\label{eq:ibd_spectrum_with_baseline}
\end{equation}
where \( \frac{dN}{dE_\nu} \) is the differential event rate,  
\( N_p \) is the number of target protons,  
\( T \) is the exposure time,  
\( L \) is the baseline,  
\( \phi(E_\nu) \) is the antineutrino flux,  
\( \sigma_{\mathrm{IBD}}(E_\nu) \) is the IBD cross section~\cite{Vogel:1999zy},  
\( \epsilon \) is the detection efficiency, and  
\( P(L) \) is the normalized baseline distribution accounting for the finite sizes of both the reactor core and the detector.
The baseline distribution \( P(L) \) was calculated based on the geometry of the detector and the reactor core.  
The overall detection efficiency \( \epsilon \) was inferred by matching the expected IBD rate from Eq.~\eqref{eq:ibd_spectrum_with_baseline} to the observed rate in the NEOS experiment (approximately 1976 events/day)~\cite{NEOS1}, and then applied to the RENE detector with appropriate scaling by the volume ratio.

The reactor antineutrino flux \( \phi(E_\nu) \) is modeled as a sum of fission contributions from four primary isotopes:
\begin{equation}
\phi(E_\nu) = \sum_i f_i S_i(E_\nu),
\label{eq:flux_sum}
\end{equation}
where \( f_i \) is the fission fraction and \( S_i(E_\nu) \) is the energy spectrum per fission of isotope \( i \in \{ ^{235}\mathrm{U}, ^{238}\mathrm{U}, ^{239}\mathrm{Pu}, ^{241}\mathrm{Pu} \} \).
The fission fractions are based on the NEOS experiment~\cite{NEOS1}.
%:
%\begin{align}
%f_{^{235}\mathrm{U}} &= 0.565, \quad
%f_{^{239}\mathrm{Pu}} = 0.304, \nonumber \\
%f_{^{238}\mathrm{U}} &= 0.076, \quad
%f_{^{241}\mathrm{Pu}} = 0.055.
%\end{align}
The isotopic spectra \( S_i(E_\nu) \) are taken from the Huber–Mueller model~\cite{StNTh1,StNTh2}, where the spectra for \({}^{235}\mathrm{U}\), \({}^{239}\mathrm{Pu}\), and \({}^{241}\mathrm{Pu}\) are based on the Huber model, while the \({}^{238}\mathrm{U}\) spectrum is taken from the Mueller calculation.
To incorporate neutrino oscillations, the flux is modified by the survival probability:
\begin{equation}
\phi_{\mathrm{osc}}(E_\nu) = \phi(E_\nu) \cdot P_{\bar{\nu}_e \rightarrow \bar{\nu}_e}(E_\nu, L).
\end{equation}

The prompt energy spectrum observed in the detector is obtained by folding the neutrino energy spectrum with the response function constructed in Section~\ref{subsec:EneRes}:
\begin{equation}
\frac{dN}{dE_{\text{prompt}}}(E_{\text{prompt}}) = \int \frac{dN}{dE_\nu}(E_\nu) \cdot R(E_{\text{prompt}}; E_\nu) \, dE_\nu.
\label{eq:reconstructed_spectrum}
\end{equation}

\color{black}

%----------------------

\begin{figure}[h]
\begin{center}
\includegraphics[scale=0.53]{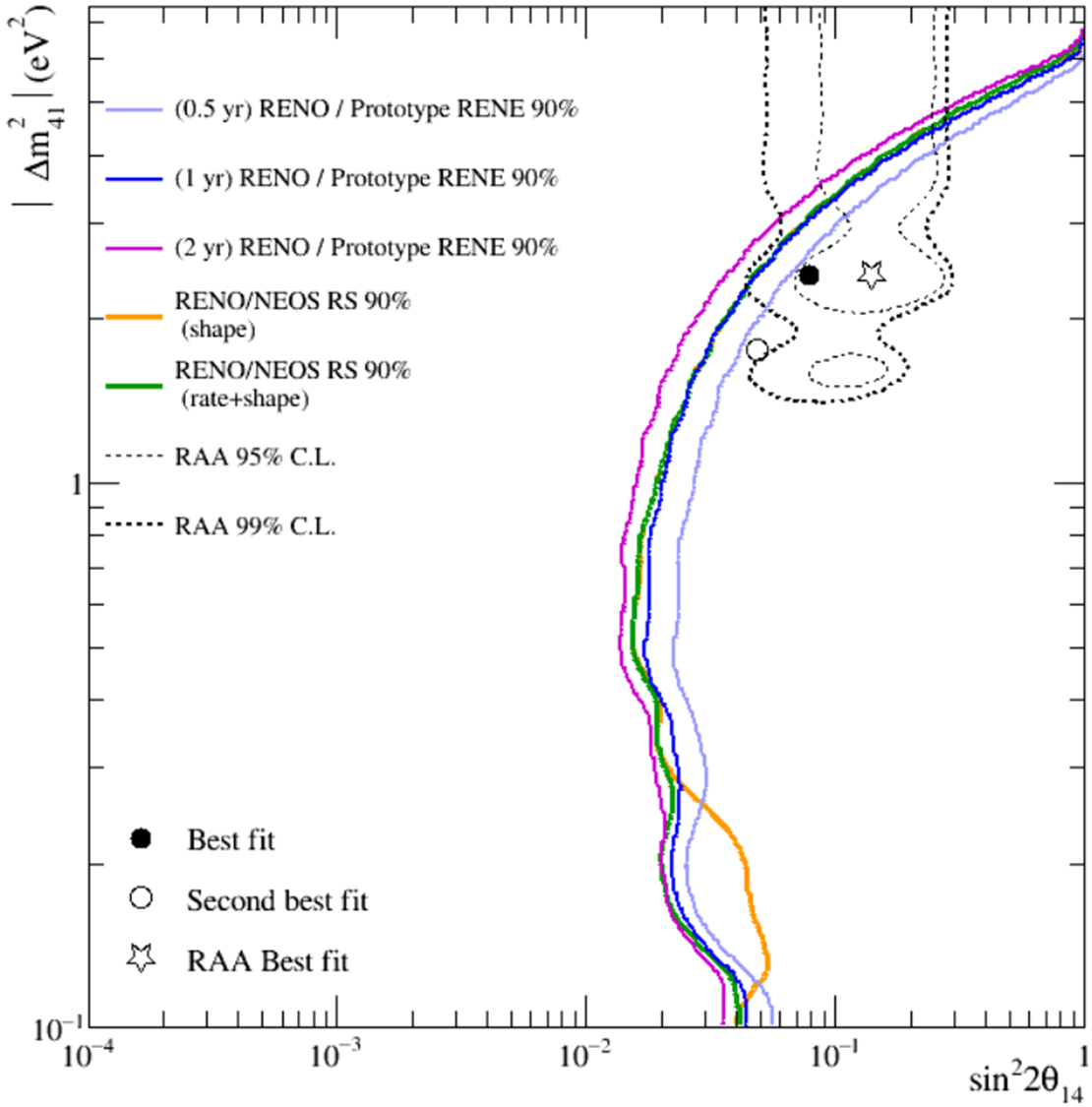}
\includegraphics[scale=0.5]{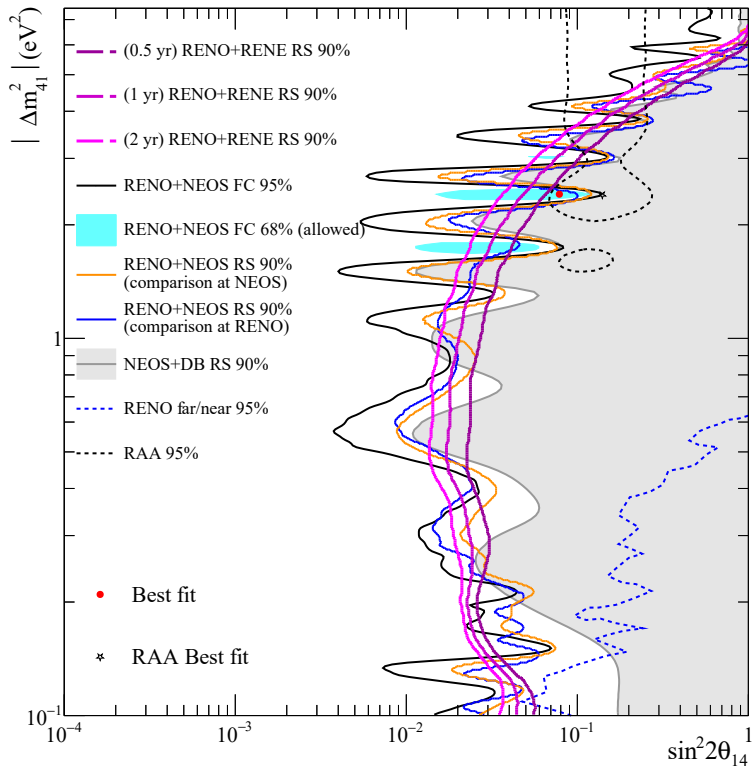}
\end{center}
\caption{\setlength{\baselineskip}{5mm} Sensitivity of parameter space derived from the Reactor Antineutrino Anomaly (RAA) and RENE (top).
Comparison of RENE, RENO, and NEOS joint analysis results (bottom).
}
\label{fig:sensitivity}
\end{figure}

\subsection{Expected Sensitivity}

The sensitivity of the RENE detector to sterile neutrino oscillations was evaluated based on the MC simulation results. 
Based on the simulated data, the RENE detector is anticipated to detect approximately 300 IBD events per day. 
This high event rate, combined with improved energy resolution, is expected to allow RENE to probe sterile neutrino oscillations 
in the $\Delta m^{2}_{14} \sim 2\,{\rm{eV}^{2}}$ region with high precision.

This analysis employs the same methodology as the joint RENO/NEOS sterile neutrino search~\cite{NEOSRENO1},  
where RENO near-detector data serve as a reference to cancel reactor flux uncertainties through relative spectral comparison.  
Accordingly, the simulated RENE spectrum is compared against RENO data to evaluate oscillation sensitivity.

The sensitivity of the detector in the parameter space was calculated using the $\chi^2$ method as described 
in eq.~\ref{eq:chi2}.
%in eq.~\ref{eq:chi22}.

\begin{equation}
\begin{aligned}
\chi^2 =& \sum_{i,j} \left( O_{\mathrm{RENO}}^i - \frac{E_{\mathrm{RENO}}^i}{E_{\mathrm{RENE}}^i} O_{\mathrm{RENE}}^i \right) V_{i,j}^{-1} \left( O_{\mathrm{RENO}}^j - \frac{E_{\mathrm{RENO}}^j}{E_{\mathrm{RENE}}^j} O_{\mathrm{RENE}}^j \right), \\
V_{i,j} =& V_{\mathrm{RENO}}^{i,j} + \left( \frac{E_{\mathrm{RENO}}^i}{E_{\mathrm{RENE}}^i} \right) \left( \frac{E_{\mathrm{RENO}}^j}{E_{\mathrm{RENE}}^j} \right) V_{\mathrm{RENE}}^{i,j}
\end{aligned}
\label{eq:chi2}
\end{equation}

Here, $O^i$ represents the observed energy spectrum, while $E^i$ denotes the expected spectrum assuming a given set of sterile neutrino oscillation parameters.
The covariance matrix $V_{ij}$ includes both statistical and systematic uncertainties from RENO and RENE detectors.  
$V_{\mathrm{RENO}}^{i,j}$ is taken from RENO measurements, and $V_{\mathrm{RENE}}^{i,j}$ is derived by rescaling the NEOS covariance matrix~\cite{NEOSRENO1} to match RENE's expected statistics.  
The NEOS matrix accounts for systematic uncertainties such as energy scale, normalization, detector response, and backgrounds including accidental and cosmic-ray events,  
thus effectively incorporating these contributions into the RENE sensitivity analysis.

The top panel of Fig.~\ref{fig:sensitivity} presents the expected detector sensitivity results highlighting the region where sterile neutrino oscillations could cause detectable distortions in the neutrino energy spectrum. The displayed sensitivity projections assume a data-taking period ranging from 0.5 to 2 years.
In this context, increasing the amount of data will improve our understanding of the target parameter space.

The projected sensitivity of the RENE experiment is expected to surpass previous results from the  RENO  and  NEOS  collaborations,  as illustrated in  Fig.~\ref{fig:sensitivity}. By employing a more compact detector design with enhanced energy resolution, RENE can provide more precise measurements of the oscillation parameters, particularly in the $\sin^{2}(2\theta_{14})$ region, where current constraints remain less stringent.

Notably, the improved sensitivity of the RENO/RENE joint analysis primarily arises from RENE's superior energy resolution rather than increased statistics.  
Although RENE has a smaller volume and thus about five times fewer events than NEOS, its approximately 4\% energy resolution at 5$\,$MeV allows for more accurate reconstruction of oscillation-induced spectral features.  
Comparisons under equal statistics but varying energy resolutions confirm that the sensitivity gain stems mainly from enhanced detector resolution.  
This underscores the advantage of the RENE detector design in sterile neutrino searches.

\newpage
\section{Project schedule and Conclusion}
\label{sec::summary_plan}
%{\color{red}Goh: Some changes applied,
%\begin{itemize}
%    \item Change section title: Summary and Plan $\to$ Project schedule and Conclusion
%    \item Switch the order, schedule first then go to the conclusion for overall summary
%    \item remove sentence "This section provides a bried summary...
%\end{itemize}
%}

%This section provides a brief summary of the RENE and outlines the overall project schedule.

\subsection{Project Schedule}
\begin{figure}[hpt]
\centering
\includegraphics[scale=0.46]{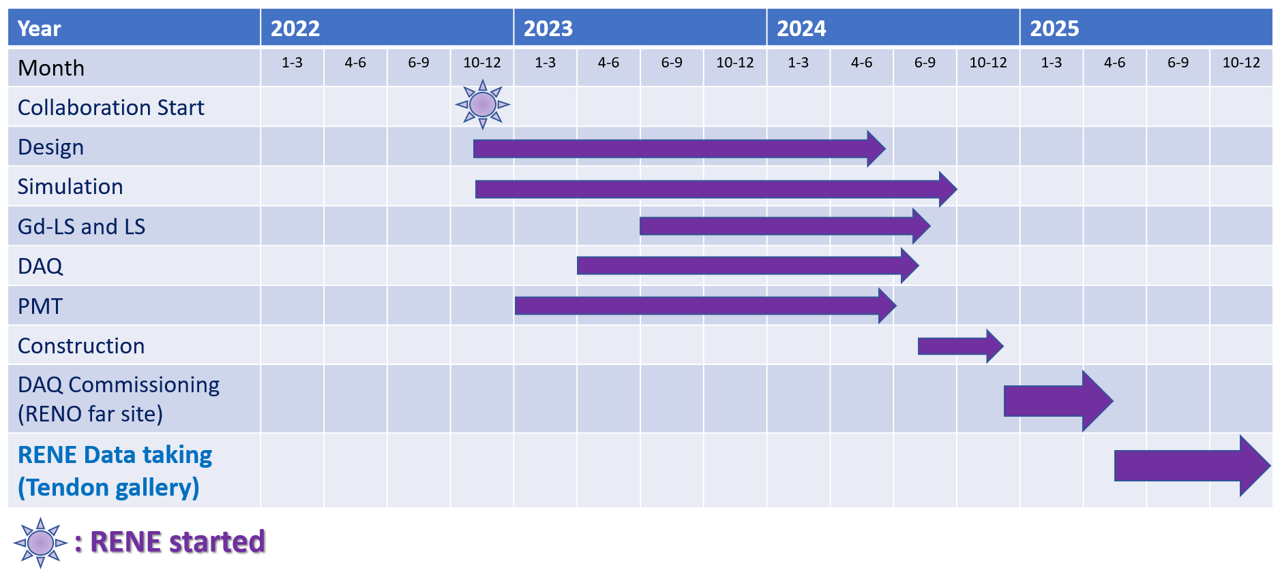}
\caption{\setlength{\baselineskip}{4mm} Timeline of RENE experiment.} 
\label{fig:plan}
\end{figure}

The RENE project has made rapid progress since the collaboration's initiation in late 2022, as presented in Fig.~\ref{fig:plan}.
The project has undergone the rapid development including project planning and detector design. Construction of the RENE detector was completed in mid-winter 2024 and commissioning is currently underway.
The processes of DAQ condition tuning, calibration, and slow monitoring are in their final stages. The RENE detector is planned for
installation at the tendon gallery of the Yeonggwang Hanbit reactor site by the end of 2024 with data-taking operation expected to commence shortly afterward.

\subsection{Conclusion}

The RENE experiment, with a baseline distance of approximately 24$\,$m, is designed to search for sterile neutrinos. This study presents a detailed technical overview of the RENE detector. The incorporation of a $\gamma$-catcher chamber in the RENE detector improves energy resolution compared to the previous NEOS experiment by enabling the capture of escaping $\gamma$-rays. This enhancement enables precise measurements of prompt energy distributions. The detector is scheduled for installation at the tendon gallery in the summer of 2025. Once operational, it is expected to detect several hundred IBD events per day from the Yeonggwang Hanbit reactor. The experiment’s sensitivity will depend on the evaluated systematic uncertainties and background levels. 
This study is anticipated to provide more precise theoretical constraints, contributing to a deeper understanding of sterile neutrinos. According to a Monte Carlo study based on improved energy resolution, it is expected that two years of data will allow a full exploration of the parameter space covered by the RENO/NEOS joint analysis.

\ack

%This research was supported by a grant and Global-Learning & Academic research institution for Master’s · PhD students, and Postdocs(LAMP) Program of the National Research Foundation of Korea(NRF) grant funded by the Ministry of Education (No. RS-2022-NR070836, RS-2022-NR069287, NRF-2020R1I1A3066835, NRF-2021R1A2C2012584, NRF-2022R1A3B1078756, RS-2023-00212787, RS-2023-00211807, and RS-2024-00442775)

This research was supported by a grant from the National Research Foundation of Korea 
(No. RS-2022-NR070836, RS-2022-NR069287, NRF-2020R1I1A3066835, NRF-2021R1A2C2012584, NRF-2022R1A3B1078756, RS-2023-00212787, RS-2023-00211807, and RS-2024-00442775).

%\bibliographystyle{unsrt}
%\bibliography{bib}

\end{document}